\documentclass[australian,english,prl, singlespace, twocolumn]{revtex4-1}
\usepackage[T1]{fontenc}
\usepackage[latin9]{inputenc}
\usepackage{xcolor}
\usepackage{amsmath}
\usepackage{amssymb}
\usepackage{graphicx}

\makeatletter
\@ifundefined{definecolor}
 {\usepackage{color}}{}
\makeatother

\makeatother

\usepackage{babel}
\begin{document}
\title{Macroscopic delayed-choice and retrocausality: quantum eraser, Leggett-Garg
and dimension witness tests with cat states}
\author{Manushan Thenabadu and M. D. Reid$^{1}$}
\affiliation{$^{1}$ Centre for Quantum Science and Technologies Theory, Swinburne
University of Technology, Melbourne 3122, Australia}
\begin{abstract}
We propose delayed choice experiments carried out with macroscopic
qubits, realised as macroscopically-distinct coherent states $|\alpha\rangle$
and $|-\alpha\rangle$. Quantum superpositions of $|\alpha\rangle$
and $|-\alpha\rangle$ are created via a unitary interaction $U(\theta)$
based on a nonlinear Hamiltonian, in analogy with polarising beam
splitters used in photonic experiments. Macroscopic delayed-choice
experiments give a compelling reason to develop interpretations not
allowing macroscopic retrocausality (MrC). We therefore consider weak
macroscopic realism (wMR), which specifies a hidden variable $\lambda_{\theta}$
to determine the macroscopic qubit value  (analogous to 'which-way'
information), independent of any future measurement setting $\phi$.
Using entangled cat states, we demonstrate a quantum eraser where
the choice to measure a which-way or wave-type property is delayed.
Consistency with wMR is possible, if we interpret the macroscopic
qubit value to be determined by $\lambda_{\theta}$ without specification
of the state at the level of order $\hbar$, where fringes manifest.
We then demonstrate violations of a delayed-choice Leggett-Garg inequality,
and of the dimension witness inequality applied to the Wheeler-Chaves-Lemos-Pienaar
experiment, where  measurements need only distinguish the macroscopic
qubit states. This negates all two-dimensional non-retrocausal models,
thereby suggesting MrC. However, one can interpret consistently
with wMR, thus avoiding conclusions of MrC, by noting extra dimensions,
and by noting that the violations require further unitary dynamics
$U$ for each system. The violations are then explained as failure
of   deterministic macroscopic realism (dMR), which specifies validity
of $\lambda_{\theta}$  \emph{prior}  to  the dynamics $U(\theta)$
determining the measurement setting $\theta$. Finally, although
there is consistency with wMR for macroscopic observations, we demonstrate
Einstein-Podolsky-Rosen-type paradoxes at a microscopic level, based
on fringe distributions.
\end{abstract}
\maketitle

\section{Introduction}

Gedanken experiments in which there is a delayed choice of measurement
motivated Wheeler and others to consider whether quantum mechanics
implied a failure of realism, or else retrocausality \cite{wheeler-retro,wheeler-retro-2,ma-zeilinger-rmp-delay-choice}.
The central argument has been presented for the two-slit experiment,
in which a photon travels through the slits exhibiting either particle
or wave-like behaviour. The observation of an interference pattern
is interpreted as wave-like behaviour, while the observation that
the photon travelled along a single path is interpreted as particle-like
behaviour. A similar proposal exists for a Mach-Zehnder (MZ) interferometer,
in which the photon travels in one or other path associated with the
outputs of a beam splitter \cite{wheeler-retro,wheeler-retro-2}.
In the delayed-choice quantum eraser \cite{delayed-choice-qubit-scully-druhl},
the decision to observe either the wave or particle-like behaviour
is delayed until after the photon has passed through the apparatus,
and the fringe distribution vanishes or emerges, conditionally on
the measurement made at the later time. Thus there is an apparently
paradoxical situation whereby it seems as though whether the photon
went through ``both slits'' or ``one slit'' can be changed by
an event (the choice of measurement) in the future.

Multiple different refinements and interpretations have been given
\cite{scully-englert-walther,kim-scully-entangled-quantum-eraser-experiment,walborn-double-slit-quantum-eraser-experiment,herzog-zeilinger-quantum-eraser-complementarity-exp,jacques-aspect-experiment-wheeler,truscott-atoms-delayed-choice,tang-exp-wheeler-delayed-choice,Ma-zeilinger-exp-causal-delayed-choice-interpretation,englert-scully-walther-double-slit-am-journ-phy,mohrhoff-delayed-choice-interpret-am-journ-phy,ingraham-nonlocality-delay-choice,ma-zeilinger-rmp-delay-choice,kastner-quantum-eraser-interpret-found-phy,la-cour-yudichak-classical-quantum-eraser-dim-wit,Ionicioiu-terno,IT-ent1,IT-ent-2,ITent3,quantum-bs-I-terno-experiment,kaiser-coudreau-milman-experiment-ent-delayed-choice,zheng-quantum-delayed-choiceBS-exp,delayed-choice-causal-model-chaves,delayed-choice-experiment-chaves,huang-delayed-choice-causal-model-compatibility,faetti-quantum-eraser-ent-interpretation-1},
but the consensus is that the original delayed-choice experiments
do not imply the need for retrocausality. The above paradox arises
only if one views the system as being either a wave or particle. The
work of \foreignlanguage{australian}{Ionicioiu} and Terno proposed
a quantum beam splitter \cite{Ionicioiu-terno}, which would place
the system in a quantum superposition of wave- and particle-like states.
An intermediate regime can be quantified, and a class of hidden variable
theories based on the assumption of either wave- or particle-like
behaviour can be negated \cite{Ionicioiu-terno,IT-ent1}. Significantly,
Chaves, Lemos and Pienaar (CLP) resolved these issues further by explicitly
constructing a two-dimensional causal model to explain the original
MZ delayed-choice experiment \cite{delayed-choice-causal-model-chaves},
thus ruling out any need for retrocausal explanations.

On the other hand, with the inclusion of an additional phase shift
in the MZ interferometer, CLP demonstrated that a two-dimensional
classical model would need to be retrocausal to explain the predicted
observations, which lead to a violation of a dimension witness inequality.
Recent experiments confirm these predictions \cite{delayed-choice-experiment-chaves,huang-delayed-choice-causal-model-compatibility}.
In their analysis, the meaning of ``non-retrocausal'' is that, in
a model which assumes realism, hidden variables $\lambda$ associated
with the preparation state are independent of any future measurement
setting, $\phi$. La Cour and Yudichak recently give a model which
is nonretrocausal, but possesses extra dimensions \cite{la-cour-yudichak-classical-quantum-eraser-dim-wit}.
Their model however is based on stochastic electrodynamics, which
is not equivalent to quantum mechanics.

In this paper, we propose and analyse \emph{macroscopic} versions
of delayed-choice experiments. Our results demonstrate that delayed-choice
paradoxes and the causal-modelling tests of CLP are evident at a macroscopic
level, beyond $\hbar$, without the need for a microscopic resolution
of measurement outcomes.  Since retrocausality is more paradoxical
at a macroscopic level, we argue that this strengthens the need to
explain the results of the experiments without invoking retrocausality.

Specifically, we follow \cite{manushan-cat-lg,manushan-bell-cat-lg,macro-bell-lg}
and map from a microscopic to a macroscopic regime, where spin qubits
$|\uparrow\rangle$ and $|\downarrow\rangle$ are realised as macroscopically
distinct coherent states $|\alpha\rangle$ and $|-\alpha\rangle$
($\alpha$ is large), that form a macroscopic qubit. The qubit values
$\eta$ of $+1$ and $-1$ corresponding to the coherent states $|\alpha\rangle$
and $|-\alpha\rangle$ can be distinguished by a measurement of the
field quadrature amplitude $X$, without the need to resolve at the
level of $\hbar$. In analogy with a polarising beam splitter (PBS)
used in the photonic experiments, superpositions of the two coherent
states (called cat states) \cite{manushan-cat-lg,manushan-bell-cat-lg,macro-bell-lg,yurke-stoler-1}
\begin{equation}
\cos\theta|\alpha\rangle+i\sin\theta|-\alpha\rangle\label{eq:sup}
\end{equation}
can be created using a unitary interaction $U(t)\equiv U(\theta)$
based on a nonlinear Hamiltonian, $H_{NL}$. The value of $t$ determines
$\theta$ and hence the probability amplitudes of the two-state superposition.
This provides a mechanism for a direct mapping from the microscopic
to macroscopic delayed-choice experiments.

To analyse quantitatively, we seek to define \emph{macroscopic retrocausality}.
In analyses of the delayed-choice experiments, the meaning of retrocausality
is intertwined with that of realism. Our approach is therefore closely
linked to that of Leggett and Garg \cite{legggarg-1} and indeed we
propose a delayed-choice version of the Leggett-Garg violation of
macro-realism, showing violation of Leggett-Garg inequalities uisng
cat states. Following Leggett and Garg, we define macroscopic realism:\emph{
Macroscopic realism} asserts a predetermination of the outcome for
a measurement of the macroscopic qubit value $\eta$ (the sign of
$X$), for the system prepared at time $t_{M}$ in a superposition
(1). The variable $\lambda_{M}$ describes the macroscopic state of
the system at the time $t_{M}$ and its value gives the outcome of
the qubit measurement $\eta$. Since the value does not require a
microscopic resolution, the validity of $\lambda_{M}$ at the given
time $t_{M}$ is a \emph{very} \emph{weak} assumption, compared to
the assumption of Bell's local hidden-variable states $\{\lambda_{i}\}$
\cite{bell-3} which give a realistic description for any measurement,
including those that are microscopically resolved. The cat-state analysis
enables a clear distinction between the macroscopic hidden variable
$\lambda_{M}$ and the more general hidden-variable states $\{\lambda_{i}\}$.

However, recent work establishes the need to also carefully consider
whether the unitary rotation $U(\theta)$ associated with the measurement
setting $\theta$ has been performed (or not) prior to $t_{M}$. This
leads to two definitions of macroscopic realism, one of which (deterministic
macroscopic realism) can be falsified \cite{macro-bell-lg,manushan-bell-cat-lg}.
Deterministic macroscopic realism (dMR) asserts a predetermination
of outcomes, at a time $t$, for multiple future choices of $\theta$
(e.g. $\theta_{1}$ and $\theta_{2}=\phi$), so that these outcomes
are given by multiple hidden variables (e.g. $\lambda_{M1}$ and $\lambda_{M2}$)
simultaneously specified at $t$.

To examine macroscopic retrocausality, we therefore consider \emph{weak
macroscopic realism} (wMR) \cite{manushan-bell-cat-lg}: Weak macroscopic
realism asserts that the system prepared at time $t_{M}$ in a superposition
(1) is in a state giving a definite outcome $\lambda_{M}$ ($\lambda_{M}$
being $+1$ or $-1$) for the macroscopic pointer qubit measurement
$\eta$. It is implicit as part of the definition that the value
$\lambda_{M}$ be independent of any future measurement setting, $\phi$.
We use the term \emph{pointer measurement}, because it is assumed
that the unitary rotation $U(\theta)$ determining the measurement
setting has already been performed, \emph{prior} to $t_{M}$ i.e.
the system has been prepared in the appropriate basis, $\theta$.

In this paper, we examine the unitary dynamics $U(t)$, showing that
at certain times $t_{M}$\emph{ }the assumption of $\lambda_{M}$
is relevant, because the system is in a two-state superposition (\ref{eq:sup}).
During the dynamics, however, the state of the system has a more general
form than (1). Extra dimensions are evident in the quantum continuous-variable
phase-space representations for the system. Thus, it is possible
to argue consistently with the Chaves-Lemos-Pienaar (CLP) analysis
that $\lambda_{M}$ is valid at the time $t_{M}$, and hence that
there is no macroscopic retrocausality. Instead, the violations of
the dimensions witness inequality and delayed-choice Leggett-Garg
inequalities reflect the failure of dMR, which (it is argued) arises
from failure of Bell-type hidden variables $\{\lambda_{i}\}$ defined
microscopically.

In summary, the main results of this paper are two-fold: First, we
demonstrate the possibility of performing macroscopic delayed-choice
tests using cat states, including those of the type previously proposed
at a microscopic level by CLP. Second, we explain how these predictions
can be viewed consistently with wMR, thus providing a counter argument
to any conclusions of macroscopic retrocausality. Lastly, although
there is no inconsistency with wMR at a macroscopic level, we point
out EPR-type paradoxes \cite{epr-1,manushan-bell-cat-lg} giving inconsistencies
with the completeness of quantum mechanics at a microscopic level,
where fringes in the distributions are evident.

\section{\emph{}summary of paper}

The paper is organised as follows. In section III, we consider a quantum
eraser experiment. In analogy with experiments using entangled states
\cite{kim-scully-entangled-quantum-eraser-experiment,kaiser-coudreau-milman-experiment-ent-delayed-choice,ma-zeilinger-rmp-delay-choice},
the system is prepared at $t_{1}=0$ in a two-mode entangled cat state
$\sim|\alpha\rangle_{a}|-\beta\rangle_{b}-|-\alpha\rangle_{a}|\beta\rangle_{b}$.
We identify the qubit value $\eta_{i}$ at time $t_{i}$ as ``which-way''
information. The qubit value for $a$ can be determined by a quadrature
measurement $X_{B}$ of the mode $b$, and the interference for system
$a$ created by interacting locally according to $U_{A}(t_{2})$ for
a specific time $t_{2}$. Similarly, one may apply $U_{B}(t_{2})$
for $b$. The loss of which-way information is identified by fringes
in the distributions of the orthogonal quadrature $P_{A}$ for $a$.
However, we conclude there is no paradox involving macroscopic retrocausality,
since the fringes are only distinguished at the level of $\hbar$.
The results can be viewed consistently with weak macroscopic realism.
Nonetheless, in Section VI we show that at the \emph{microscopic}
level of $\hbar$, EPR-type paradoxes can be constructed (similar
to those discussed in \cite{macro-coherence-paradox,manushan-bell-cat-lg}),
based on the fringe pattern.

We turn to macroscopic paradoxes, in Section IV, by presenting tests
where it is only necessary to measure the macroscopic qubit value
$\eta$. We show violation of a Leggett-Garg inequality, where one
measures $\eta_{i}$ at three times $t_{i}$ ($t_{3}>t_{2}>t_{1}$).
The violation reveals failure of the joint assumptions of \emph{weak
macroscopic realism} (wMR) and \emph{noninvasive measurability} (called
\emph{macrorealism}). Noninvasive measurability asserts that one can
determine the value of $\lambda_{i}$ for the system satisfying wMR,
in a way that does not affect the future $\lambda_{j}$ ($j>i$).
In the present proposal, the measurement of $\lambda_{i}$ ($i=1,2$)
of $a$ is justified to be noninvasive because it is performed on
the spacelike separated system $b$ and, furthermore, the choice of
which measurement ($\lambda_{1}$ or $\lambda_{2}$) to make  is
delayed, until after $t_{3}$. A natural interpretion is that the
measurement of $\lambda_{2}$ (or $\lambda_{1}$) disturbs the dynamics
to affect the result for $\lambda_{3}$, therefore violating macrorealism.
Since there is a delayed choice, this suggests macroscopic retrocausality.
In Section V, we follow the Chaves-Lemos-Pienaar experiment, demonstrating
violation of the dimension witness inequality for cat states, thus
falsifying all two-dimensional non-retrocausal models \cite{delayed-choice-causal-model-chaves}.
This consolidates the work of \cite{manushan-bell-cat-lg}, which
outlined the possibility of delayed-choice experiments with cat states.

To counter conclusions of macroscopic retrocausality, in Section IV.C
we give an interpretation of the violations of the delayed-choice
Leggett-Garg inequalities that is consistent with wMR: The apparent
macroscopic retrocausality comes about because of the entanglement
with the meter system $b$ at the time $t_{2}$, and the macroscopic
nonlocality associated with the dynamics of the unitary rotations,
when such rotations occur for \emph{both} systems after the time $t_{2}$.
Using phase-space depictions of $P(X_{A},X_{B})$, we identify extra
dimensions not present in the two-dimensional non-retrocausal models.
We show that the violation of the Leggett-Garg inequalities certifies
a failure of \emph{deterministic macroscopic realism} (dMR), but not
wMR (which is a weaker assumption than dMR). A similar explanation
is given in Section V to explain the violation of the dimension witness
inequality. 

\section{Delayed-choice quantum eraser with entangled cat states}

\subsection{Set-up}

We begin by presenting an analogue of the delayed-choice quantum eraser
experiment, for cat states. We consider the variant that uses two
spatially separated systems $A$ and $B$. The overall system is prepared
at time $t_{1}=0$ in the entangled cat Bell state \cite{cat-bell-wang-1}
\begin{equation}
|\psi_{Bell}(t_{1})\rangle=\mathcal{N}\{|\alpha\rangle|-\beta\rangle-|-\alpha\rangle|\beta\rangle\}\label{eq:supt1}
\end{equation}
where $|\alpha\rangle$ and $|\beta\rangle$ are coherent states for
single-mode systems $A$ and $B$. We take $\alpha$ and $\beta$
to be real, positive and large. Here, $\mathcal{N}=\frac{1}{\sqrt{2}}\{1-\exp(-2\left|\alpha\right|^{2}-2\left|\beta\right|^{2})\}^{-1/2}$
is the normalisation constant.

For each system, one may measure the field quadrature phase amplitudes\textcolor{black}{{}
$\hat{X}_{A}={\color{red}{\color{blue}{\color{black}\frac{1}{\sqrt{2}}}}}(\hat{a}+\hat{a}^{\dagger})$,
$\hat{P}_{A}={\color{red}{\color{blue}{\color{black}\frac{1}{i\sqrt{2}}}}}(\hat{a}-\hat{a}^{\dagger})$,
$\hat{X}_{B}=\frac{1}{\sqrt{2}}(\hat{b}+\hat{b}^{\dagger})$ and $\hat{P}_{A}={\color{red}{\color{blue}{\color{black}\frac{1}{i\sqrt{2}}}}}(\hat{a}-\hat{a}^{\dagger})$,
which are defined in a rotating frame, with units so that $\hbar=1$
\cite{yurke-stoler-1}. The boson destruction mode operators for modes
$A$ and $B$ are denoted by $\hat{a}$ and $\hat{b}$, respectively.
The outcome $X_{A}$ of the measurement $\hat{X}_{A}$ distinguishes
between the states $|\alpha\rangle$ and $|-\alpha\rangle$, and similarly
$\hat{X}_{B}$ distinguishes between the states $|\beta\rangle$ and
$|-\beta\rangle$. We define the outcome of the measurement }$\hat{S}^{(A)}$\textcolor{black}{{}
to be $S^{(A)}=+1$ if $X_{A}>0$, and $-1$ otherwise. Similarly,
the outcome of the measurement }$\hat{S}^{(B)}$ \textcolor{black}{is
$S^{(B)}=+1$ if $X_{B}>0$, and $-1$ otherwise.} $S$ is identified
as the spin of the system i.e. the qubit value $\eta$.

The coherent states of $A$ and $B$ become orthogonal in the limit
of large $\alpha$ and $\beta$, in which case the superposition (\ref{eq:supt1})
maps onto the two-qubit Bell state 
\begin{equation}
|\psi_{Bell}\rangle=\frac{1}{\sqrt{2}}(|+\rangle_{a}|-\rangle_{b}-|-\rangle_{a}|+\rangle_{b})\label{eq:sup-bell}
\end{equation}
At time $t_{1}$, the outcomes for $S^{(A)}$ and $S^{(B)}$ are anti-correlated.
Therefore, one may infer the outcome for $S^{(A)}$ noninvasively
by measuring $X_{B}$, and hence $S^{(B)}$. 

We present an analogy with the delayed-choice quantum eraser based
on the photonic versions of the state (\ref{eq:sup-bell}). In the
photonic version, the next step is that the photon of system $A$
propagates through two slits, or else through a 50/50 beam splitter
($BS1$) with two equally probable output paths as in a Mach-Zehnder
(MZ) interferometer. If a single photon is incident on $BS1$, this
creates a superposition e.g. for mode $A$, the state $|+\rangle_{a}$
is transformed to 
\begin{equation}
|\psi\rangle_{a,2}=\frac{1}{\sqrt{2}}(|+\rangle_{a,2}+i|-\rangle_{a,2})\label{eq:su-bs1}
\end{equation}
where $|+\rangle_{a,2}$ and $|-\rangle_{a,2}$ refer to the photon
in paths designated $+$ or $-$ in the MZ interferometer. In the
original quantum eraser, the measurement of which-way information
is made by measuring whether the system is $+$ or $-$. This is done
by recombining the paths using a second beam splitter $BS2$, which
is set to be fully transmitting so that the paths are not mixed. An
alternative choice is that $BS2$ is similar to $BS1$ with a 50\%
transmittivity, which restores the state $|+\rangle$, the photon
appearing only at one of the output paths, indicating interference.

In the cat-state gedanken experiment, the superposition (\ref{eq:su-bs1})
is achieved by a unitary interaction $U(t)$ for a particular choice
$t=t_{3}$. We consider $t_{2}<t_{3}$ in the next Section. After
preparation at the time $t_{1}$, the systems $A$ and $B$ evolve
independently according to the local unitary transformations $U_{A}(t_{a})$
and $U_{B}(t_{b})$, defined by
\begin{equation}
U_{A}(t_{a})=e^{-iH_{NL}^{(A)}t_{a}/\hbar},\ \ U_{B}(t_{b})=e^{-iH_{NL}^{(B)}t_{b}/\hbar}\label{eq:state5}
\end{equation}
where
\begin{equation}
H_{NL}^{(A)}=\Omega\hat{n}_{a}^{k},\ \ H_{NL}^{(B)}=\Omega\hat{n}_{b}^{k}\label{eq:ham_NL}
\end{equation}
\textcolor{black}{Here $t_{a}$ and $t_{b}$ are the times of evolution
at each site, $k$ is a positive integer, $\hat{n}_{a}=\hat{a}^{\dagger}\hat{a}$
and $\hat{n}_{b}=\hat{b}^{\dagger}\hat{b}$, and $\Omega$ is a constant.
We take $k=2$; or else $k>2$ and $k$ is even. As the systems evolve,
the spin for each can be measured at a given time. We denote the value
of spin $S^{(A)}$ after an interaction time $t_{a}=t_{i}$ to be
$S_{i}^{(A)}$, and the value of the spin $S^{(B)}$ after the interaction
time $t_{b}=t_{j}$ to be $S_{j}^{(B)}$.} The dynamics of the unitary
evolution (\ref{eq:state5}) is well known \cite{yurke-stoler-1,collapse-revival-bec-2,collapse-revival-super-circuit-1}.
If the system $A$ is prepared in a coherent state $|\alpha\rangle$,
then after at time  $t_{a}=t_{3}=\pi/2\Omega$, the state of the
system $A$ is \cite{manushan-cat-lg,manushan-bell-cat-lg,macro-bell-lg}
\begin{eqnarray}
U_{\pi/4}^{(A)}|\alpha\rangle & = & e^{-i\pi/4}\{\cos\pi/4|\alpha\rangle+i\sin\pi/4|-\alpha\rangle\}\label{eq:state3}
\end{eqnarray}
where $U_{\pi/4}^{(A)}=U_{A}(\pi/2\Omega)$. A similar transformation
$U_{\pi/4}^{(B)}$ is defined at $B$ for $t_{b}=t_{3}=\pi/2\Omega$.
We note the state (\ref{eq:state3}) maps onto (\ref{eq:su-bs1}).
The generation of the superposition (\ref{eq:state3}) using $k=2$
has been reported in \cite{collapse-revival-bec-2,collapse-revival-super-circuit-1}.
The system $A$ in the superposition (\ref{eq:state3}) exhibits
interference fringes in the distribution $P(P_{A})$ for $\hat{P}_{A}$
\cite{yurke-stoler-1}.

According to the premise \emph{weak macroscopic realism} (wMR) defined
in the Introduction, at the time $t_{2}$ the system (\ref{eq:state3})
may be regarded as being in one or other of two macroscopically distinguishable
states ($\varphi_{+}$ and $\varphi_{-}$) which have a definite value
$+1$ or $-1$ for the outcome $S_{3}^{(A)}$. While it might be
tempting to identify the states $\varphi_{+}$ and $\varphi_{-}$
as being $|\alpha\rangle$ and $|-\alpha\rangle$, this would be a
full microscopic identification of the states in quantum terms. The
states $\varphi_{+}$ and $\varphi_{-}$ are not be specified to this
precision. The states $\varphi_{+}$ and $\varphi_{-}$ correspond
to distinct values of the \emph{macroscopic} observable $S_{3}^{(A)}$
only. The determination of the value of $S_{3}^{(A)}$ gives the ``which-way''
information in the quantum eraser experiment. If one is able to design
an appropriate macroscopic observable (similar to $S_{3}^{(A)}$)
for the two-slit and MZ scenarios, then the assumption of wMR is analogous
to the interpretation that the particle goes through one slit or the
other in the double slit experiment, or goes through one path or the
other, in the MZ interferometer. This assumption however, does \emph{not}
specify the system to be in \emph{either} state $|+\rangle_{a,2}$
or $|-\rangle_{a,2}$.

If one evolves for a time of $t_{3}=\pi/2\Omega$ at both sites, then
the final state is

\begin{eqnarray}
|\psi_{Bell}(t_{3})\rangle & = & U_{\pi/4}^{(A)}U_{\pi/4}^{(B)}|\psi_{Bell}(t_{1})\rangle\nonumber \\
 & = & \mathcal{N}e^{-i\pi/4}(|\alpha\rangle|-\beta\rangle-|-\alpha\rangle|\beta\rangle)\label{eq:qe}
\end{eqnarray}
which is a Bell state. At the time $t_{3}$, the spin $S_{3}^{(A)}$
of system $A$ can be inferred by measuring $S_{3}^{(B)}$, which
is anticorrelated with the spin $S_{3}^{(A)}$ at $A$. This gives
the which-way information of system $A$ at time $t_{3}$, analogous
to measuring through which slit or path the photon went through in
the original quantum eraser set-ups. Only the absolute interaction
times $t_{a}$ and $t_{b}$ at each site are relevant to the correlation
however, and it is hence possible to \emph{delay} interaction at $B$
until a time $t_{4}$, after the system at $A$ has already interacted.

With this method of measurement of $S_{3}^{(A)}$, the system $A$
has not been directly measured. One can thus make a measurement of
$\hat{P}_{A}$ at the time $t_{3}$. The system (being coupled to
$B$) can be detected as being in one or other state, $\varphi_{+}$
or $\varphi_{-}$, giving $+$ or $-$ outcomes for $S_{3}^{(A)}.$
Which-way information is present and, consistent with that, the distribution
$P(P_{A})$ shows no fringes. This is seen in Figure 1, where we plot
the conditional distributions $P(X_{A})_{\pm}$ and $P(P_{A})_{\pm}$
given the outcome $\pm$ for $X_{B}$ at $B$, as evaluated from the
joint distributions $P(X_{A},X_{B})$ and $P(P_{A},X_{B})$. The distribution
$P(P_{A})_{\pm}$ for an outcome $P_{A}$ for the measurement $\hat{P}_{A}$
is a Gaussian centred at $0$ with no fringes present, consistent
with that of the coherent state $|\pm\alpha\rangle$ \cite{yurke-stoler-1}.
\begin{figure}[t]
\begin{centering}
\includegraphics[width=0.45\columnwidth]{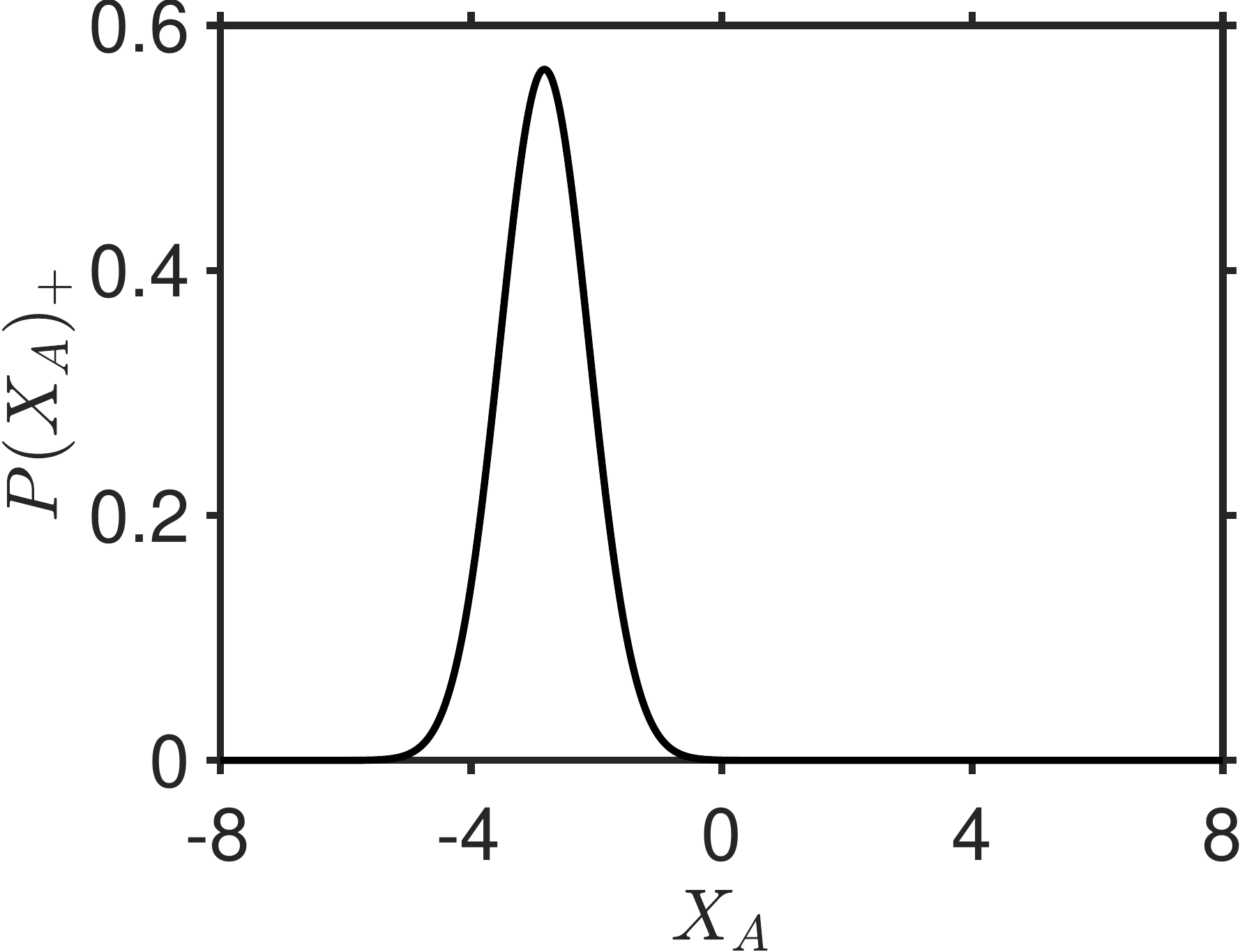}\ \ \includegraphics[width=0.45\columnwidth]{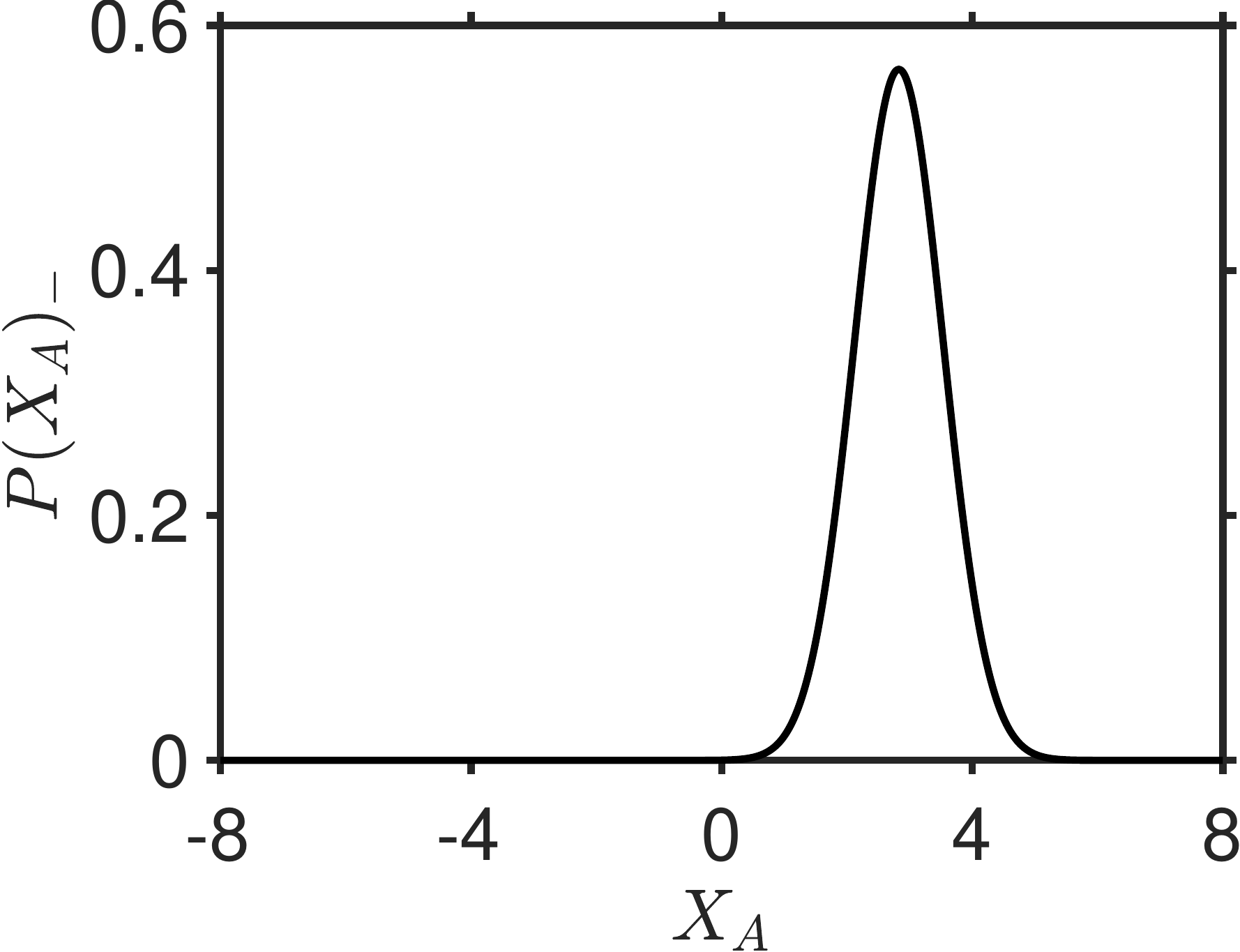}
\par\end{centering}
\medskip{}

\begin{centering}
\includegraphics[width=0.45\columnwidth]{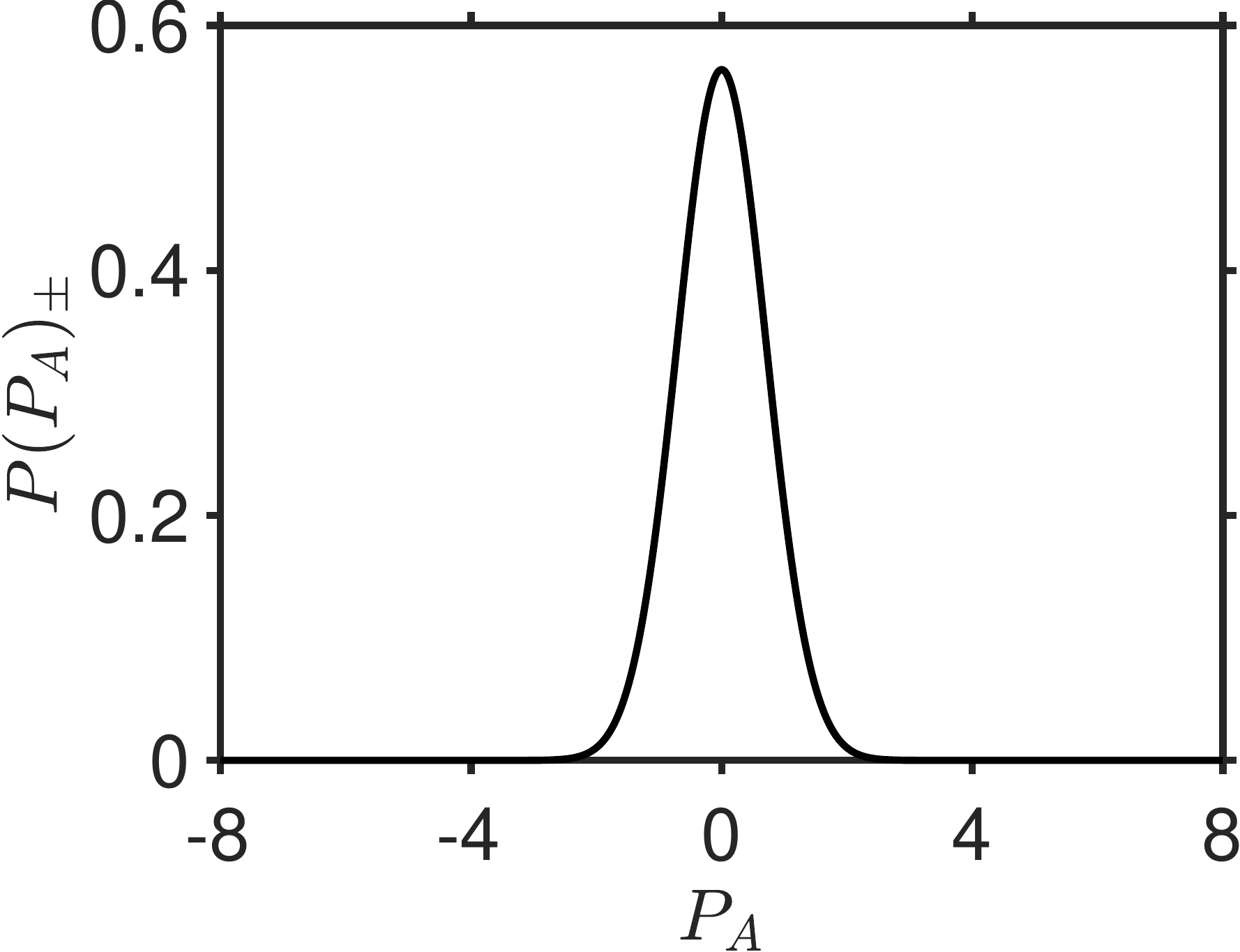}
\par\end{centering}
\begin{centering}
\par\end{centering}
\caption{Top: Plots of $P(X_{A})_{\pm}$ and $P(P_{A})_{\pm}$ for the system
$A$ at time $t_{3}$, when which-way information is present. The
$P(P_{A})_{\pm}$ show no fringes. Here, $\alpha=\beta=2$.\textcolor{red}{}\textcolor{blue}{}\textcolor{red}{\label{fig:qe-which-way}}}
\end{figure}

On the other hand, one may take $t_{a}=t_{3}$ and $t_{b}=0$, so
that there is no local unitary intertaction at $B$. Alternatively,
one may evolve both sites according to $t_{a}=t_{b}=t_{3}$, and then
perform a local unitary transformation $U_{B}(t_{2})^{-1}=(U_{\pi/4}^{(B)})^{-1}$
at $B$, to transform the system $B$ ``back'' to the initial state
of $B$ at time $t_{1}$. Which-way information about $A$ at $t_{3}$
is then absent. The state of the combined systems at this time $t_{4}>t_{3}$
is
\begin{eqnarray}
|\psi(t_{4})\rangle & = & \mathcal{N}\{U_{\pi/4}^{(A)}|\alpha\rangle|-\beta\rangle-U_{\pi/4}^{(A)}|-\alpha\rangle|\beta\rangle\}\label{eq:aftertransP}
\end{eqnarray}
If the final stage of the spin measurement $B$ is made at time $t_{4}$,
the result will give either $S^{(B)}(t_{4})=1$ or $-1$. From the
anti-correlation of (\ref{eq:supt1}), $S^{(B)}(t_{4})$ is interpreted
as a measurement of the initial value of $-S_{1}^{(A)}$, and hence
knowledge of that state of system $A$ at that time, $t_{1}$. If
the outcome of $S^{(B)}(t_{4})$ is $\mp1$ then, assuming the limit
where $|-\beta\rangle$ and $|\beta\rangle$ are orthogonal states
(i.e. large $\beta$), the system $A$ is projected into the superposition
state
\begin{equation}
U_{\pi/4}|\pm\alpha\rangle=e^{-i\pi/4}\{\cos\pi/4|\pm\alpha\rangle+i\sin\pi/4|\mp\alpha\rangle\}\label{eq:supt2-1-2}
\end{equation}
This is the state of the local system $A$ at time $t_{2}$ (see eqn
(\ref{eq:state3})), conditioned on the initial state of $A$ at time
$t_{1}$ being $|\pm\alpha\rangle$. Thus, if one measures $P(P_{A})$
conditional on the result of $-S^{(B)}(t_{4})=S_{1}^{(A)}$, the fringes
are recovered. We find\textcolor{red}{}
\begin{equation}
P(P_{A})_{\pm}=\frac{e^{-P_{A}^{2}}}{\sqrt{\pi}}\{1\mp\sin(2\sqrt{2}P_{A}|\alpha|)\}\label{eq:supfringep}
\end{equation}
where $P(P_{A})_{+}$ and $P(P_{A})_{-}$ is the distribution for
$P_{A}$ conditional on the result $+1$ or $-1$ for $S_{1}^{(A)}$,
respectively. The distributions (Figure 2) show fringes, indicative
of the system $A$ at time $t_{3}$ being in the superposition (\ref{eq:supt2-1-2}),
and indicative of the loss of which-way information.\textcolor{red}{{}
}\textcolor{blue}{}\textcolor{blue}{}\textcolor{blue}{}\textcolor{blue}{}
\begin{figure}[t]
\begin{centering}
\includegraphics[width=0.45\columnwidth]{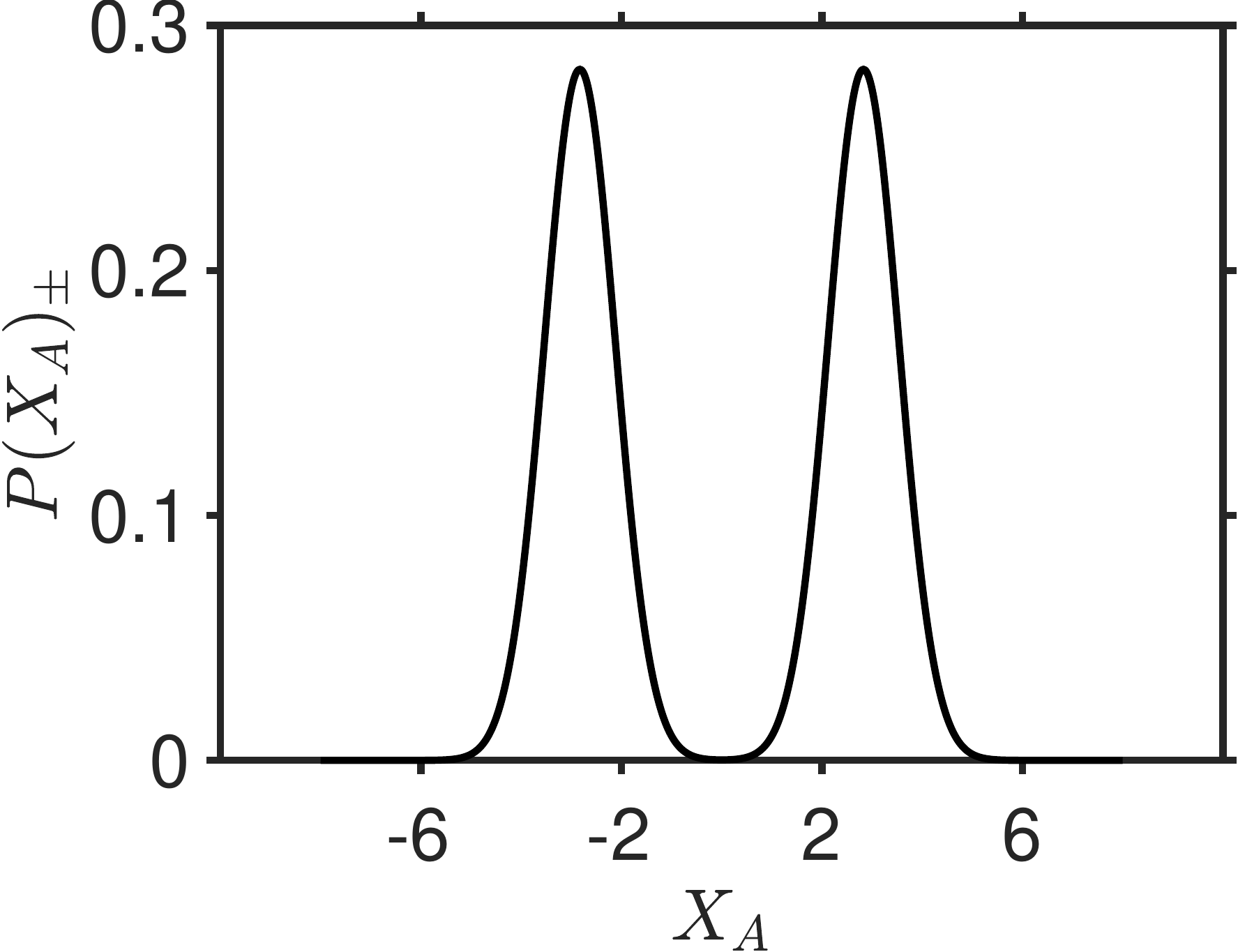}
\par\end{centering}
\medskip{}

\begin{centering}
\includegraphics[width=0.45\columnwidth]{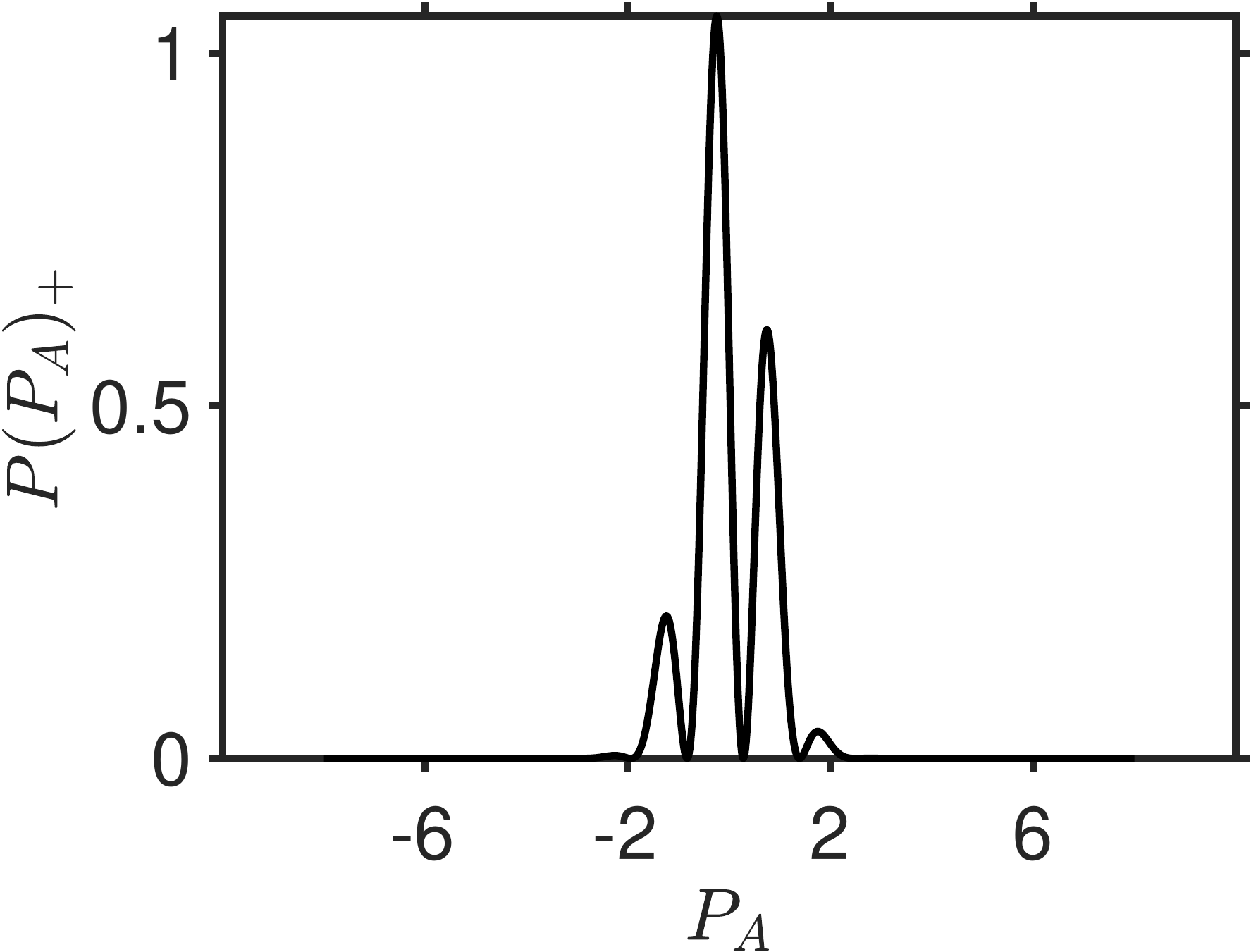}\ \ \includegraphics[width=0.45\columnwidth]{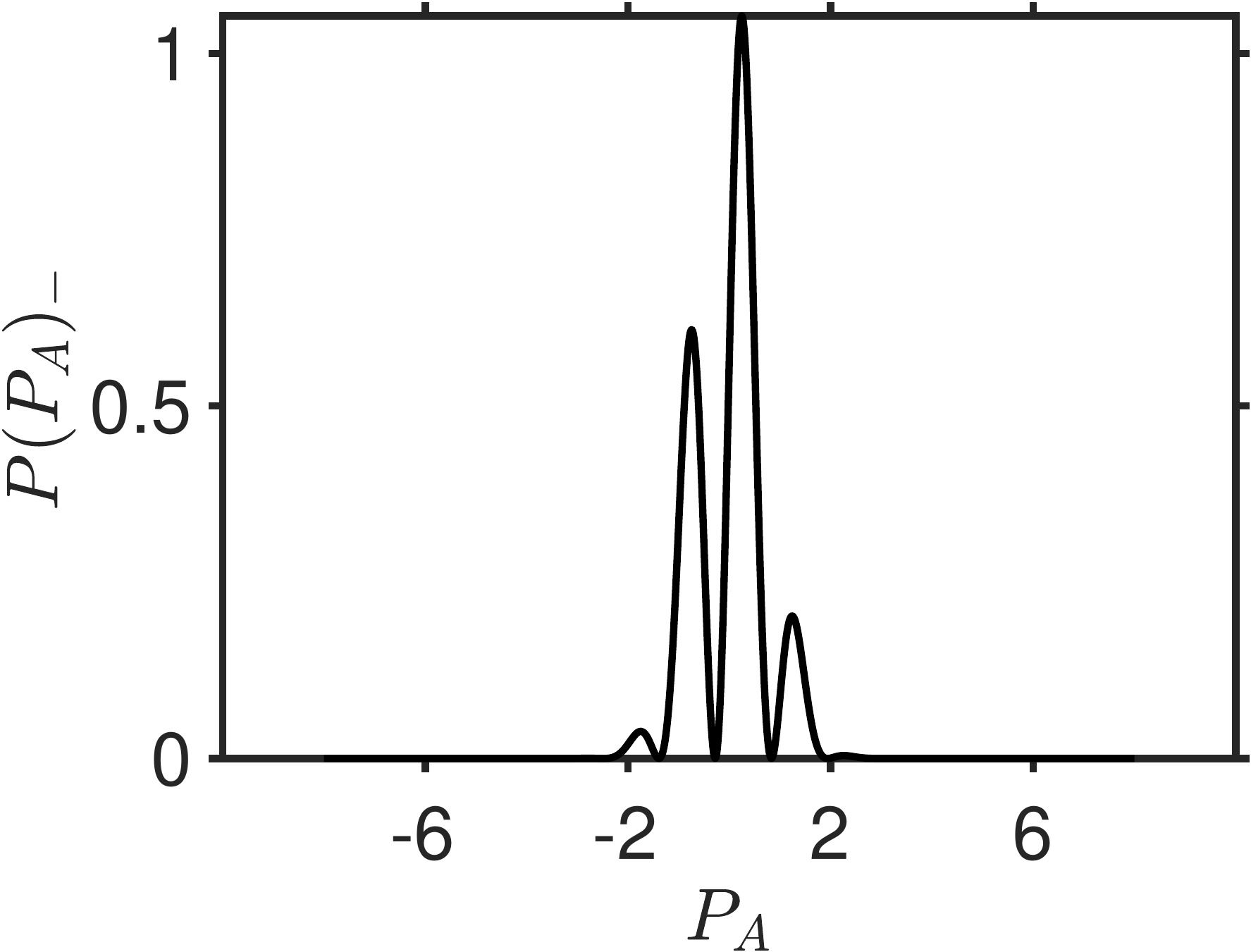}
\par\end{centering}
\begin{centering}
\par\end{centering}
\caption{Top: Plots of $P(X_{A})_{\pm}$ and $P(P_{A})_{\pm}$ of the system
$A$ at time $t_{2}$, where the outcome for $S_{4}^{(B)}=S_{A}^{(1)}$
is (left) $+1$, and (right) $-1$. The which-way information is lost,
and the system $A$ is in the superposition (\ref{eq:supt2-1-2}).
Here, $\alpha=\beta=2$.\textcolor{red}{}\textcolor{blue}{\label{fig:QE-fringe}}}
\end{figure}

The accurate calculation of the conditional probabilities $P(P_{A})_{\pm}$,
without the simplistic assumption of a projection into a definite
coherent state at $A$ on measurement at $B$, gives\textcolor{red}{}
\begin{align}
P(P_{A})_{\pm} & =\frac{2\mathcal{N}^{2}e^{-P_{A}^{2}}}{\sqrt{\pi}}\biggr\{1-e^{-2|\beta|^{2}}\cos(2\sqrt{2}P_{A}|\alpha|)\nonumber \\
 & \mp\sin(2\sqrt{2}P_{A}|\alpha|)erf(\sqrt{2}|\beta|)\biggl\}\label{eq:full-fringe}
\end{align}
where $erf$ is the error function. The plots are indistinguishable
from those of the approximate result for $\beta>1$, the limit $\beta\rightarrow\infty$
being the limit of an ideal measurement. The calculations in Figures
1 and 2 are based on evaluation of the joint distribution $P(P_{A},X_{B})$
(refer to \cite{macro-bell-lg}).\textcolor{red}{}\textcolor{blue}{}

\subsection{Interpretation in terms of wMR}

As summarised in the Introduction, the delayed choice experiment has
been interpreted as suggesting retrocausality. The decision to observe
either the particle-like behaviour (which-way information) or the
wave-like behaviour (fringes) of system $A$ is made at the later
time $t_{4}$ (at $B$). This appears to retrospectively change the
system $A$ at time $t_{3}$ from being in ``one or other state''
($\varphi_{1}$ \emph{or} $\varphi_{2}$; $|\alpha\rangle$ or $|-\alpha\rangle$)
to being ``in both states'' (since the observation of fringes in
$P(P_{A})$ is often interpreted to suggest the system $A$ was in
``both states'', $|\alpha\rangle$ and $|-\alpha\rangle$). As
explained in \cite{delayed-choice-causal-model-chaves}, there is
no requirement to assume retrocausality for the MZ delayed-choice
experiment. The experiment described for cat states maps onto the
qubit experiment for large $\alpha$, $\beta$, and gives a similar
conclusion for the macroscopic qubits.

The macroscopic version of the quantum eraser is informative, because
with the introduction of the \emph{macroscopic} hidden variable, $\lambda_{i}$,
it allows us to separate the macroscopic from the microscopic behavior.
We consider compatibility with the assumption of weak macroscopic
realism (wMR) $-$ that the system $A$ at the time $t_{2}$ is in
a state with a definite value $\lambda_{3}^{(A)}$ which corresponds
to the outcome of a measurement $S_{3}^{(A)}$, should it be performed.
Here, there is \emph{no} attempt to define the \emph{quantum} state
associated with that predetermination, so that predictions for other
more microscopic measurements (and hence other hidden variables that
determine those predictions) are not relevant. Thus, wMR does not
postulate that the system is in one or other state $|\alpha\rangle$
or $|-\alpha\rangle$. In fact, we see there is \emph{no negation
of wMR}, because the fringes are only evident at the microscopic level
of $\hbar$ (here $\hbar\sim1$). The gedanken experiment is consistent
with wMR. In that sense, the system always displays a particle-like
behaviour.

The assumption of weak macroscopic realism (wMR) \emph{if} applied
to the double-slit experiment would be that the particle has a position
constraining it to go through a definite slit even when fringes are
observed (provided the slit does not restrict the position to of order
$\hbar$ or less). For the definition of wMR, the predictions for
other more precise position or momentum measurements of order $\hbar$
are not relevant. A similar interpretation of wMR for the MZ experiment
is that the photon/ particle takes one or other path with a macroscopic
uncertainty, but is not defined to be in one or other state $|+\rangle_{a,2}$
and $|-\rangle_{a,2}$.

The interpretation based on wMR suggests a lack of completeness of
the description at the microscopic level. This can be clarified further.
Indeed, if wMR holds, then it is possible to show that EPR-type paradoxes
exist at the \emph{microscopic} level. The EPR-type arguments indicate
an \emph{incompleteness} of a quantum state description if compatible
with wMR, as explained in \cite{manushan-bell-cat-lg}, and will be
discussed further in Section VI.

\section{Delayed-choice Leggett-Garg test of macrorealism}

In this section, we consider the delayed choice experiment in the
form of a Leggett-Garg test of macrorealism using entangled cat states.
The advantage of the Leggett-Garg test is that all relevant measurements
are macroscopic, distinguishing between the two macroscopically distinct
coherent states. This contrasts with the quantum eraser proposal,
where the paradoxical effects are inferred by the measurement of finely
resolved fringes.

\subsection{Set-up}

At time $t_{1}$, the system is prepared in the entangled cat state
$|\psi_{Bell}(t_{1})\rangle$ of eqn (\ref{eq:supt1}). The spatially
separated systems $A$ and $B$ dynamically evolve according to the
unitary interactions (\ref{eq:state5}) where $k=4$. We consider
three times $t_{1}=0$, $t_{2}=\pi/4\Omega$ and $t_{3}=\pi/2\Omega$
(Figure 3). If the system at $A$ were prepared in a coherent state
$|\alpha\rangle$, then at the later time $t_{a}=t_{2}=\pi/4\Omega$,
the state of the system $A$ at time $t_{2}$ is in the asymmetric
superposition \cite{manushan-cat-lg,macro-bell-lg,manushan-bell-cat-lg}
\begin{eqnarray}
U_{\pi/8}^{(A)}|\alpha\rangle & = & e^{-i\pi/8}\{\cos\pi/8|\alpha\rangle+i\sin\pi/8|-\alpha\rangle\}\nonumber \\
\label{eq:state3-3}
\end{eqnarray}
where $U_{\pi/8}^{(A)}=U_{A}(\pi/4\Omega)$.  A similar transformation
$U_{\pi/8}^{(B)}$ is defined at $B$ for $t_{b}=t_{2}=\pi/4\Omega$.
If one evolves for a time of $t_{2}=\pi/4\Omega$ at both sites, then
the final state is

\begin{eqnarray}
|\psi_{Bell}(t_{2},t_{2})\rangle & = & U_{\pi/8}^{(A)}U_{\pi/8}^{(B)}|\psi_{Bell}(t_{1})\rangle\nonumber \\
 & = & \mathcal{N}e^{-i\pi/4}(|\alpha\rangle|-\beta\rangle-|-\alpha\rangle|\beta\rangle)\nonumber \\
\label{eq:qe-1}
\end{eqnarray}
The values of the macroscopic spins after the interaction time $t_{2}$
at each site are denoted $S_{2}^{(A)}$ and $S_{2}^{(B)}$. The spin
$S_{2}^{(A)}$ of system $A$ can be inferred by measuring $S_{2}^{(B)}$
which is anticorrelated with the spin at $A$.
\begin{figure}[t]
\begin{centering}
\includegraphics[width=1\columnwidth]{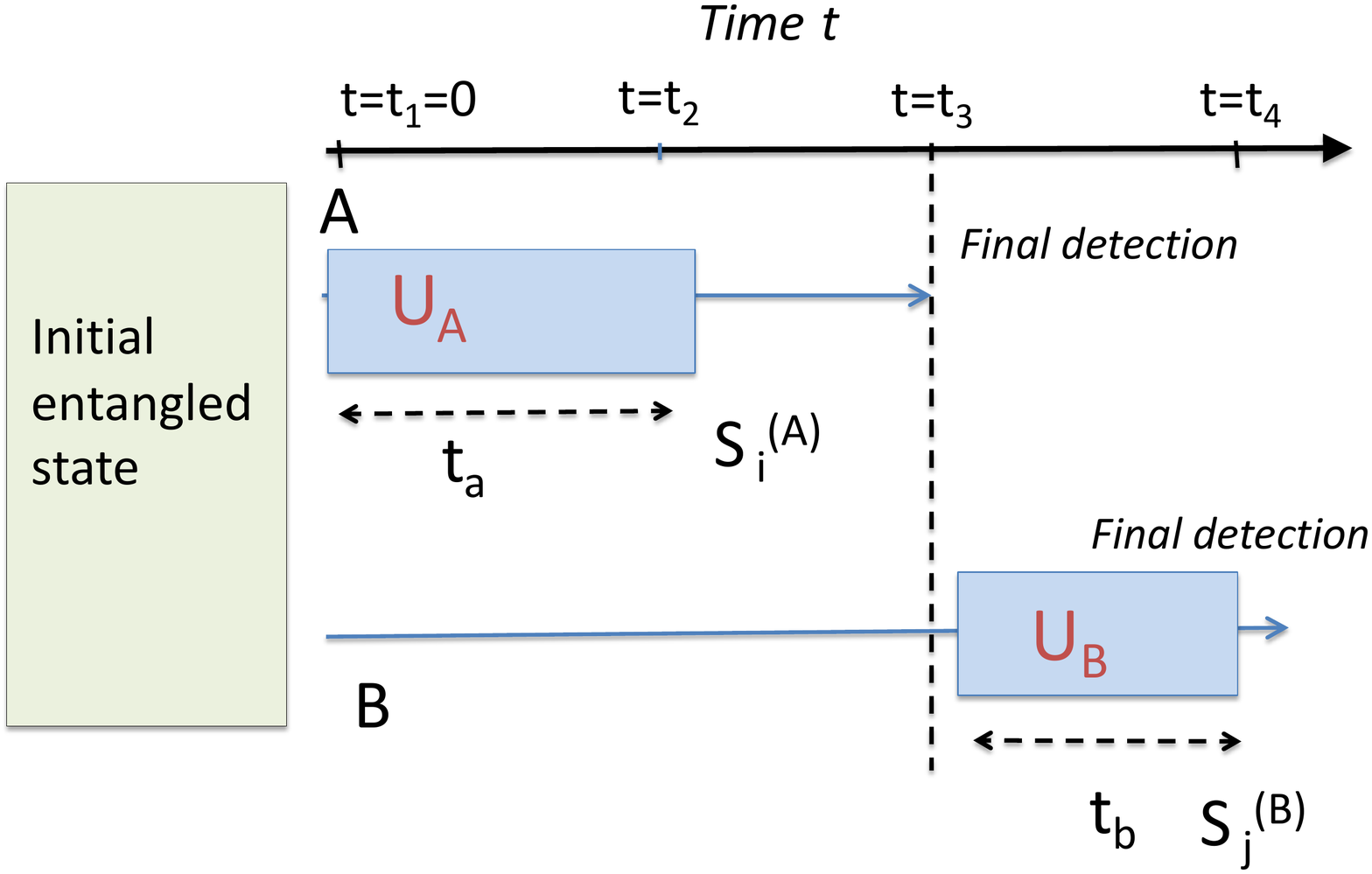}
\par\end{centering}
\caption{Sketch of the set-up for the delayed choice Leggett-Garg test. The
system is prepared in the two-mode entangled cat state $|\psi_{Bell}(t_{1})\rangle$
at the time $t_{1}=0$, with the modes spatially separated. Independent
local unitary interactions $U_{A}$ and $U_{B}$ take place at sites
$A$ and $B$ respectively, with time settings $t_{a}$ and $t_{b}$.
The times at $A$ are selected as either $t_{a}=t_{2}=\pi/4\Omega$
or $t_{a}=t_{3}=\pi/2\Omega$ and the final detection enables measurement
of $S_{2}^{(A)}$ or $S_{3}^{(A)}$ respectively. At $B$, one selects
either $t_{b}=t_{1}=0$ or $t_{b}=t_{2}=\pi/4\Omega$, the final detection
enabling measurement of $S_{1}^{(B)}$ or $S_{2}^{(B)}$. The outcomes
of $S_{1}^{(B)}$and $S_{2}^{(B)}$ are anticorrelated with the outcomes
of $S_{1}^{(A)}$ and $S_{2}^{(A)}$ respectively, if measured. In
the delayed choice experiment, the interaction at $B$ is delayed
until after the final detection at $A$, at time $t_{3}$. Hence,
the measurement of $S_{1}^{(B)}$ (or $S_{2}^{(B)}$) allows inference
of the past value of $S_{1}^{(A)}$ (or $S_{2}^{(A)}$).\label{fig:Sketch-general}}
\end{figure}

On the other hand, one may choose to evolve at $A$ for a time $t_{a}=t_{2}=\pi/4\Omega$,
but \emph{not} at the site $B$, so that $t_{b}=0$. The state
after these interactions is
\begin{eqnarray}
|\psi(t_{2},t_{1})\rangle & = & \mathcal{N}\{U_{\pi/8}^{(A)}|\alpha\rangle|-\beta\rangle-U_{\pi/8}^{(A)}|-\alpha\rangle|\beta\rangle\}\nonumber \\
\label{eq:aftertransP-2}
\end{eqnarray}
If the final readout stage of the spin measurement $B$ is made at
time $t_{4}$ (Figure 3), the result will give a value $S^{(B)}(t_{4})\equiv S_{1}^{(B)}=\pm1$.
From $|\psi_{Bell}(t_{1})\rangle$ (eqn (\ref{eq:supt1})), the value
of $S^{(B)}(t_{4})$ is anticorrelated with the initial value of $S_{1}^{(A)}$,
if we had chosen $t_{a}=t_{1}=0$. Therefore the measurement at $B$
is interpreted as a measurement of $S_{1}^{(A)}$. If the outcome
of $S^{(B)}(t_{4})$ is $\mp1$ then (assuming $|\beta\rangle$ and
$|-\beta\rangle$ are orthogonal) from (\ref{eq:aftertransP}) we
see that the system $A$ is reduced to the superposition state
\begin{equation}
U_{\pi/8}|\pm\alpha\rangle=e^{-i\pi/8}\{\cos\pi/8|\pm\alpha\rangle+i\sin\pi/8|\mp\alpha\rangle\}\label{eq:supt2-1-2-1}
\end{equation}
This is the state of the local system $A$ at time $t_{2}$ (see eqn
(\ref{eq:state3})), conditioned on the initial state of $A$ at time
$t_{1}$ being $|\pm\alpha\rangle$. The value of $S_{2}^{(A)}$ can
be measured directly at $A$. This combination of interactions therefore
allows measurement of both $S_{2}^{(A)}$ and $S_{1}^{(A)}$.

Alternatively, we may evolve the system $A$ for a time $t_{a}=t_{3}=\pi/2\Omega$,
while not evolving at $B$ ($t_{b}=t_{1}=0$). This gives 
\begin{eqnarray}
|\psi(t_{3},t_{1})\rangle & = & \mathcal{N}\{U_{\pi/4}^{(A)}|\alpha\rangle|-\beta\rangle-U_{\pi/4}^{(A)}|-\alpha\rangle|\beta\rangle\}\nonumber \\
\label{eq:aftertransP-2-2}
\end{eqnarray}
where $U_{\pi/4}|\pm\alpha\rangle$ is given by eqn (\ref{eq:state3}).
The spin $S_{3}^{(A)}$ can be measured directly at $A$. Measurement
of $S^{(B)}(t_{4})\equiv S_{1}^{(B)}$  at $B$ gives the inferred
result for the measurement $S_{1}^{(A)}$. This allows measurement
of both $S_{3}^{(A)}$ and $S_{1}^{(A)}$.

Alternatively, one may select $t_{b}=t_{2}=\pi/4\Omega$ at $B$.
According to (\ref{eq:qe-1}), the measurement at $B$ then allows
measurement of $S_{2}^{(A)}$. If one evolves at $A$ for a time $t_{a}=t_{3}=\pi/2\Omega$,
then this combination of interactions allows measurement of both $S_{2}^{(A)}$
and $S_{3}^{(A)}$.

The set-up (Figure 3) allows for a delayed choice of the measurement
of either $S_{1}^{(A)}$ or $S_{2}^{(A)}$, by delaying the choice
at $B$ to measure either $S_{1}^{(B)}$ or $S_{2}^{(B)}$. This amounts
to a delay in the choice to interact the system $B$ for a time $t_{b}=0$,
or else to interact system $B$ for a time $t_{b}=t_{2}$. This choice
can be delayed until a time well after the time $t_{3}$, and well
after the final detection (given by the measurement and readout of
$X_{A}$) takes place at $A$.

\begin{figure}[t]
\includegraphics[width=0.45\columnwidth]{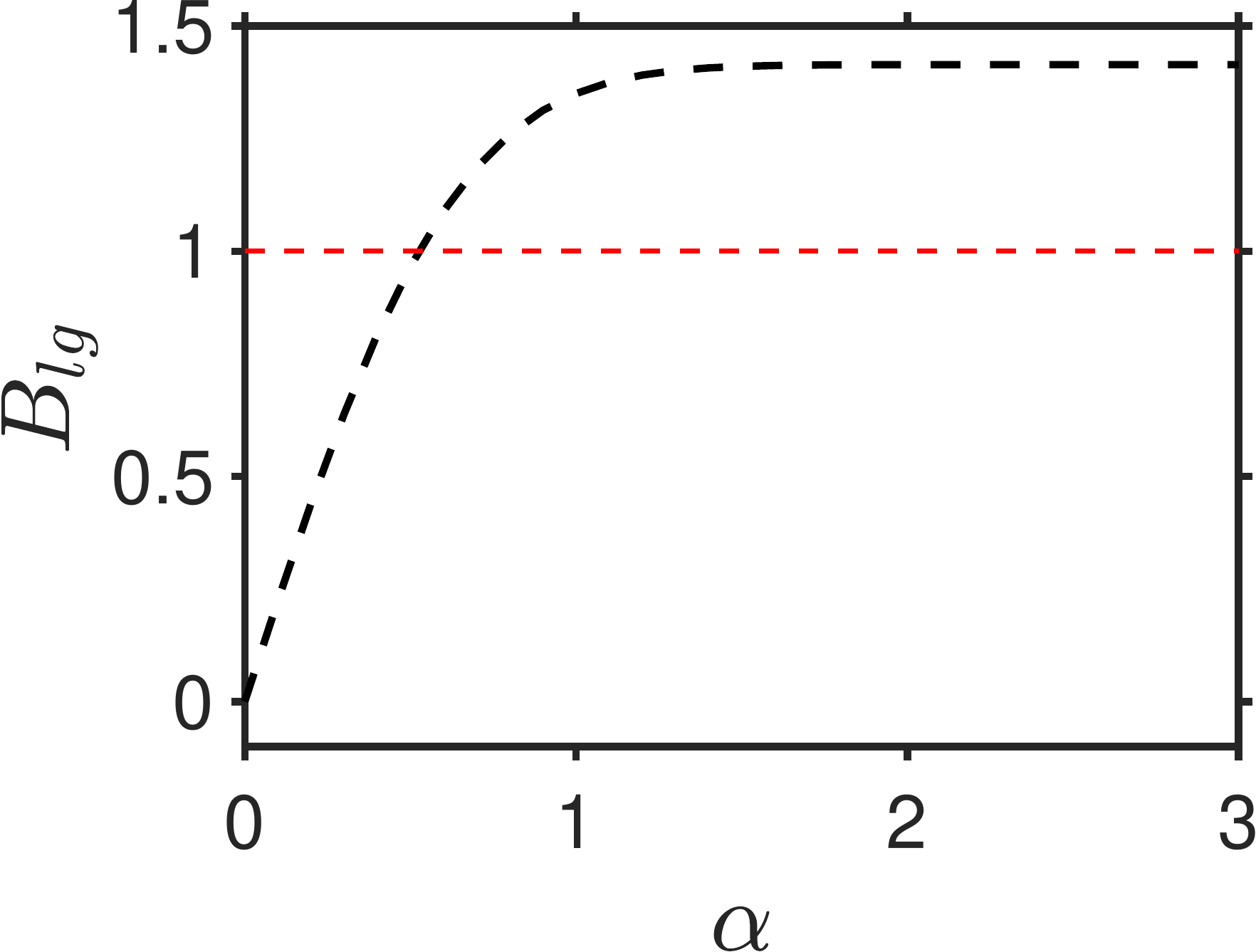}\ \ \includegraphics[width=0.45\columnwidth]{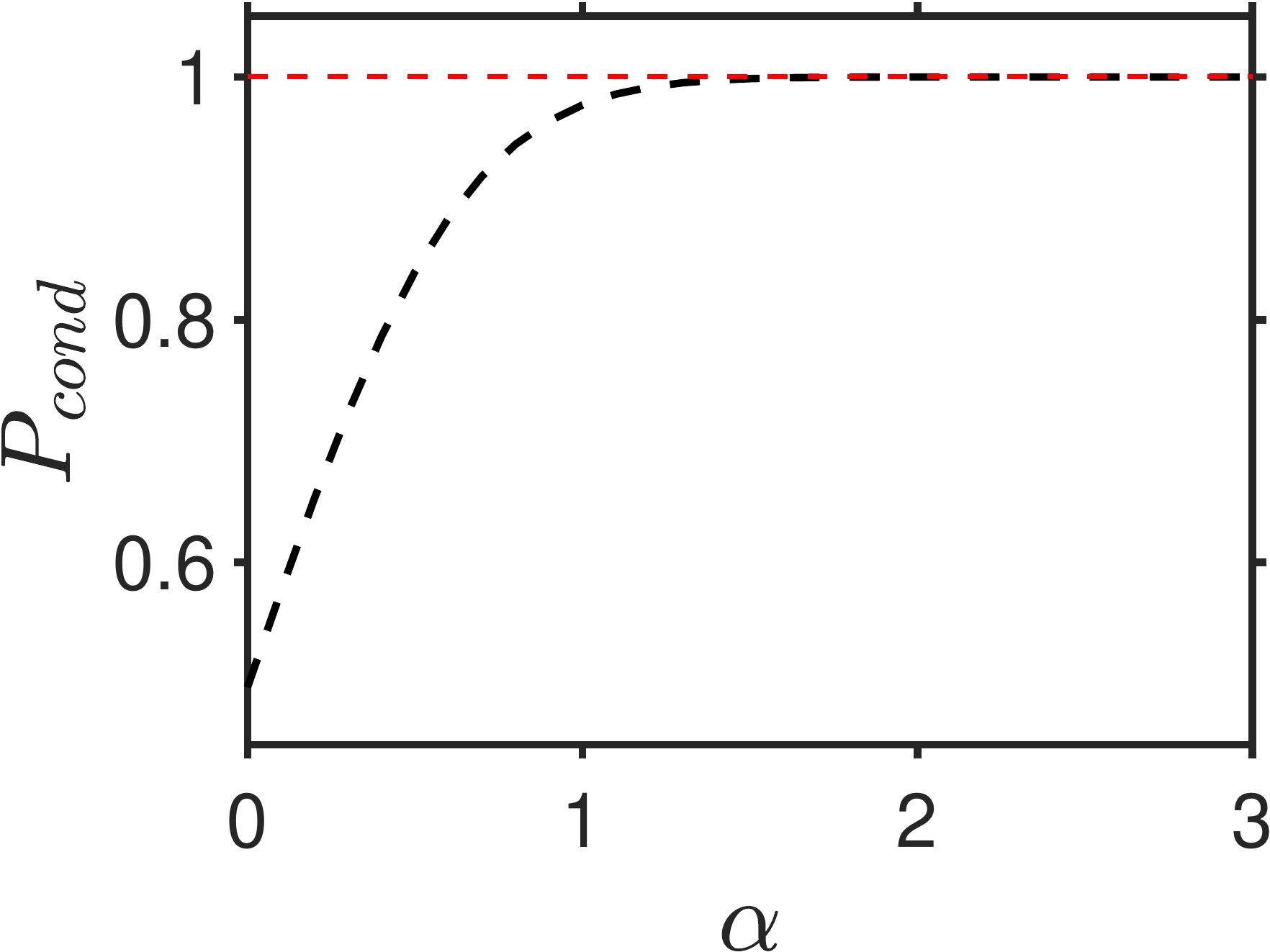}

\caption{Violation of the Leggett-Garg inequality (\ref{eq:lg-ineq}). We
plot $B_{lg}=-\{\langle S_{1}^{(B)}S_{2}^{(A)}\rangle-\langle S_{1}^{(B)}S_{3}^{(A)}\rangle+\langle S_{2}^{(B)}S_{3}^{(A)}\rangle\}$
 versus $\alpha$ for the state $|\psi_{Bell}(t_{1})\rangle$ (\ref{eq:supt1}),
with $\beta=2$. Violation is obtained when $B_{lg}>1$. The verification
of $\langle S_{i}^{(A)}S_{j}^{(A)}\rangle=-\langle S_{i}^{(B)}S_{j}^{(A)}\rangle$
for $i=1,2$ is given by the conditional distribution $P_{cond}$
defined as $P_{cond}=P(S_{i}^{(A)}=1|S_{i}^{(B)}=-1)$ as shown. \textcolor{red}{}\textcolor{blue}{\label{fig:Violation-lg}}}
\end{figure}

\subsection{Leggett-Garg inequality and violations}

We now summarise the Leggett-Garg test of macrorealism for this system
\cite{manushan-bell-cat-lg}. The definition of macrorealism involves
two assumptions: macroscopic realism and noninvasive measurability
(NIM). For our purposes, we take the definition of macroscopic realism
to be that of weak macroscopic realism (wMR) defined in the Introduction:
This asserts that the system given by (1) is in a state with a definite
prediction for the macroscopic spin $S^{(A)}$, $+1$ \emph{or} $-1$.
The system can then be assigned the hidden variable $\lambda$, the
value of $\lambda$ being $+1$ or $-1$, which determines the result
of the measurement $S^{(A)}$ should it be performed. Macrorealism
also implies NIM, that the value of $\lambda$ can be measured with
negligible affect on the subsequent macroscopic dynamics of the system.

For measurements of spin $S_{j}^{(A)}$ made on a single system
$A$ at consecutive times $t_{1}<t_{2}<t_{3}$, macrorealism implies
the Leggett-Garg inequality \cite{weak-solid-qubits-williams-jordan,jordan_kickedqndlg2-2,legggarg-1}\textcolor{red}{}
\begin{equation}
B_{lg}={\color{black}{\color{black}\langle S_{1}^{(A)}S_{2}^{(A)}\rangle+\langle S_{2}^{(A)}S_{3}^{(A)}\rangle-\langle S_{1}^{(A)}S_{3}^{(A)}\rangle}\leq1}\label{eq:lg-ineq}
\end{equation}
As shown in \cite{manushan-bell-cat-lg,macro-bell-lg}, the cat system
of Section IV.A is predicted to violate this inequality (Figure 4),
meaning that macrorealism is falsified. While other Leggett-Garg inequalities
have been proposed (e.g. \cite{nst,halliwell-lg-multidimension,legggarg-1}),
this particular inequality is useful where measurements are made on
entangled subsystems. The approach we give in this paper uses spatial
separation \emph{and} delayed-choice to justify noninvasiveness, since
the measurements of $S_{1}^{(A)}$ and $S_{2}^{(A)}$ can be made
on system $B$. The approach can be applied to other macroscopic superposition
states, such as NOON states \cite{noon-dowling,bognoon-1} using the
local unitary interaction given in \cite{macro-bell-lg}. We comment
that violations of Leggett-Garg inequalities have been predicted and
tested for a range of superposition states (e.g. \cite{emary-review,experiment-lg-2,lgexpphotonweak-1-2,goggin-1,massiveosci-1-1-1,Mitchell-1-1,lauralg-1,leggett-garg-recent-1,halliwell-leggett-garg-double-slit,pan-leggett-garg-weak-value-interference-exp,NSTmunro-1-1,dressel-bell-hybrid})
and alternative procedures exist to justify NIM. 
\begin{figure}[t]
\begin{centering}
\includegraphics[width=1\columnwidth]{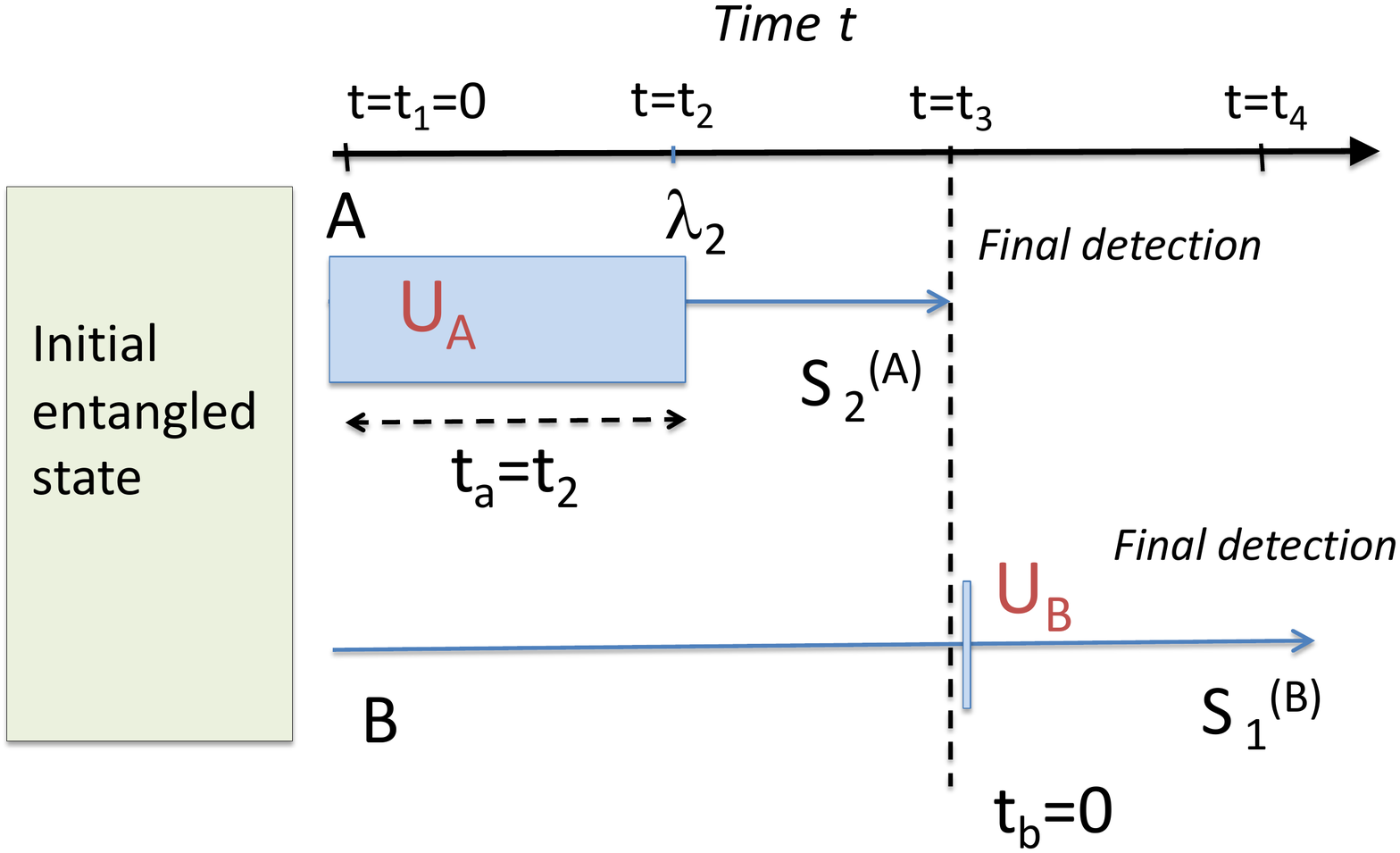}
\par\end{centering}
\begin{centering}
\includegraphics[width=1\columnwidth]{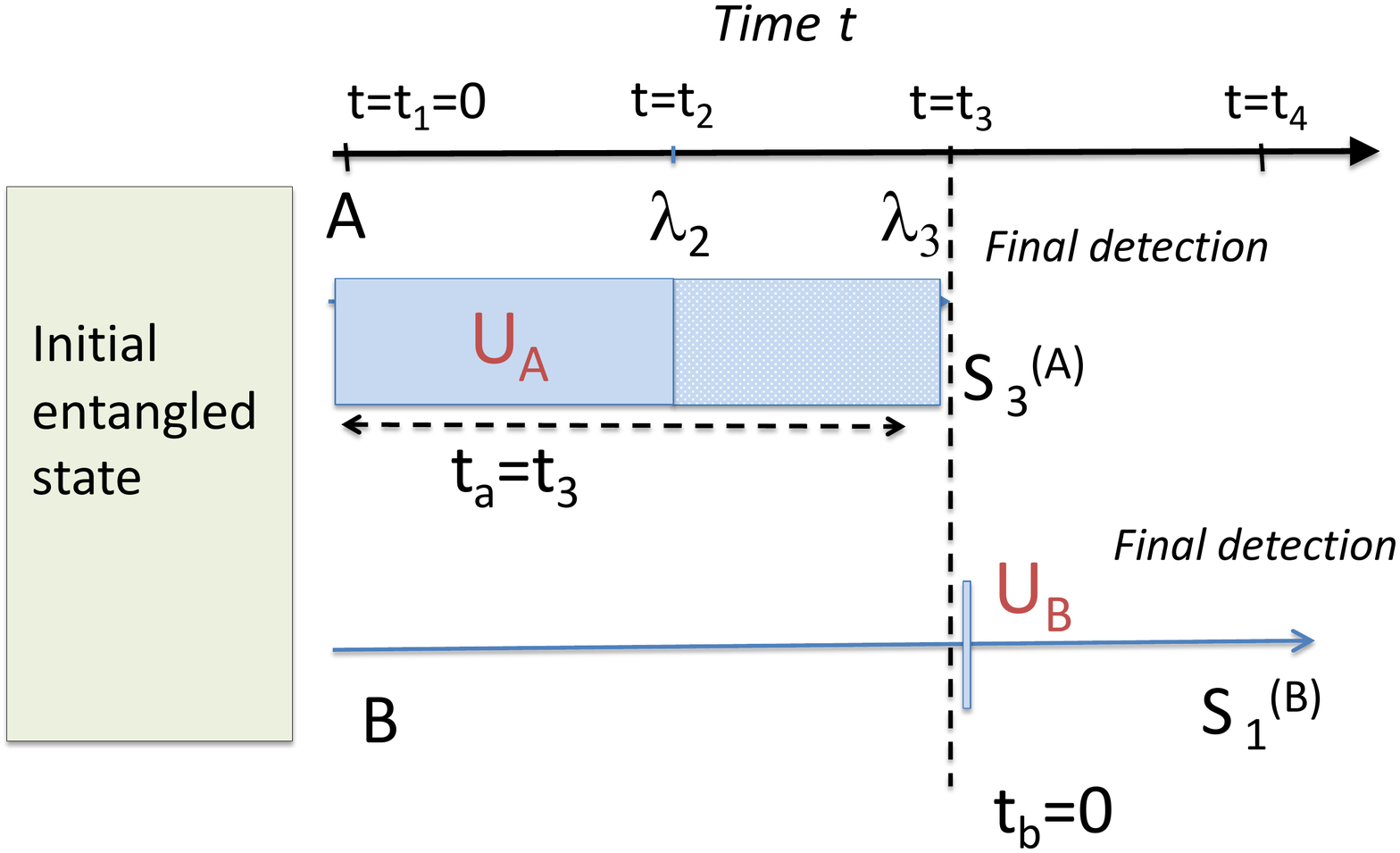}
\par\end{centering}
\caption{Sketch of the set-up for the delayed choice Leggett-Garg test. Notation
is as for Figure 3. The top (lower) sketch shows measurement of $\langle S_{2}^{(A)}S_{1}^{(B)}\rangle$
($\langle S_{3}^{(A)}S_{1}^{(B)}\rangle$). These measurements give
the values of $\langle S_{2}^{(A)}S_{1}^{(A)}\rangle$ and $\langle S_{3}^{(A)}S_{1}^{(A)}\rangle$,
based on the anticorrelation $S_{1}^{(B)}=-S_{1}^{(A)}$. For this
measurement, there is no unitary interaction (rotation) at $B$. The
predictions for the relevant distributions are given in Figure 7 (top).
The results here are indistinguishable from those of an initial
non-entangled state $\rho_{mix}$ (compare Figure 8 (top)).\label{fig:Sketch-single}}
\end{figure}

We summarise the measurements enabling a test of the inequality (\ref{eq:lg-ineq}),
as in Figures 5 and 6. As we have seen, the value of $S_{1}^{(A)}$
or $S_{2}^{(A)}$ of system $A$ can be inferred noninvasively by
measurement of the anti-correlated spin $S_{1}^{(B)}$ or $S_{2}^{(B)}$.
The result for the moment $\langle S_{1}^{(A)}S_{2}^{(A)}\rangle$
is determined by a direct measurement of $S_{2}^{(A)}$ at time $t_{2}$,
and an inferred measurement of $S_{1}^{(A)}$ by measuring $S_{1}^{(B)}$
at $B$ (Figure 5). The moment $\langle S_{1}^{(A)}S_{3}^{(A)}\rangle$
is measured similarly (Figure 5).

\textcolor{black}{The quantum prediction for $\langle S_{1}^{(A)}S_{2}^{(A)}\rangle$
is based on the assumption that }the measurement of $S_{1}^{(B)}$
projects the system $A$ into \textcolor{black}{one or other state,
}$|\alpha\rangle$ or $|-\alpha\rangle$. The prediction is then \textcolor{black}{$\langle S_{1}^{(A)}S_{2}^{(A)}\rangle=\cos(\pi/4)$,
based on the evolution time of $t_{2}$ at $A$ (see eqn (\ref{eq:supt2-1-2-1})).
The moment $\langle S_{1}^{(A)}S_{3}^{(A)}\rangle$ is evaluated similarly,
and from eqn (\ref{eq:state3}) we see the prediction is $\langle S_{1}^{(A)}S_{3}^{(A)}\rangle=\cos(\pi/2)=0$.}
\begin{figure}[t]
\begin{centering}
\includegraphics[width=1\columnwidth]{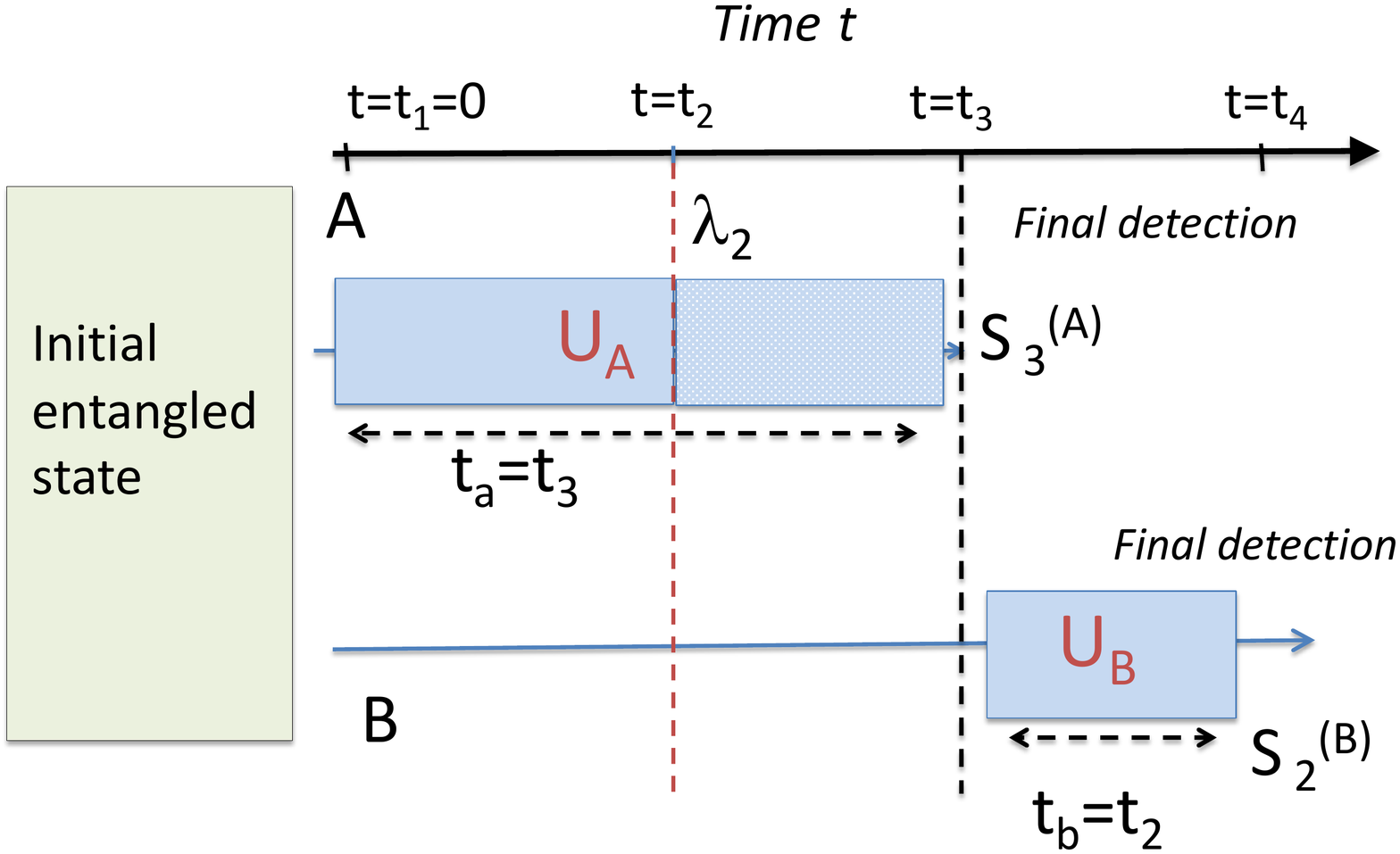}
\par\end{centering}
\caption{Sketch of the set-up for the Leggett-Garg test. Notation is as for
Figure 3. The sketch depicts measurement of $\langle S_{3}^{(A)}S_{2}^{(B)}\rangle$,
which based on the anticorrelation $S_{2}^{(B)}=-S_{2}^{(A)}$ gives
the value for $\langle S_{3}^{(A)}S_{2}^{(A)}\rangle$. The predictions
for the relevant distributions are given in Figure 7 (lower). The
results at time $t_{4}$ are macroscopically different from those
obtained if the state at time $t_{2}$ is non-entangled (compare Figure
8 (lower)). The results are inconsistent with local hidden variable
models and the premise of deterministic macroscopic realism.\label{fig:Sketch-double}}
\end{figure}

\begin{figure*}[t]
\begin{centering}
\par\end{centering}
\begin{centering}
\par\end{centering}
\begin{centering}
\includegraphics[width=1.8\columnwidth]{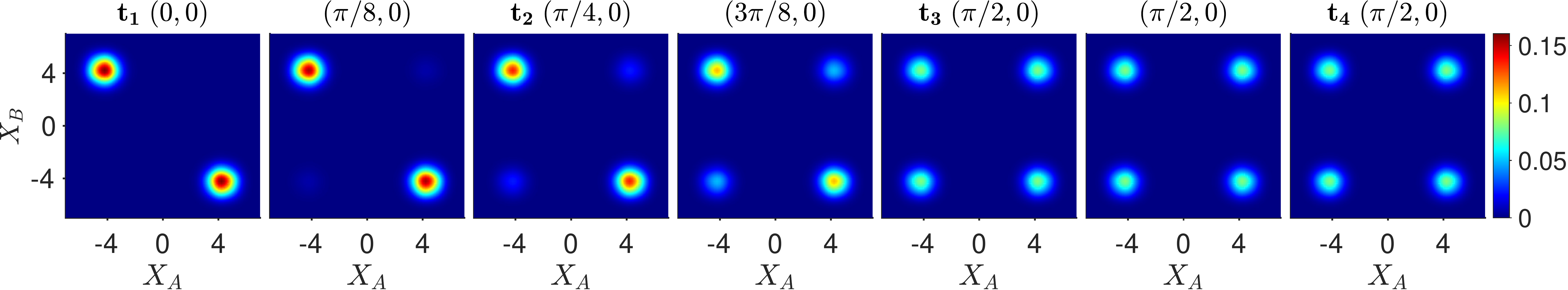}
\par\end{centering}
\medskip{}

\begin{centering}
\includegraphics[width=1.8\columnwidth]{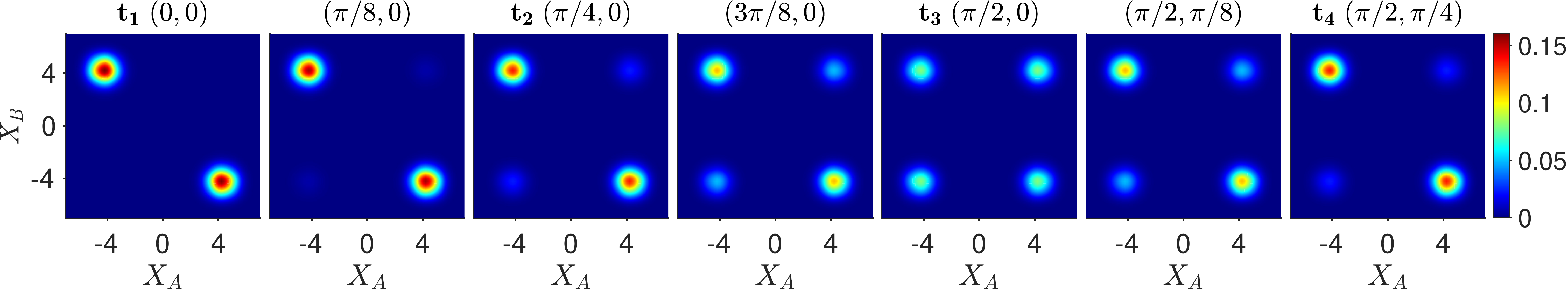}
\par\end{centering}
\begin{centering}
\par\end{centering}
\caption{Contour plots of $P(X_{A},X_{B})$ showing the dynamics as the state
$|\psi_{Bell}(t_{1})\rangle$ evolves through the three measurement
sequences of the Leggett-Garg test in the delayed-choice gedanken
experiment depicted in Figures 5 and 6. Here, we go from time $t=t_{1}$
(far left), through to $t=t_{2}$ (third picture from left), $t=t_{3}$
(fifth picture from left) and, finally, $t=t_{4}$ (far right). The
systems evolve locally according to $H_{NL}^{(A/B)}$ for interaction
times $t_{a}$ and $t_{b}$ given by $(t_{a},t_{b})$ in units of
$\Omega^{-1}$.  Top: The sequence to infer $S_{1}^{(A)}$ by
delayed measurement of $S_{1}^{(B)}$, enabling measurement of  $\langle S_{1}^{(B)}S_{3}^{(A)}\rangle=-\langle S_{1}^{(A)}S_{3}^{(A)}\rangle$
(final picture), as in Figure 5 (lower). The sequence to measure $\langle S_{1}^{(B)}S_{2}^{(A)}\rangle=-\langle S_{1}^{(A)}S_{2}^{(A)}\rangle$
uses $t_{a}=t_{2}=\pi/4$ as in Figure 5 (top) and ends with the third
picture of the sequence. Lower: The sequence to infer $S_{2}^{(A)}$
by measurement of $S_{2}^{(B)}$, enabling measurement of $\langle S_{2}^{(B)}S_{3}^{(A)}\rangle=-\langle S_{2}^{(A)}S_{3}^{(A)}\rangle$
(final picture) as in Figure 6. Here, $t_{1}=0$, $t_{2}=\pi/4$
and $t_{3}=\pi/2$. $\alpha=\beta=3$.\textcolor{red}{}\textcolor{blue}{\label{fig:evolution-ent}}}
\end{figure*}
\textcolor{black}{For $\langle S_{2}^{(A)}S_{3}^{(A)}\rangle$, one
would measure $S_{2}^{(B)}$ to determine the anticorrelated $S_{2}^{(A)}$,
and measure $S_{3}^{(A)}$ directly at $A$ (Figure 6). The prediction
for $\langle S_{2}^{(A)}S_{3}^{(A)}\rangle$ is based on the assumption
that the system $A$ is in either }$|\alpha\rangle$ or $|-\alpha\rangle$,
at time $t_{2}$ (\textcolor{black}{or else, that the measurement
of $S_{2}^{(B)}$ projects $A$ to one of these states). }The subsequent
evolution for a time $\Delta t=\pi/8$ then leads to the prediction
of $\langle S_{2}^{(A)}S_{3}^{(A)}\rangle=\cos(\pi/4)$ (refer eqn
\textcolor{black}{(\ref{eq:state3-3}})). This gives violation of
the inequality (\ref{eq:lg-ineq}), the left side being $\sqrt{2}$.

The above calculations assume large $\beta$ (and hence orthogonal
$|\beta\rangle$ and $|-\beta\rangle$) so that one may justify the
assumption that the system $A$ at times $t_{1}$ and $t_{2}$ is
projected into one or other of the states $|\alpha\rangle$ or $|-\alpha\rangle$
once the measurement at $B$ is performed. To evaluate accurately
requires evaluation of the joint distributions $P(X_{A},X_{B})$ for
the different times of interaction $t_{a}$ and $t_{b}$. For large
$\alpha$ and $\beta$, the simplistic result is indeed recovered,
for all $\alpha,$$\beta>1$. The precise results were calculated
in \cite{manushan-bell-cat-lg}, and are given in Figures 4. The results
agree with the moments above, predicting violation of the inequality,
for $\alpha>1$. The plots of $P(X_{A},X_{B})$ for the various times
of evolution are given in Figure 7.

The violation of the inequality (\ref{eq:lg-ineq}) implies falsification
of macrorealism. We note that the measurements $S_{i}^{(A)}$and $S_{j}^{(B)}$
are macroscopic in the sense that one needs only to distinguish between
the two macroscopically separated peaks of the distributions $P(X_{A},X_{B})$
(Figure 7). Here, the meaning of ``macroscopic'' refers to a separation
in phase space of quadrature amplitudes $X$ by an arbitrary amount
($\alpha\rightarrow\infty$).

\begin{figure*}[t]
\begin{centering}
\par\end{centering}
\begin{centering}
\includegraphics[width=1.8\columnwidth]{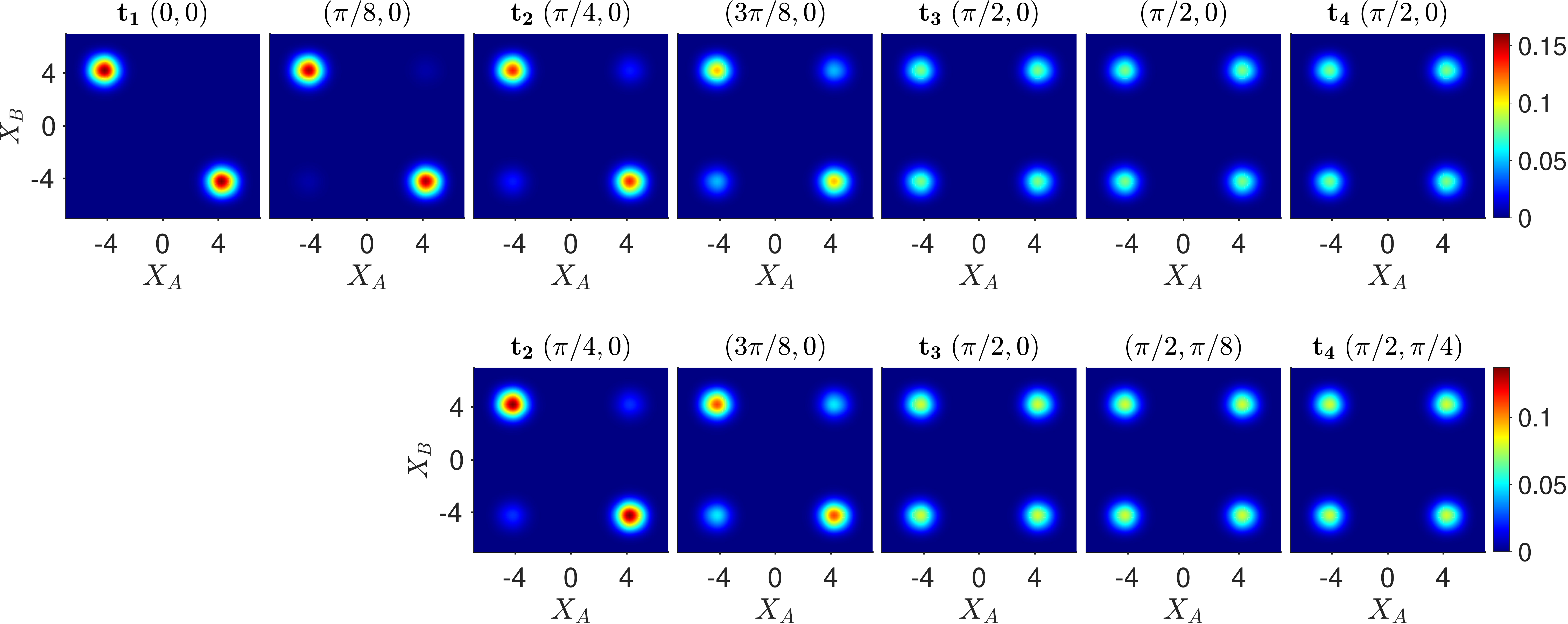}
\par\end{centering}
\begin{centering}
\par\end{centering}
\caption{Contour plots of $P(X_{A},X_{B})$ showing the dynamics as the \emph{non-entangled}
state $\rho_{mix}$ evolves through the same measurement sequences
given in Figure 7. Notation as for Figure 7. Top: The sequence evolves
as in Figure 5 (lower) with a unitary rotation at site $A$ only.
Athough starting with $\rho_{mix}$ at time $t_{1}$, the sequence
is indistinguishable from that given by the top sequence in Figure
7 for the entangled state $|\psi_{Bell}(t_{1})\rangle$. Lower: We
assume the system evolves as for Figure 6 and the lower sequence of
Figure 7 with two unitary rotations, one at $A$ and one at $B$,
but starting from a nonentangled state $\rho_{mix}$ at the time $t=t_{2}$.
Although indistinguishable at the initial time $t_{2}$, the final
picture at $t=t_{4}$ ($(\pi/2,\pi/4)$) differs macroscopically from
that of the entangled state (compare with the lower sequence in Figure
7). \textcolor{red}{}\textcolor{blue}{\label{fig:evolution-non-ent}}}
\end{figure*}

\subsection{Interpretation without macroscopic retrocausality}

As explained above, macrorealism involves two assumptions: weak macroscopic
realism (wMR) and noninvasive measurability. If we assume the validity
of wMR, then we would conclude that noninvasive measurability fails:
the measurement of the spin $S_{i}^{(B)}$ of $B$ disturbs the result
for the spin $S_{j}^{(A)}$of $A$ ($j>i$). However, since the measurements
are made at $B$ after the state of $A$ at the time $t_{3}$ is measured,
this conclusion would seem to suggest a macroscopic retrocausal effect,
where which measurement is made at $B$ alters the past value of $\lambda_{i}$
at $A$. In Section IV, we rigorously clarify the nature of this apparent
retrocausality, by examining the dimension witness test proposed in
\cite{delayed-choice-causal-model-chaves}.

Here, we examine further, by analysing how the dynamics pictured in
Figure 7 provides an interpretation that avoids the conclusion of
macroscopic retrocausality. First, it is useful to compare with the
dynamics of a non-entangled state (Figure 8) \textcolor{blue}{}
\begin{equation}
\rho_{mix}=\frac{1}{2}\{|\alpha\rangle|-\beta\rangle\langle\alpha|\langle-\beta|+|-\alpha\rangle|\beta\rangle\langle-\alpha|\langle\beta|\}\label{eq:mix}
\end{equation}
The non-entangled cat state is consistent with the first Leggett-Garg
premise of weak macroscopic realism (wMR), since each system can be
viewed as being in one or other of two macroscopically distinct coherent
states at time $t_{1}$. We first note that there is \emph{no distinguishable
difference} between the predictions $P(X_{A},X_{B})$ for the entangled
($|\psi_{Bell}(t_{1})\rangle$) and non-entangled ($\rho_{mix}$)
states, at the level of the macroscopic outcomes (compare the first
plot of the top sequences in Figures 7 and 8). A distinction exists,
but at order $\hbar e^{-|\alpha|^{2}}$, invisible on the plots.

It is seen that where one measures $S_{1}^{(B)}$, the predictions
$P(X_{A},X_{B})$ for the two systems beginning with the entangled
($|\psi_{Bell}(t_{1})\rangle$) and non-entangled ($\rho_{mix}$)
states\emph{ remain indistinguishable} (compare the top sequences
of Figures 7 and 8). This corresponds to there being no rotation (unitary
evolution) at site $B$ (Figure 5). A distinction in fact exists,
but this is at the microscopic level of order $\hbar e^{-|\alpha|^{2}}$,
invisible on the plots \cite{manushan-bell-cat-lg}.

There is a \emph{macroscopic} difference however for the evolution
where one measures $S_{2}^{(B)}$, which involves \emph{two} unitary
rotations after $t_{2}$, one at each site, as depicted in Figure
6. This is seen by comparing the lower sequences of Figure 7 and Figure
8. Here, if one starts with a non-entangled state $\rho_{mix}$ at
time $t_{2}$ (Figure 8 (lower)), then even though the joint probabilities
$P(X_{A},X_{B})$ are indistinguishable at $t_{2}$, the joint probabilities
differ macroscopically after the evolution involving rotations at
both sites (compare the last pictures in the lower sequences).

We conclude that the violation of macrorealism and the apparent retrocausality
arises from the measurement of $\langle S_{2}^{(A)}S_{3}^{(A)}\rangle$,
as depicted in Figure 6. The scenario of Figure 5 is consistent with
macrorealism, since it can be modelled by evolution of $\rho_{mix}$.

\subsubsection{Weak macroscopic realism: the pointer measurement}

To interpret without macroscopic retrocausality, we aim to show consistency
with the assumption of weak macroscopic realism (wMR). We first examine
this assumption more closely, along the lines given in \cite{manushan-bell-cat-lg}.

Let us suppose the systems $A$ and $B$ are prepared at a time $t_{j}$
in a macroscopic superposition $|\psi_{pointer}\rangle$ of states
with definite outcomes for \emph{pointer} measurements $\hat{S}_{j}^{(A)}$
and $\hat{S}_{j}^{(B)}$. In this paper, the example of such a superposition
is
\begin{equation}
|\psi_{Bell}\rangle=\mathcal{N}(|\alpha\rangle|-\beta\rangle-|-\alpha\rangle|\beta\rangle)\label{eq:cat-1-3-2-2}
\end{equation}
where $\alpha,$$\beta\rightarrow\infty$. The premise wMR asserts
that the system $A$ at the time $t_{j}$ is\emph{ }in one or other
of two macroscopic states $\varphi_{+}$ and $\varphi_{-}$, for
which the result of the spin measurement $S_{j}^{(A)}$ (given by
the sign of the coherent amplitude) is determined to be $+1$ or $-1$
respectively. Hence, the system $A$ at time $t_{j}$ may be described
by the macroscopic hidden variable $\lambda_{j}^{(A)}$. The value
of $\lambda_{j}^{(A)}$ is fixed as either $+1$ or $-1$ at the particular
time $t_{j}$, prior to the pointer measurement, and is independent
of any future measurement. By the pointer measurement, it is meant
that the measurement can be made as a final quadrature detection,
$X_{A}$, with \emph{no further unitary rotation} $U_{A}$ necessary.
Weak macroscopic realism does not mean that prior to the measurement
of spin $S^{(A)}$ the system is in the state $|\alpha\rangle$ or
$|-\alpha\rangle$, or indeed in any quantum state $-$ since the
quantum states are microscopically specified, giving predictions for
all measurements that might be performed on $A$. For the entangled
state $|\psi_{Bell}\rangle$, similar assumptions apply to system
$B$.

For the bipartite system depicted in Figures 5 and 6, wMR is to be
consistent with a form of macroscopic locality. ``Macroscopic locality
of the pointer'' was summarised in \cite{manushan-bell-cat-lg} and
asserts that the value of the macroscopic hidden variable $\lambda_{j}^{(A)}$
for the system $A$ cannot be changed by any spacelike separated event,
or measurement at the system $B$ that takes place at time $t\geq t_{j}$
e.g. it cannot be changed by a future event at $B$. In this interpretation,
the system $A$ at each time $t_{i}$ ($i=1,2,3$) is in one or other
of states $\varphi_{i,+}$ or $\varphi_{i,-}$ with a definite value
$+1$ or $-1$ of spin $S_{i}^{(A)}$. The premise ``macroscopic
locality of the pointer'' is to be distinguished from the stronger
assumption, macroscopic locality, introduced in \cite{manushan-bell-cat-lg}.
``Macroscopic locality'' assumes locality to apply to spacelike-separated
measurement events, but here the measurement setting for system $A$
is not necessarily established, so that $A$ is \emph{not necessarily
prepared in the pointer basis}. This allows for the possibility of
a further unitary rotation at $A$, before the final detection of
$X_{A}$. The premise of weak macroscopic realism is thus not contradicted
by the violation of the macroscopic Bell inequalities reported in
\cite{macro-bell-lg,manushan-bell-cat-lg}.

\subsubsection{Delayed collapse and unitary rotation at one site only: consistency
with wMR}

The dynamics indicates consistency with wMR. We focus on two features,
explained in Ref. \cite{manushan-bell-cat-lg}: delayed collapse and
the single rotation.

Let us suppose that at the time $t_{j}$ the dynamics $U^{(A)}$ for
a pointer measurement $S_{j}^{(A)}$ has taken place at $A$. The
final detection (the ``collapse'' or ``projection'') stage of
the measurement $S_{j}^{(A)}$ at $A$ \emph{can be delayed} for an
infinite time, and there is no change in the \emph{macroscopic} joint
probabilities $P(X_{A},X_{B})$.  The result is true even where there
is a unitary rotation $U_{B}$ at the site $B$ after the time $t_{j}$:
the joint probabilities $P(X_{A},X_{B})$ do not depend on whether
the final detection at $A$ is before or after the unitary evolution
$U_{B}$. The full calculations are given in \cite{manushan-bell-cat-lg}
and show that while there are differences in the final distributions,
these differences are negligible, of order $\hbar e^{-|\alpha|^{2}}$.
This supports the wMR assumption, that for the \emph{pointer} measurement
($S_{j}^{(A)}$ in this case), the result is determined by $\lambda_{j}^{(A)}$
at the time $t_{j}$ $-$ we can consider $\lambda_{j}^{(A)}$ as
fixed.

There is also consistency with wMR for the dynamics given by Figures
5 and 7 (top), where there is \emph{no unitary rotation} at the site
$B$ (after $t_{1}$). Comparing Figures 7 (top) and 8 (top), we see
that the macroscopic dynamics of the sequences for $\langle S_{1}^{(B)}S_{3}^{(A)}\rangle$
and $\langle S_{1}^{(B)}S_{2}^{(A)}\rangle$, which involve only one
unitary rotation (at $A$), are identical to those of the non-entangled
state $\rho_{mix}$, and hence are consistent with wMR. The macroscopic
probabilities for the sequences with a rotation at one site only are
also consistent with those of a local hidden variable theory i.e.
the final outcomes at $A$ and $B$ can be interpreted as being due
to a local interaction at $A$.

\subsubsection{Failure of deterministic macroscopic realism: unitary rotation at
both sites}

The violations of the Leggett-Garg inequality can be shown to arise
as a failure of deterministic macroscopic realism (dMR), as studied
in \cite{manushan-bell-cat-lg,macro-bell-lg}. This premise (different
to wMR) asserts a predetermined outcome for the measurement \emph{prior}
to the unitary rotation $U$ that determines the measurement setting.
Where one has \emph{two} unitary rotations, one at each site, after
the time $t_{j}$, as in Figure 6, there is no longer consistency
with the predictions of $\rho_{mix}$.

Let us consider the scenario of Figure 6, at time $t_{2}$. The value
of $\lambda_{2}^{(A)}$ is predetermined according to wMR, for the
pointer measurement $S_{2}^{(A)}$. However, one may also consider
the outcome of a measurement $S_{3}^{(A)}$ at the later time, made
by applying a rotation $U^{(A)}(\pi/4)$ and then measuring $X_{A}$.
If we assume dMR, then this latter outcome can also be regarded as
predetermined, and we can assign the hidden variable $\lambda_{3}^{(A)}$
to the system at the time $t_{2}$. Similarly, assuming dMR, one may
assign variables $\lambda_{2}^{(B)}$ and $\lambda_{3}^{(B)}$ to
system $B$, at time $t_{2}$.

Extending this argument, the premise dMR would imply simultaneous
values for the outcomes at time $t_{1}$ regardless of the future
unitary dynamics required to make the actual measurements, and would
hence imply the Leggett-Garg inequality (\ref{eq:lg-ineq}). Similarly,
the macroscopic Bell inequality studied in \cite{macro-bell-lg,manushan-bell-cat-lg}
would apply. We have show in Section IV.B that the Leggett-Garg inequality
is violated, indicating failure of dMR. Similarly, the macroscopic
Bell inequality derived in \cite{macro-bell-lg,manushan-bell-cat-lg}
is violated. This implies that dMR is (predicted to be) falsified.

\subsubsection{Explanation}

The apparent retrocausal effect can be explained as arising from the
failure of deterministic macroscopic realism. The failure of dMR may
also be viewed as a macroscopic Bell nonlocality, as discussed in
\cite{manushan-bell-cat-lg,macro-bell-lg}. We argue however that
the gedanken experiment is consistent with weak macroscopic realism.

We explain further. First, examining Figure 7 for the Leggett-Garg
violations, we see that the macroscopic dynamics of the sequences
for $\langle S_{1}^{(B)}S_{3}^{(A)}\rangle$ and $\langle S_{1}^{(B)}S_{2}^{(A)}\rangle$
(Figure 5) involving only one unitary rotation are identical to those
of the non-entangled state $\rho_{mix}$, and hence are consistent
with wMR. We next consider measurement of $\langle S_{2}^{(A)}S_{3}^{(A)}\rangle$.
In measuring $\langle S_{2}^{(A)}S_{3}^{(A)}\rangle$ via $\langle S_{2}^{(B)}S_{3}^{(A)}\rangle$,
as in the lower sequence of Figure 7, the system at $A$ is \emph{entangled}
with $B$ at time $t_{2}$. An interpretation consistent with wMR
is possible, since the measurement of $\langle S_{2}^{(B)}S_{3}^{(A)}\rangle$
involves \emph{two} rotations after the time $t_{2}$, one at $A$
and one at $B$ (as in Figure 6). This double rotation gives rise
to \emph{macroscopic nonlocality }(violations of a macroscopic Bell
inequality ) i.e. to the failure of deterministic macroscopic realism
\cite{macro-bell-lg,manushan-bell-cat-lg}.
\begin{figure}[t]
\begin{centering}
\includegraphics[width=1\columnwidth]{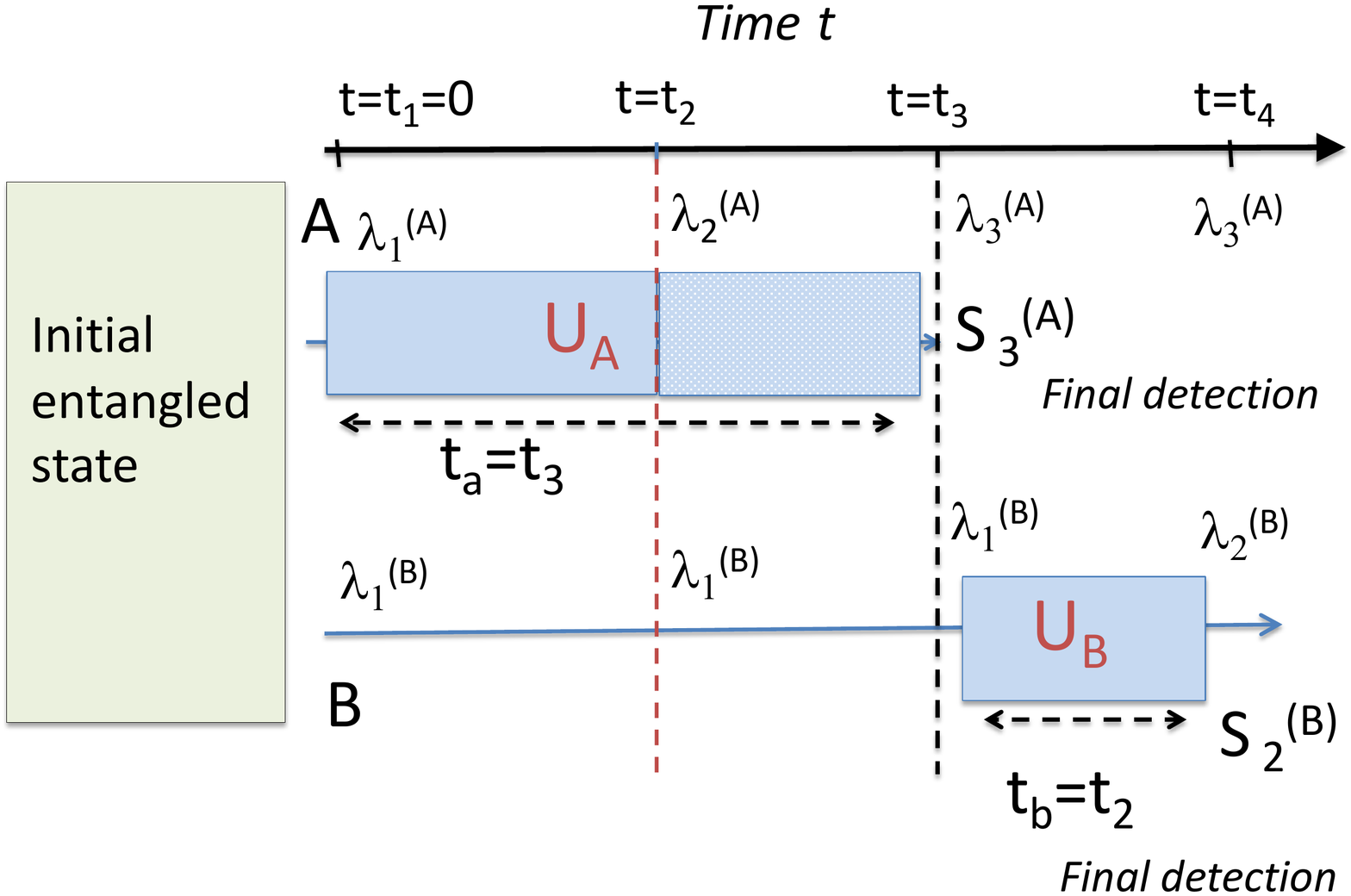}
\par\end{centering}
\caption{Consistency with weak macroscopic realism is possible for the time
sequence of Figure 6. The system at time $t_{1}$ has valid hidden
variables $\lambda_{1}^{(A)}$ and $\lambda_{1}^{(B)}$, being indistinguishable
from $\rho_{mix}$. At time $t_{2}$, system $A$ has valid $\lambda_{2}^{(A)}$
and $\lambda_{1}^{(A)}$, the value of $\lambda_{1}^{(A)}$ being
given by the pointer measurement on $B$ at the time $t_{2}$. Similarly,
system $B$ at time $t_{2}$ has valid $\lambda_{1}^{(B)}$ and $\lambda_{2}^{(B)}$.
At time $t_{3}$, the system $A$ has valid $\lambda_{3}^{(A)}$ and
$\lambda_{1}^{(A)}$, since the value of $\lambda_{1}^{(A)}$ can
be given by the pointer measurement at $t_{3}$ on $B$. At time $t_{3}$,
system $B$ has valid $\lambda_{1}^{(B)}$ and $\lambda_{3}^{(B)}$
(because $\lambda_{3}^{(B)}$ can be inferred from $\lambda_{3}^{(A)}$).
At time $t_{4}$, system $A$ has valid $\lambda_{3}^{(A)}$ and $\lambda_{2}^{(A)}$.\label{fig:schematic-lambda-wmr}}
\end{figure}

The validity of weak macroscopic realism can then be argued as follows
(Figure 9). Following Figure 6, the system $A$ at times $t_{1}$
and $t_{2}$ can indeed be represented by the hidden variables $\lambda_{1}^{(A)}$
and $\lambda_{2}^{(A)}$ (meaning that the pointer measurements of
$S_{1}^{(A)}$ and $S_{2}^{(A)}$ have predetermined outcomes), because
the predictions for pointer measurement $S_{1}^{(A)}$ and $S_{2}^{(A)}$
are identical with those arising from $\rho_{mix}$ (there has been
a rotation at one site, $A,$ only). This is also true of the system
$B$ at time $t_{1}$: it can be described by a $\lambda_{1}^{(B)}$,
for the reason that the predictions are indistinguishable from those
of $\rho_{mix}$.

At time $t_{2}$, $A$ can also be consistently represented by a hidden
variable $\lambda_{1}^{(A)}$, because the value $S_{1}^{(B)}$ at
$B$ is determinable by a pointer measurement, without further rotation.
Also, because of the correlation with $S_{2}^{(A)}$, one would conclude
$\lambda_{2}^{(B)}$ can be assigned to the state $B$ at the time
$t_{2}$, because the outcome after the unitary evolution $U^{(B)}(\pi/4)$
is predetermined. However, it is \emph{not} the case that at time
$t_{2}$ the outcome of $S_{3}^{(A)}$ is predetermined (if $U^{(A)}(\pi/4)$
would be performed), because dMR fails. Hence, at time $t_{2}$, it
is not true that the hidden variable $\lambda_{3}$ can be assigned
to the state at $A$, because the unitary rotation $U^{(A)}(\pi/4)$
has not been performed. Regardless, this does not imply failure
of wMR, because the dynamics associated with $U^{(A)}(\pi/4)$ is
in the future of $t_{2}$. 

On the other hand, if the unitary rotation $U^{(B)}(\pi/2)$ that
precedes the measurement $S_{3}^{(B)}$ \emph{is} performed \emph{prior}
to the time $t_{2}$ at $B$, then the state at $A$ at time $t_{2}$
can be assigned $\lambda_{3}^{(A)},$ but can no longer be assigned
$\lambda_{1}^{A)}$ at that time $t_{2}$. This interpretation allows
for macroscopic Bell nonlocal effects when there are unitary rotations
at both sites, but is also consistent with weak macroscopic realism
(wMR) and hence does not indicate macroscopic retrocausality.

\section{Dimension Witness test}

We next follow the approach of Chaves, Lemos and Pienaar (CLP) \cite{delayed-choice-causal-model-chaves},
by demonstrating violation of the dimension witness inequality \cite{Dimension-Witness-Brunner,Quantum-Dimension-Witness,bowles-dimension-test-exp,ahrens-exp-dimension-test-nat-phys,Yu-causal-model-exp,delayed-choice-experiment-chaves}.
Here, one considers two-dimensional models and, within this framework,
confirms the failure of all non-retrocausal models. Our results extend
beyond those of CLP because the conclusions of retrocausality apply
to the macroscopic qubits $|\alpha\rangle$ and $|-\alpha\rangle$
where $\alpha$ is large, for which the binary outcomes of the relevant
measurements are distinguishable beyond $\hbar$. This test makes
concrete the apparent retrocausality discussed in Section IV.C, and
elucidates how this can be interpreted as due to the limitation of
the assumption of two-dimensional hidden variable model.

\begin{figure}[t]
\begin{centering}
\includegraphics[width=0.6\columnwidth]{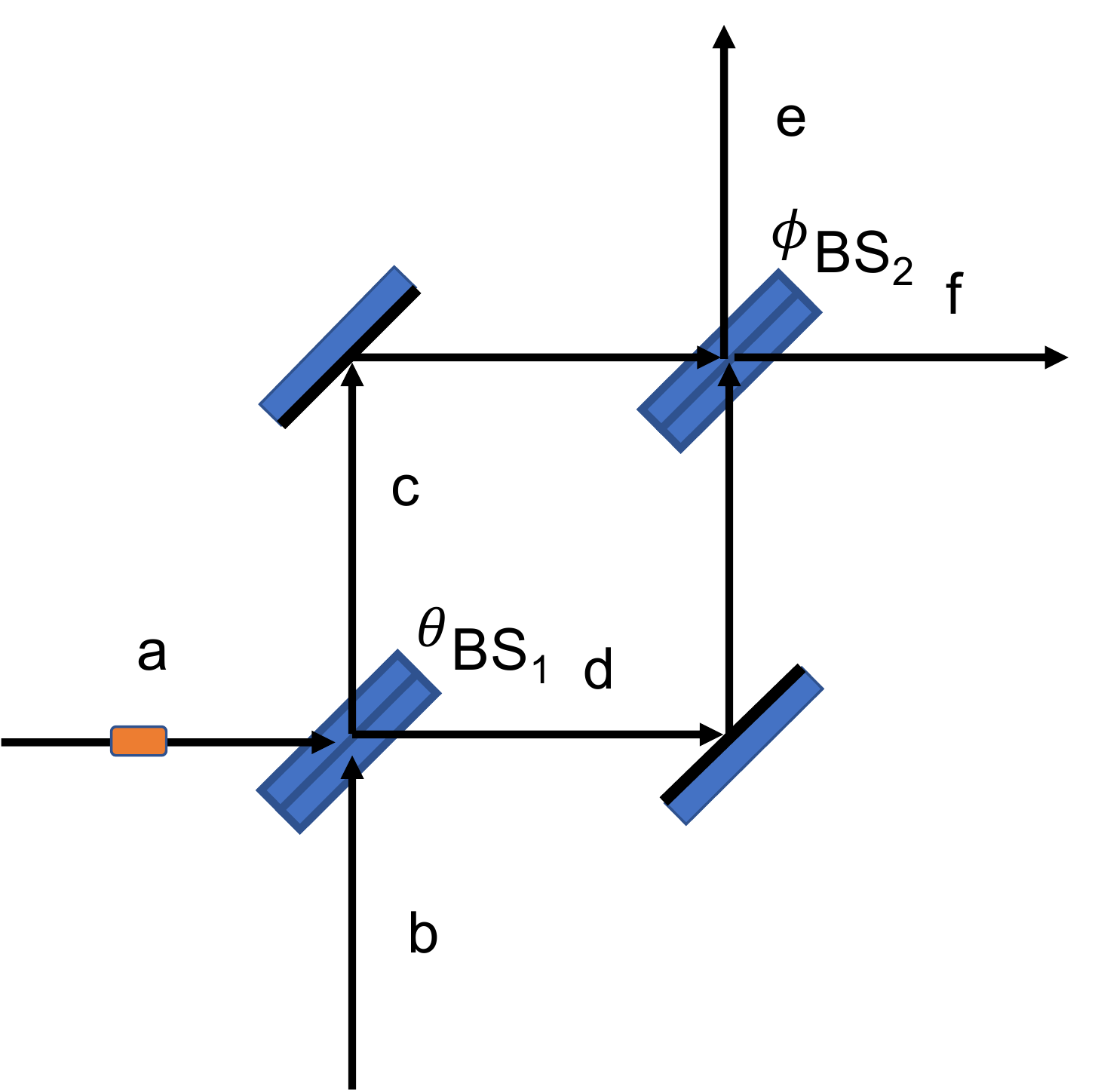}
\par\end{centering}
\caption{Schematic of Wheeler-CLP delayed choice experiment. A single boson
two-mode state $|1\rangle_{a}|0\rangle_{b}$ is incident on the first
beam splitter ($BS_{1}$). The first beam splitter introduces a variable
reflectivity given by $\theta$ with output modes $c$ and $d$. These
two modes are again recombined at a second beam splitter $BS_{2}$
to produce final output modes $e$ and $f$ with variable transformation
angle $\phi$.\label{fig:A-single-boson-beam-splitter}\textcolor{blue}{}}
\end{figure}

We first consider the Wheeler-CLP delayed-choice experiment performed
with tunable beam splitters i.e. with a variable reflectivity. A single
boson is incident on the beam splitter, so that the input system is
the two-mode state $|1\rangle_{a}|0\rangle_{b}$ (Figure \ref{fig:A-single-boson-beam-splitter}).
The two modes ($c$ and $d$) at the outputs of the beam splitter
have boson operators
\begin{eqnarray}
\hat{c} & = & \hat{a}\cos\theta-\hat{b}\sin\theta\nonumber \\
\hat{d} & = & \hat{a}\sin\theta+\hat{b}\cos\theta\label{eq:trans-1}
\end{eqnarray}
After the beam splitter, the state of the field in the interferometer
is 
\begin{equation}
|\psi\rangle_{p}=a^{\dagger}|0\rangle_{a}|0\rangle_{b}=\cos\theta|1\rangle_{c}|0\rangle_{d}+\sin\theta|0\rangle_{c}|1\rangle_{d}\label{eq:state-bs-p}
\end{equation}
This is the preparation state, prepared at time $t_{1}$. The fields
pass through the interferometer, and are recombined at a second beam
splitter to produce final output modes $e$ and $f$. The beam splitter
transformation 
\begin{eqnarray}
\hat{e} & = & \hat{c}\cos\phi-\hat{d}\sin\phi\nonumber \\
\hat{f} & = & \hat{c}\sin\phi+\hat{d}\cos\phi\label{eq:trans2}
\end{eqnarray}
constitutes the measurement, and gives the final state
\begin{eqnarray}
|\psi\rangle_{m} & = & \cos(\theta-\phi)|1\rangle_{e}|0\rangle_{f}+\sin(\theta-\phi)|0\rangle_{e}|1\rangle_{f}\label{eq:soln-delay-choice-wclp}
\end{eqnarray}
The binary outcomes $|1\rangle_{c}|0\rangle_{d}$ and $|0\rangle_{0}|1\rangle_{d}$
are denoted $b=1$ and $b=-1$ respectively. The expectation value
for $b$ is $E(\theta,\phi)=\cos^{2}(\theta-\phi)-\sin^{2}(\theta-\phi)=\cos\left(2(\theta-\phi)\right)$.
Certain choices of angles $\theta$ and $\phi$ will violate the dimension
witness inequality, as we show below.

\begin{figure}[t]
\begin{centering}
\includegraphics[width=1\columnwidth]{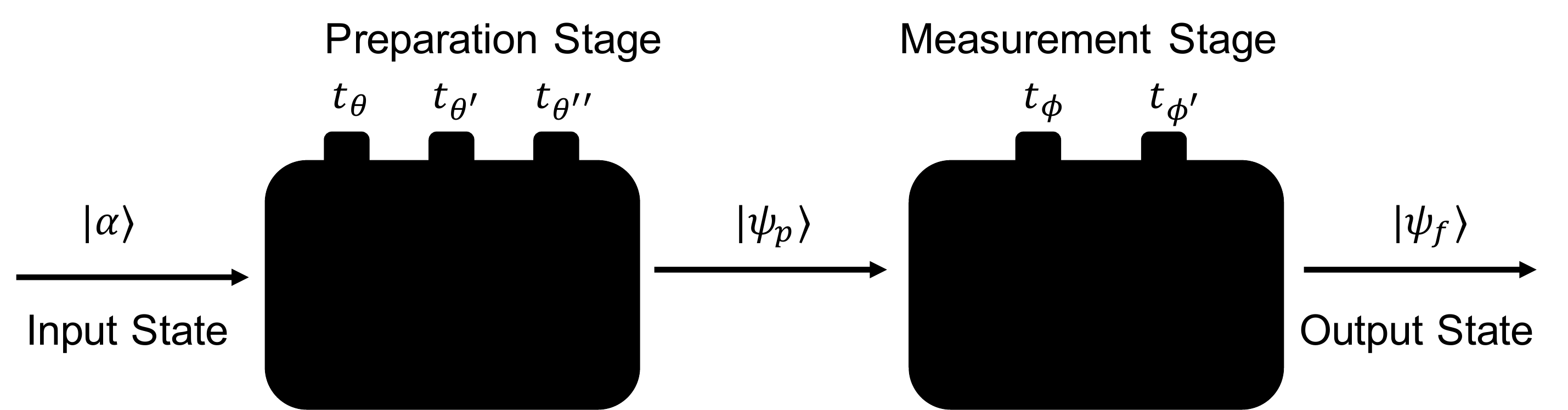}
\par\end{centering}
\caption{The set up for a macroscopic Wheeler-CLP delayed-choice experiment
where we make use of the cat-state dynamics for a prepare and measure
scenario. An initial input of $|\alpha\rangle$ undergoes a nonlinear
interaction $H_{NL}$ for a time $t_{\theta}$ at the preparation
stage of the system corresponding to the first beam splitter $BS1$.
A second interaction $H_{NL}$ is applied for a time $t_{\phi}$ at
the measurement stage which corresponds to the second beam splitter
$BS2$. We make use of a dimension test on the final output to demonstrate
failure of two-dimensional non-retrocausal models for the macroscopic
system. \label{fig: Macro-Cat-delayed}\textcolor{blue}{}}
\end{figure}

We map the above scheme onto a macroscopic system using the cat-state
dynamics as shown by Figure \ref{fig: Macro-Cat-delayed}. The input
state is $|\alpha\rangle$. The nonlinear interaction $H_{NL}$ replaces
the beam splitter, and for certain choices of interaction time $t_{\theta}=m\pi/8$
where $m$ is an integer prepares the system in the superposition

\begin{equation}
|\psi\rangle_{p}=e^{-i\varphi}(\cos\theta|\alpha\rangle+i\sin\theta|-\alpha\rangle)\label{eq:two-state-sup-psi_p}
\end{equation}
where $\theta=t_{\theta}/2$ and $\varphi=t_{\theta}/2$ is a phase
factor. This is proved in the Appendix C. The measurement stage corresponding
to the second beam splitter consists of a second interaction $H_{NL}$
applied for a time $t_{\phi}$, so that
\[
|\alpha\rangle\rightarrow|\alpha\rangle_{t}=e^{-i\varphi_{2}}(\cos\phi|\alpha\rangle+i\sin\phi|-\alpha\rangle)
\]
\begin{equation}
|-\alpha\rangle\rightarrow|-\alpha\rangle_{t}=e^{-i\varphi_{2}}(\cos\phi|-\alpha\rangle+i\sin\phi|\alpha\rangle)\label{eq:trans-coh}
\end{equation}
for certain choices of $\phi$. The final state after the interaction
is\textcolor{red}{{} }
\begin{eqnarray}
|\psi\rangle_{f} & = & e^{-iH_{NL}t_{\phi}/\hbar}|\psi\rangle_{p}\nonumber \\
 & = & e^{-i(\varphi+\varphi_{2})}(\cos\theta(\cos\phi|\alpha\rangle+i\sin\phi|-\alpha\rangle)\nonumber \\
 &  & +i\sin\theta(\cos\phi|-\alpha\rangle+i\sin\phi|\alpha\rangle))\nonumber \\
 & = & e^{i\eta}(\cos(\theta+\phi)|\alpha\rangle+i\sin(\theta+\phi)|-\alpha\rangle)\label{eq:trans-macro}
\end{eqnarray}
where $\eta$ is a phase factor. \textcolor{red}{}Identifying $b=1$
as outcome $|\alpha\rangle$ and $b=-1$ as outcome $|-\alpha\rangle$,
we obtain the results 
\begin{equation}
E(\theta,\phi)=\cos(2(\theta+\phi))\label{eq:soln-dw}
\end{equation}
similar to the modified Wheeler-CLP delayed choice experiment.\textcolor{red}{{}
}It is emphasized that the expression for $E(\theta,\phi)$ is
only true for certain values of $\theta$ and $\phi$, where (\ref{eq:trans-coh})
holds.

The set-up is an example of a prepare and measure scenario considered
by CLP \cite{delayed-choice-causal-model-chaves}. In their notation,
the first measurement setting $t_{\theta}$ is denoted $\theta$ and
the second $t_{\phi}$ is denoted by $\phi$. They derived a dimension
witness inequality (DWI) that is satisfied for nonretrocausal models
of no more than two dimensions.\textcolor{red}{{} }In our notation,
this inequality for the preparation settings $\theta$, $\theta'$,
$\theta''$ and the measurement settings $\phi$, $\phi'$ is

\begin{align}
I_{DW}={\color{black}{\color{black}\Bigl|E(\theta,\phi)+E(\theta,\phi')+E(\theta',\phi)\ \ \ \ \ }}\nonumber \\
-E(\theta',\phi')-E(\theta'',\phi)\Bigr| & \leq3\label{eq:dwi-2}
\end{align}
\textcolor{red}{}where here $E(\theta,\phi)=\cos(2(\theta+\phi))$.
The $t_{\theta}$ and $t_{\phi}$ denote the time settings at the
respective beam splitter interactions $H_{NL}$. If we violate DWI,
then this indicates failure of all non-retrocausal classical two-dimensional
models, suggesting the implication of retrocausality if we are to
view the system as observing a two-dimensional classical realist model.
For a classical two-dimensional model, one would conclude that the
choice of measurement $\phi$ affects the earlier state.

The inequality DWI (\ref{eq:dwi-2}) also follows from the assumptions
of macrorealism. Let us suppose the system to be in one or other of
two states $\varphi_{+}$ and $\varphi_{-}$ (such as $|\alpha\rangle$
and $|-\alpha\rangle$) that will give outcomes $+1$ and $-1$ for
the measurement of the macroscopic value $S$ at the times $t_{p}$
and $t_{m}$. Here, $S$ is the sign of $X_{A}$, as defined in Sections
III and IV. Then one may assign hidden variables $\lambda_{p}=\pm1$
and $\lambda_{m}=\pm1$ to the system at each of these times, the
value $+1$ ($-1$) denoting that the outcome for $S$ will be $+1$
($-1$) respectively. If we assume one may measure the value of $\lambda_{p}$
without affecting the value of $\lambda_{m}$ at the later time (and
vice versa), then the expectation value defined as $E(\theta,\phi)=\langle\lambda_{p}\lambda_{m}\rangle$
will satisfy the DW inequality. This is readily proved by calculating
the averages allowing for all possible combinations of values $\pm1$
for $\lambda_{p}$ and $\lambda_{m}$.

It is known that for the solution $E(\theta,\phi)=\cos(2(\theta-\phi))$
given by eq. (\ref{eq:soln-dw}), violation of the DW inequality is
possible, the maximum value for $I_{DW}$ being $I_{DW}=1+2\sqrt{2}=3.8284$.
The angle choices are $\theta=\pi/8$, $\theta'=3\pi/8$, $\theta''=-\pi/4$,
$\phi=\pi/4$, $\phi'=0$ \cite{delayed-choice-causal-model-chaves}.
In the macroscopic case where the solution is $E(\theta,\phi)=\cos(2(\theta+\phi))$,
we select $\theta=\pi/8$, $\theta'=3\pi/8$, $\theta''=7\pi/4$,
$\phi=7\pi/4$, $\phi'=0$.\textcolor{blue}{{} }For these angle choices,
the two-state solution (\ref{eq:trans-coh}) holds (refer Appendix
C), as necessary for a macroscopic two-state test. The maximum violation
$I_{DW}=1+2\sqrt{2}$ is possible for this angle choice. We may
also select $\theta=\pi/4$, $\theta'=\pi/2$, $\theta''=7\pi/8$,
$\phi=13\pi/8$, $\phi'=15\pi/8$.\textcolor{red}{}\textcolor{blue}{}\textcolor{red}{}

\begin{figure}[t]
\begin{centering}
\includegraphics[width=0.8\columnwidth]{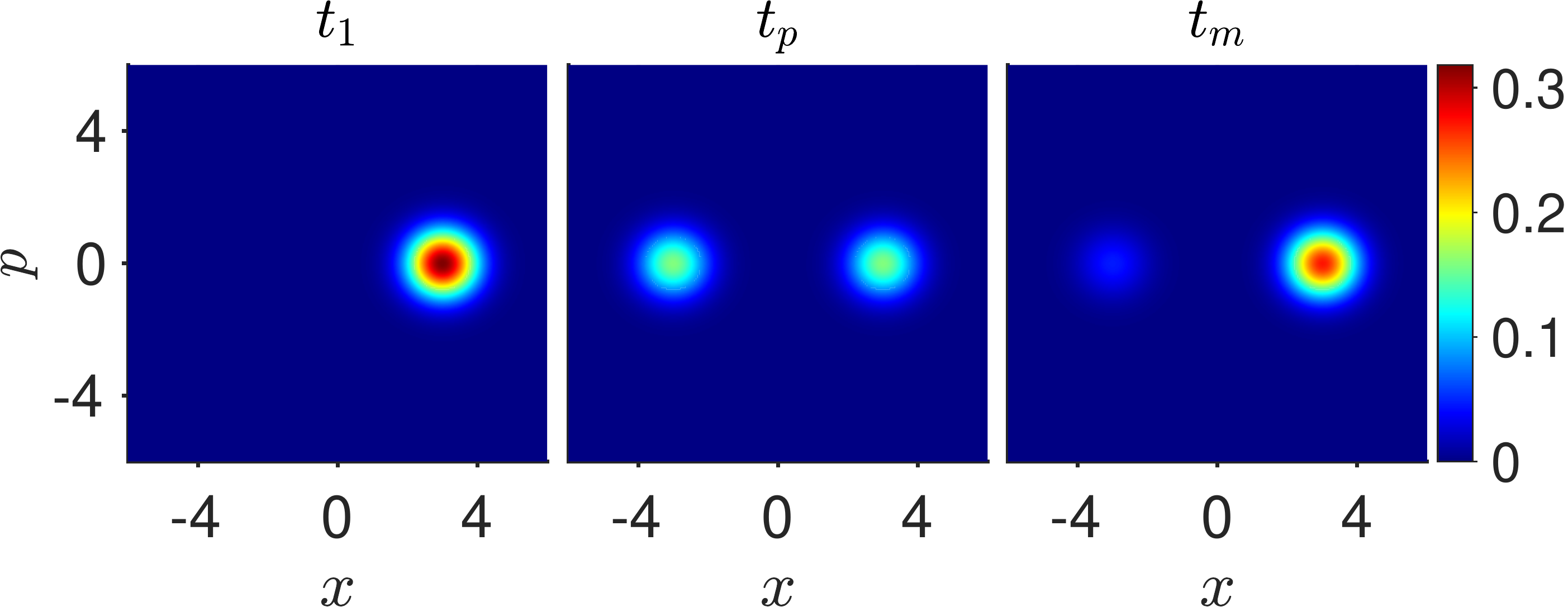}
\par\end{centering}
\smallskip{}
\smallskip{}

\begin{centering}
\includegraphics[width=0.8\columnwidth]{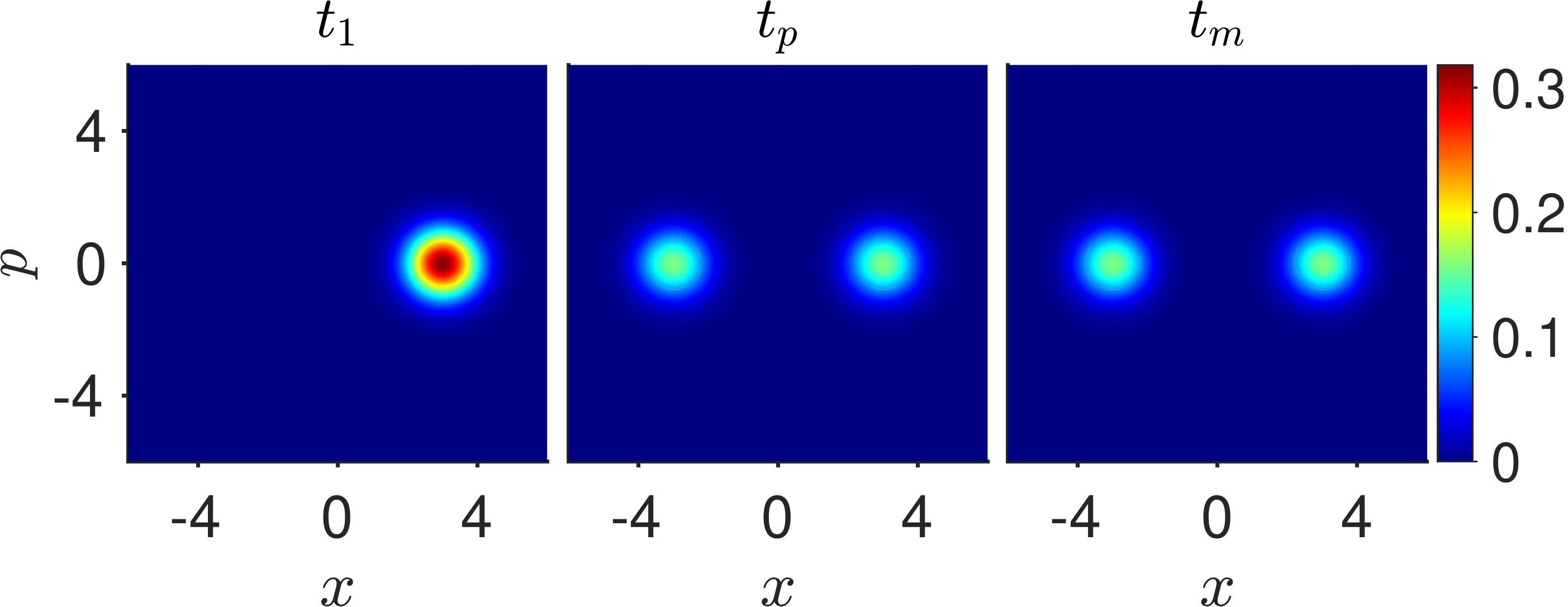}
\par\end{centering}

\caption{\textcolor{brown}{}Contour plots for the $Q$ function $Q(x,p)$
as the system of Figure 11 evolves from the coherent state $|\alpha\rangle$
at time $t_{1}=0$. (\emph{Top}) In this sequence, the first interaction
$H_{NL}$ acts for a time $t_{2}=t_{\theta}$, preparing the system
in the two-state superposition $|\psi\rangle_{p}$ (eq. (\ref{eq:two-state-sup-psi_p}))
at the time $t_{2}=t_{p}$. This is followed by a second interaction
$H_{NL}$ for a time $t_{\phi}$ to produce a final state at time
$t_{3}=t_{m}$. Here, $\theta=\pi/4$ and $\phi=-\pi/8$. \textcolor{red}{}
(\emph{Lower}) The lower sequence depicts the system prepared at the
time $t_{p}$ in a mixture of states $|\alpha\rangle$ and $|-\alpha\rangle$.
This occurs if the system is measured at that time, in such a way
that the system collapses to the mixture. The system then evolves
according to $H_{NL}$ for a time $t_{\phi}$ to produce the final
state at time $t_{3}=t_{m}$.\textcolor{blue}{\label{fig:q-function-catstate-delay}}}
\end{figure}

In Figure \ref{fig:q-function-catstate-delay}, we plot the $Q$
function for the state of the system at the times $t_{0}$, $t_{p}$
and $t_{m}$. The $Q$ function is defined as
\begin{equation}
Q(x,p)=\frac{1}{\pi}\langle\alpha_{0}|\rho|\alpha_{0}\rangle\label{eq:q}
\end{equation}
where $|\alpha_{0}\rangle$ is a coherent state, and $\alpha_{0}=x+ip$.\textcolor{blue}{{}
}The two-state dynamics is evident, as the system evolves under the
action of $H_{NL}$. The $H_{NL}$ provides the rotation into the
superposition state, in analogy to the beam splitter interaction.
Also plotted in Figure 12 is the $Q$ function where the system at
the time $t_{p}$ is prepared in a mixture of $|\alpha\rangle$ and
$|-\alpha\rangle$. This applies where the system in the superposition
at $t_{p}$ is measured, so that an experimentalist may determine
which of the states the system was in at the time $t_{p}$. In fact,
the $Q$ function for the superposition (top graph) differs from that
of the mixture (lower graph) by terms of order $e^{-|\alpha|^{2}}$.
For $\alpha>1$, this difference is not visually noticeable on the
scale of the plots. It is noted however that after the subsequent
rotation $H_{NL}^{\phi}$ ($\phi\neq0$), the $Q$ functions provided
from the superposition (top graph at time $t_{m}$) and the mixture
(lower graph at $t_{m}$) are \emph{macroscopically} distinguishable.

The $Q$ function $Q(x,p)$ corresponds to anti-normally ordered moments,
and hence does not directly correspond to the measured probabilities
for $x$ and $p$ at the microscopic level of $\hbar$. However, at
the macroscopic level where one distinguishes between the two states
$|\alpha\rangle$ and $|-\alpha\rangle$, the $Q$ function accurately
depicts the relative probabilities i.e. the weighting of the two peaks
as pictured in the plots corresponds to the relative probabilities
for the binary outcomes, $b=1$ and $b=-1$. The extra terms of order
$e^{-|\alpha|^{2}}$are negligible.

The violation of the dimension witness inequality indicates failure
of two-dimensional non-retrocausal models. This is not inconsistent
with the non-retrocausal interpretation given by Section IV.C, because
the phase space dynamics relies on a \emph{continuum} of values for
$X$ and $P$. At time $t_{2}$ there is no distinction between the
\emph{macroscopic} depictions $Q(x,p)$ for the superposition and
mixed state (compare also the pictures at $t_{2}$ for the lower sequences
of Figures 7 and 8). Yet, there are differences of order $\hbar e^{-|\alpha|^{2}}$.
It is due to these \emph{microscopic} differences between the superposition
(entangled) and mixed (non-entangled) states, evident in the full
phase-space distribution at $t_{2}$, that there is a different dynamics,
leading to a macroscopic difference in $E(\theta,\phi)$ at the later
time $t_{3}$.\textcolor{blue}{}

\section{Weak macroscopic realism and EPR paradoxes at a microscopic level}

In the previous sections, we show how to realise macroscopic
paradoxes involving Leggett-Garg and dimension witness inequalities.
While there is a contradiction between deterministic macroscopic realism
(dMR) and quantum mechanics for these paradoxes, inconsistency with
weak macroscopic realism (wMR) is not demonstrated at this macroscopic
level. However, inconsistencies arise at the microscopic level.

In this section, we show that at a \emph{microscopic }level where
measurements resolve at the level of $\hbar$, the premises of wMR
and local causality give EPR-type paradoxes \cite{epr-1}. This implies
that there is inconsistency between each of these premises and the
\emph{completeness} of quantum mechanics. EPR paradoxes involving
local causality have been illustrated previously for macroscopic superpositions
of type \foreignlanguage{australian}{\cite{laura-decoh-steer-josa,macro-pointer-interpretation-jphysa}}
\begin{equation}
|\psi\rangle=\frac{1}{\sqrt{2}}(|\alpha\rangle|\uparrow\rangle+|-\alpha\rangle|\downarrow\rangle)\label{eq:cat}
\end{equation}
often taken as an example of a ``Schrodinger cat'' state \cite{cats-brune,frowis-rmp,cats-monroe-1}.
The approach here is similar, since for large $\beta$, the coherent
states $|\beta\rangle$ and $|-\beta\rangle$ are orthogonal qubits.

\subsection{EPR paradox using local causality}

We consider the bipartite system prepared in the Bell state
\begin{equation}
|\psi_{Bell}\rangle=\frac{1}{\sqrt{2}}(|\alpha\rangle|-\beta\rangle-|-\beta\rangle|\alpha\rangle)\label{eq:bell-epr-1}
\end{equation}
at time $t_{2}$, as for (\ref{eq:qe}). The original EPR argument
shows incompatability between the premise of local realism and the
completeness of quantum mechanics \cite{epr-1}. The EPR argument
was generalised to allow for imperfect correlation between the two
sites in \cite{epr-reid-2}, including for spin systems in \cite{bohm-crit,rmpepr-2}.
Here, we apply this generalisation to illustrate the paradox for the
entangled Bell cat state.

The EPR argument considers the prediction for $X_{A}$, given a measurement
at $B$. A measurement of $S_{2}^{(B)}$ at $B$ will ``collapse''
system $A$ to the quantum state $|\alpha\rangle$ or $|-\alpha\rangle$,
implying a variance $(\Delta X_{A})^{2}=1/2$ for $A$, conditioned
on the result for $S_{2}^{(B)}$. We write this conditional variance
as $\Delta_{inf}^{2}X_{A}\equiv(\Delta_{inf}X_{A})^{2}=1/2$, the
variance for the inference of $X_{A}$ given the measurement at $B$.
\begin{figure}[t]
\begin{centering}
\includegraphics[width=0.45\columnwidth]{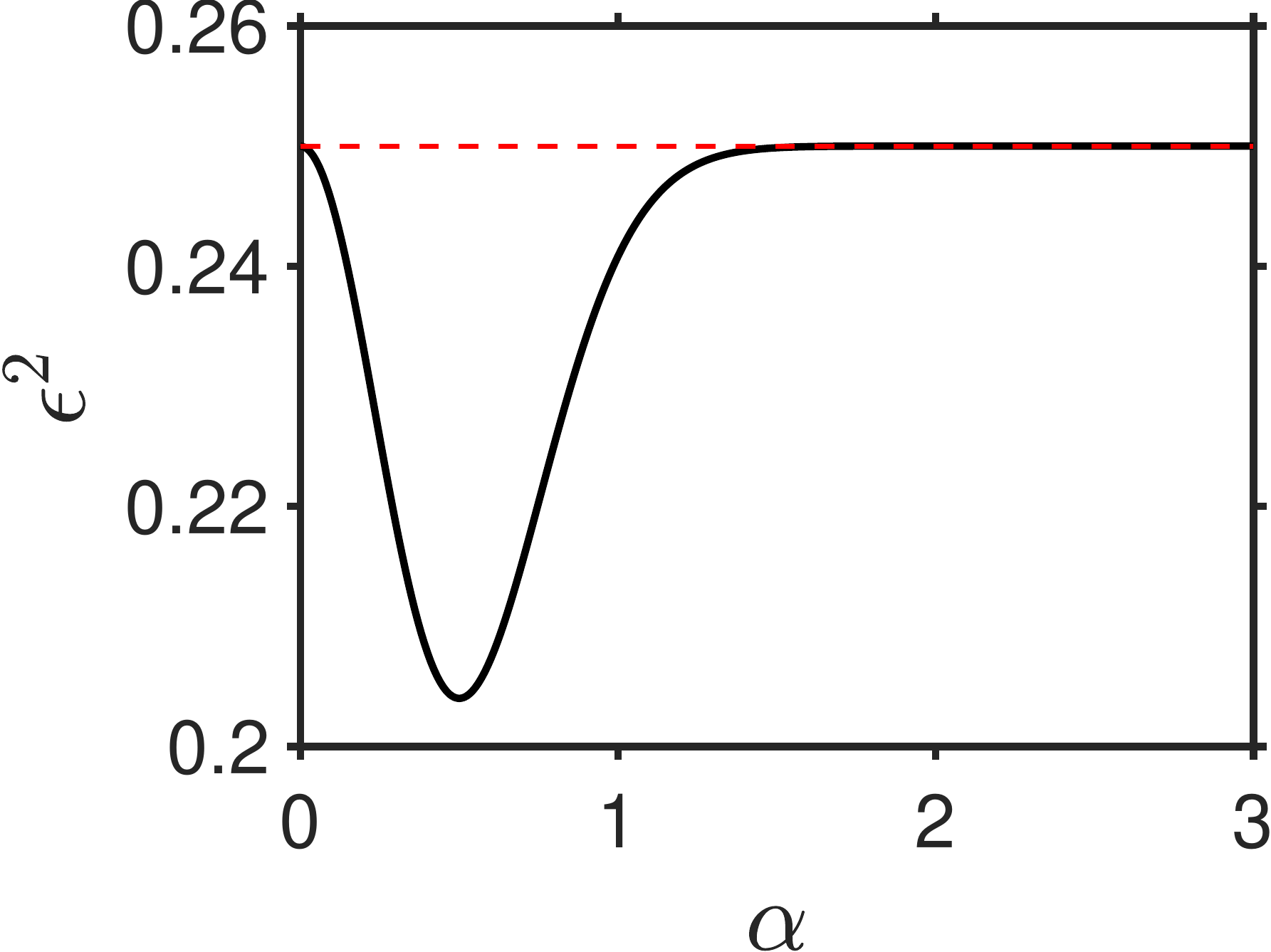}\ \ \includegraphics[width=0.45\columnwidth]{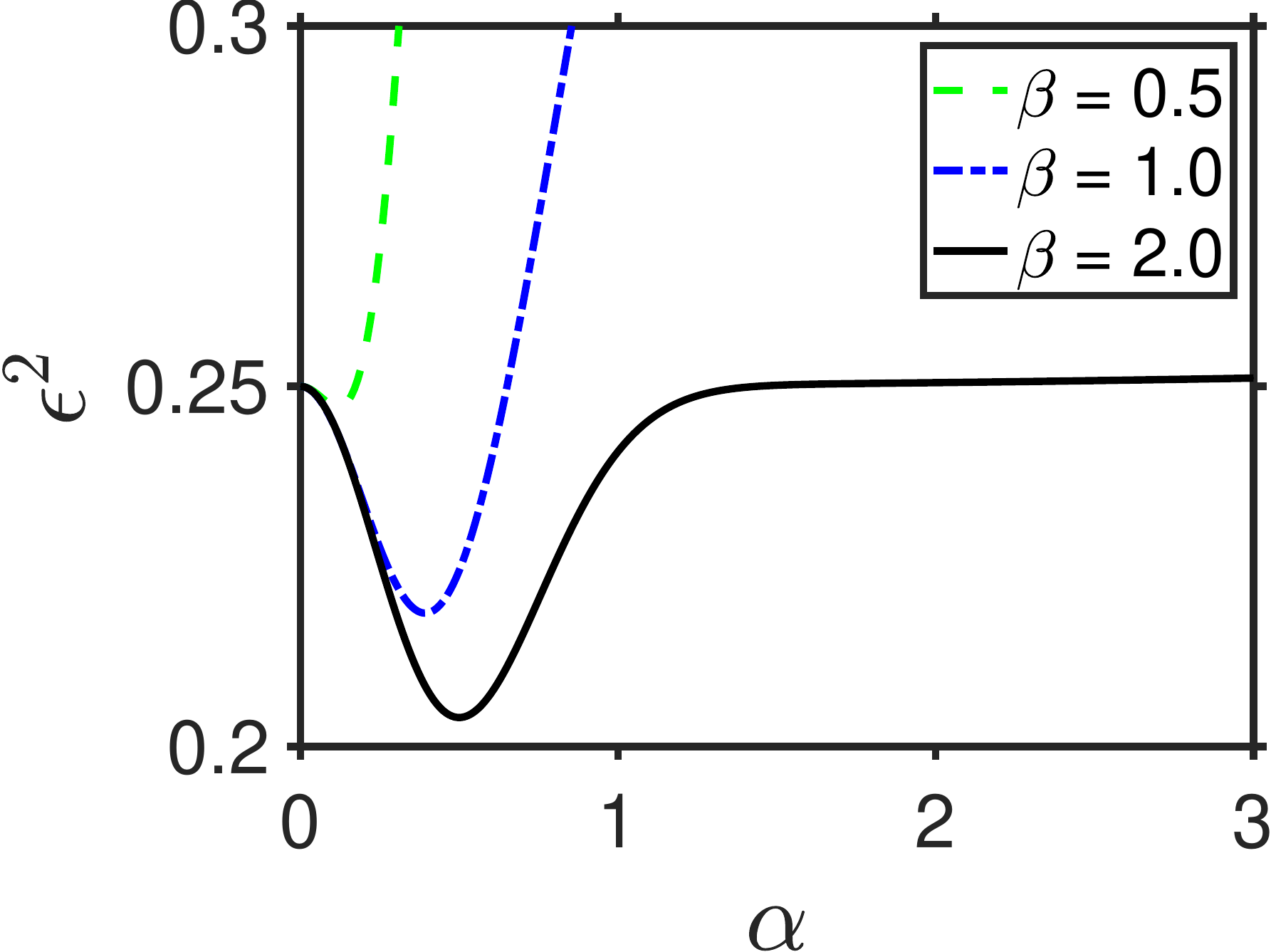}
\par\end{centering}
\caption{Plots showing the violation of the macroscopic EPR inequality (\ref{eq:eprinf}).
We first plot (left) $\mathcal{E}^{2}$ given by eqn (\ref{eq:var-p-simple-epr}).
The same result is given for  $\mathcal{E}_{M}^{2}$ which defined
the macroscopic paradox, eq. (\ref{eq:macro-epr}). \textcolor{red}{}The
second plot (right) shows the full calculation for $\mathcal{E}^{2}$
given by eqs. (\ref{eq:var-epr-xfull}) and (\ref{eq:var-epr-pfull})
based on the proposed method to measure $\beta$ using $X_{B}$, which
assumes $\beta$ to be sufficiently large. Here, we show $\mathcal{E}^{2}$
versus $\alpha$ for $\beta=0.5$, $\beta=1$ and $\beta=2.$ \textcolor{blue}{\label{fig:epr-var}}}
\end{figure}

 The EPR argument then considers the prediction for $P_{A}$ of system
$A$ at time $t_{2}$, as can be inferred from a measurement made
at $B$. Here, we propose that the measurement made at $B$ be given
by $U_{B}(t_{2})^{-1}$ followed by a measurement of $S_{2}^{(B)}$
(the sign of $\hat{X_{B}}$). The state after the transformation
$U_{B}(t_{2})^{-1}$ is (\ref{eq:aftertransP}), and the measurement
of $S_{2}^{(B)}$ allows an inference of the value of $P_{A}$, of
system $A$ at time $t_{2}$. The measurement of $S_{2}^{(B)}$ at
$B$ ``collapses'' system $A$ to either $U_{\pi/8}^{(A)}|\alpha\rangle$
or $U_{\pi/8}^{(A)}|-\alpha\rangle$. Following the method of \cite{epr-reid-2},
the inferred statistics is thus given by $U_{\pi/8}^{(A)}|\alpha\rangle$
or $U_{\pi/8}^{(A)}|-\alpha\rangle$, which are superpositions (\ref{eq:supt2-1-2})
of $|\alpha\rangle$ or $|-\alpha\rangle$, and for which the conditional
distributions are $P_{+}(P_{A})$ and $P_{-}(P_{A})$ of eqn (\ref{eq:supfringep})
respectively. These distributions show fringes, and have the variance
$\Delta_{inf}^{2}P_{A}$ for $P$. This variance of the inferred value
for $P_{A}$ is \cite{macro-pointer-interpretation-jphysa}
\begin{equation}
\Delta_{inf}^{2}P_{A}=\frac{1}{2}-|\alpha|^{2}e^{-4|\alpha|^{2}}\label{eq:var-p-simple-epr}
\end{equation}
The level of combined inference is 
\begin{equation}
\varepsilon=\Delta_{inf}X_{A}\Delta_{inf}P_{A}<\frac{1}{2}\label{eq:eprinf}
\end{equation}
which is below the value for the uncertainty principle, $\Delta X_{A}\Delta P_{A}\geq\frac{1}{2}$,
thus implying an EPR paradox \cite{epr-reid-2}.

It is also known that the observation of (\ref{eq:eprinf}) demonstrates
an EPR steering \cite{hw-1-2,eric-2,uola-steer-review}. If Bell's
premise of \emph{local causality} is assumed valid, the condition
(\ref{eq:eprinf}) is paradoxical because it implies that the system
$A$ cannot be specified as being in any mixture of localised \emph{quantum}
states $\varphi_{+}$ or $\varphi_{-}$ (since such states would need
to violate the uncertainty principle) \cite{hw-1-2,eric-2,uola-steer-review}.
This negates the hypothesis that the system of (\ref{eq:bell-epr-1})
can be regarded as being in either $|\alpha\rangle$ or $|-\alpha\rangle$
(or indeed in any $\varphi_{+}$ or $\varphi_{-}$ if these are to
be quantum states) in a way that is consistent with local causality.
\textcolor{blue}{}The original EPR paradox assumes local realism,
a more specific form of local causality useful when one has perfectly
correlated results for both conjugate measurements.

\begin{figure}[t]
\begin{centering}
\par\end{centering}
\begin{centering}
\includegraphics[width=0.45\columnwidth]{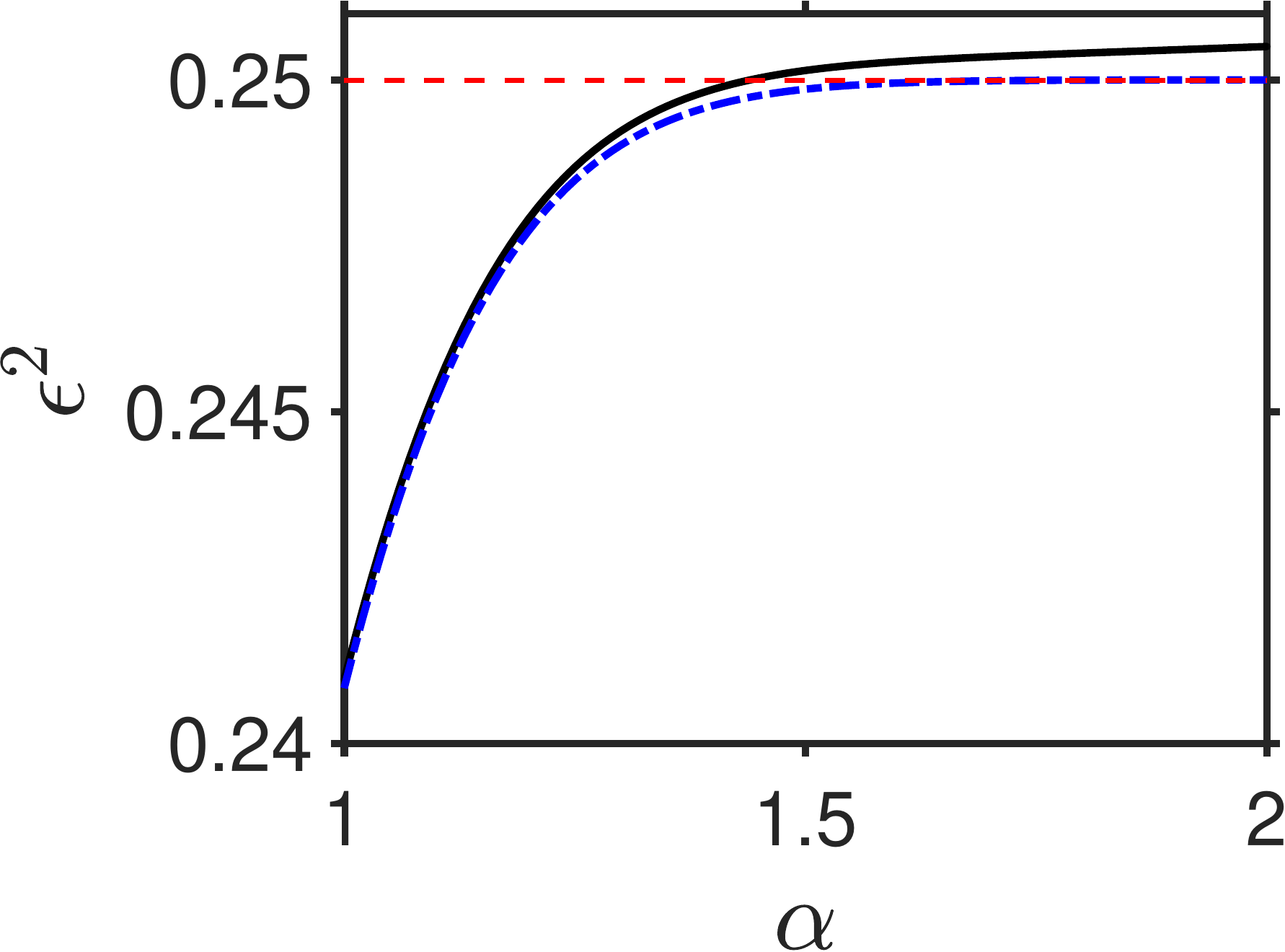}\ \ \includegraphics[width=0.45\columnwidth]{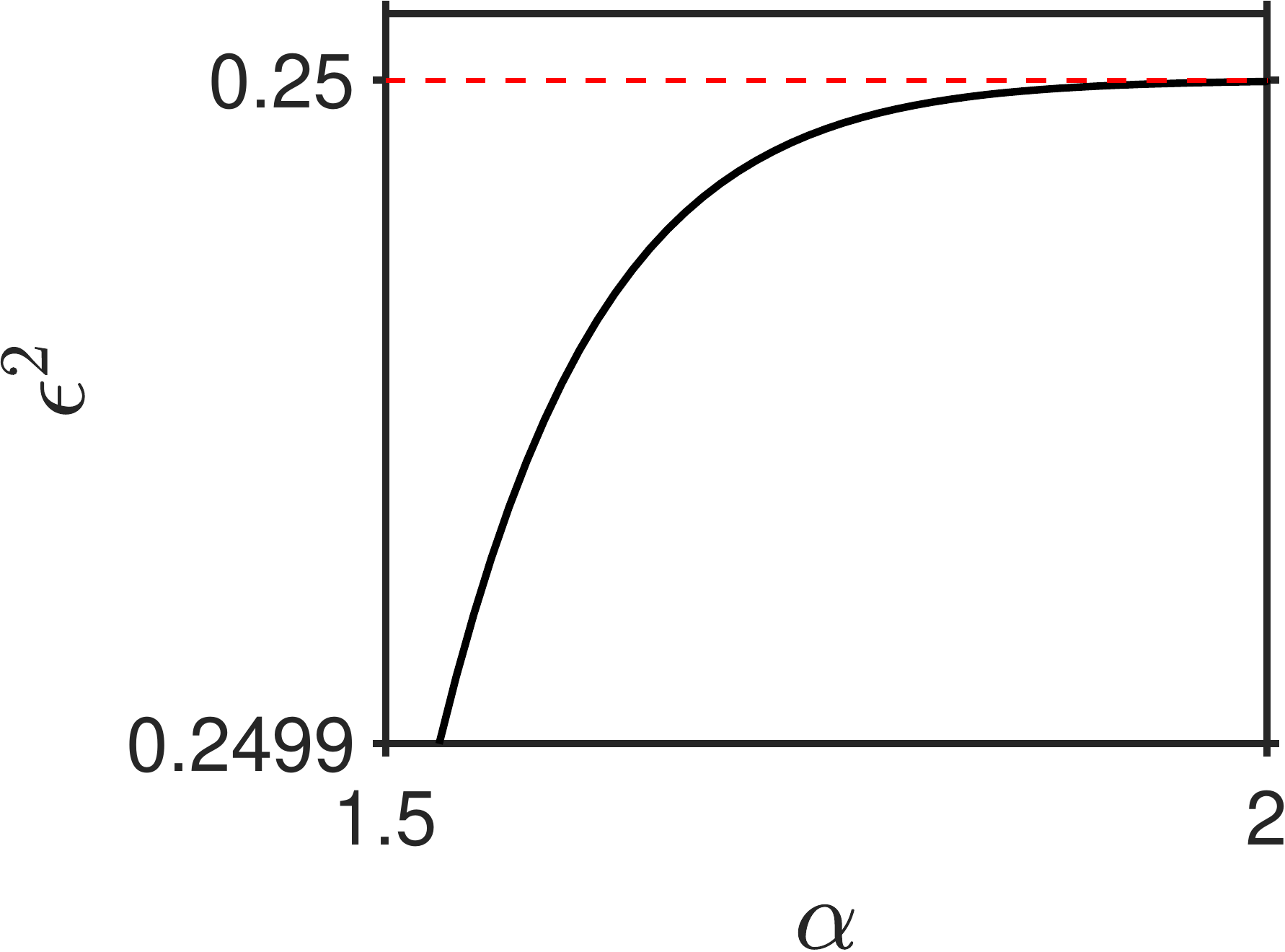}
\par\end{centering}
\caption{Plots showing the violation of the macroscopic EPR inequality (\ref{eq:eprinf}).
Notation as for Figure 13. Here, we give a close-up of results for
the full calculation showing the cut-off values of $\beta$ needed
for $\mathcal{E}^{2}<1/4$, for larger $\alpha$.\textcolor{blue}{\label{fig:epr-var-close-up}}}
\end{figure}
In the above prediction for the EPR inequality, it is assumed that
an idealised measurement at $B$ ``collapses'' system $A$ into
one or other of the coherent states. In a more rigorous analysis,
we evaluate the conditional statistics for system $A$ using the specific
proposal for the measurement at $B$, where the sign of $X_{B}$ is
measured, as in the calculations of Section IV. This gives for the
state (\ref{eq:qe})\textcolor{red}{{} }an inference variance in
$X_{A}$ of
\begin{align}
\Delta_{inf}^{2}X_{A} & =\frac{1}{2}+\frac{2\left|\alpha\right|^{2}}{\left(1-e^{-2\left|\alpha\right|^{2}-2\left|\beta\right|^{2}}\right)}-\frac{2\left|\alpha\right|^{2}erf(\sqrt{2}\left|\beta\right|)^{2}}{\left(1-e^{-2\left|\alpha\right|^{2}-2\left|\beta\right|^{2}}\right)^{2}}\nonumber \\
\label{eq:var-epr-xfull}
\end{align}
The full details are given in the Appendix. Similarly, the inferred
variance for $P$ is calculated assuming the state (\ref{eq:aftertransP}).
We find
\begin{eqnarray}
{\normalcolor {\normalcolor \Delta_{inf}^{2}P_{A}}} & = & {\color{blue}{\color{black}\frac{1}{2}+}{\color{black}\frac{2|\alpha|^{2}}{{\color{black}\left(e^{2|\alpha|^{2}+2|\beta|^{2}}-1\right)}}-\frac{|\alpha|^{2}erf(\sqrt{2}|\beta|)^{2}}{\{e^{2|\alpha|^{2}}-e^{-2|\beta|^{2}}\}^{2}}}}\nonumber \\
\label{eq:var-epr-pfull}
\end{eqnarray}
In the limit of large $\beta$, where the measurement becomes ideal,
we see that $\Delta^{2}X_{A}\rightarrow1/2$ and $\Delta^{2}P_{A}$
reduces to (\ref{eq:var-p-simple-epr}), consistent with the arguments
above.  Figures 13 and 14 plot $\varepsilon^{2}$
for varying $\beta$. The results become indistinguishable from the
ideal case for larger $\beta$.

\subsection{EPR paradox based on weak macroscopic realism}

The original EPR paradox argues the incompleteness of quantum mechanics
based on the assumption of local realism, or local causality, as above.
As explained in \cite{manushan-bell-cat-lg}, one may also argue
an EPR paradox based on the validity of weak macroscopic realism.
We summarise this result, for the purpose of comparison.

The cat state for system $A$ is the superposition $c_{_{+}}|\alpha\rangle+ic_{-}|-\alpha\rangle$
(for $\alpha$ large), where $c_{+}/c_{-}$ is real. Weak macroscopic
realism postulates that the system $A$ in such a state is actually
in one or other state $\varphi_{+}$ and $\varphi_{-}$ for which
the value of the macroscopic spin $S_{2}^{(A)}$ is determined. The
spin $S_{2}^{(A)}$ is measured from the quadrature amplitude$X_{A}$
(as the sign of $X_{A}$). The distribution $P(X_{A})$ for $X_{A}$
gives two distinct Gaussian hills, each hill with variance $(\Delta X_{A})^{2}=1/2$
\cite{yurke-stoler-1}. Following \cite{manushan-bell-cat-lg}, one
may specify the variance of $X_{A}$ for the states $\varphi_{+}$
and $\varphi_{-}$. We denote the specified variances as $(\Delta X_{A})_{+}^{2}$
and $(\Delta X_{A})_{-}^{2}$ respectively. With the assumption that
$\varphi_{+}$ and $\varphi_{-}$ are to be \emph{quantum} states,
the Heisenberg uncertainty relation $(\Delta X_{A})(\Delta P_{A})\geq1/2$
applies to each state. Then, as explained in \cite{macro-coherence-paradox},
for the ensemble of systems in a classical mixture of states $\varphi_{+}$
and $\varphi_{-}$, it is readily proved that $(\Delta X_{A})_{ave}(\Delta P_{A})\geq1/2$,
where $(\Delta X_{A})_{ave}^{2}=P_{+}(\Delta X_{A})_{+}^{2}+P_{-}(\Delta X_{A})_{-}^{2}$,
$P_{+}+P_{-}=1$ and $P_{\pm}\geq0$. The violation of
\begin{equation}
\varepsilon_{M}\equiv(\Delta X_{A})_{ave}(\Delta P_{A})\geq1/2\label{eq:macro-epr}
\end{equation}
will therefore imply incompatibility of weak macroscopic realism with
the completeness of quantum mechanics, since in this case the states
$\varphi_{+}$ and $\varphi_{-}$ cannot be represented as quantum
states. Since here $(\Delta X_{A})_{ave}\rightarrow1/2$ (or more
precisely $(\Delta X_{A})_{ave}\ngtr1/2$), we find the inequality
(\ref{eq:macro-epr}) is violated for $(\Delta P_{A})^{2}<1/2$. This
is the case for the Leggett-Garg gedanken experiment, where the distribution
$P(P_{A})$ at times $t_{2}$ and $t_{3}$ is given by eqn (\ref{eq:supfringep}).
The variance is\textcolor{red}{{} }\cite{macro-pointer-interpretation-jphysa}
\begin{equation}
(\Delta P_{A})^{2}=\frac{1}{2}-\alpha^{2}e^{-4\alpha^{2}}\label{eq:varp-cat}
\end{equation}
The violation is plotted in Figure 13.

\subsection{Discussion}

In conclusion, if one assumes weak macroscopic realism (wMR) for the
state in a superposition of $|\alpha\rangle$ and $|-\alpha\rangle$,
then the fringe distributions shown in Figure 2 do \emph{not} indicate
that the system cannot be regarded as having a definite value for
the macroscopic spin (as sometimes interpreted). Rather, the fringes
signify that those states $\varphi_{+}$ or $\varphi_{-}$ which would
have definite macroscopic spin values (if defined consistently with
wMR) cannot be given as \emph{quantum} states. There is an incompleteness
of quantum mechanics, if wMR is to be valid.

The original EPR paradox concluded inconsistency between local realism
and the completeness of quantum mechanics \cite{epr-1}. Bell later
showed that local realism itself can be falsified \cite{bell-3}.
Similarly, the EPR paradox of Section VI.A shows inconsistency between
local causality (at the level of $\hbar$) and the completeness of
quantum mechanics. However, the assumption of local causality itself
has been falsified, based on Bell theorems \cite{bell-rmp-review,bell-later-theorems},
thereby apparently resolving the paradox. By contrast, the EPR-type
paradox explained in Section VI.B is not readily resolved in the same
manner. This paradox shows inconsistency between wMR and the completeness
of quantum mechanics \cite{manushan-bell-cat-lg}. However, there
is to date no obvious way to falsify wMR. The paradox involving weak
macroscopic realism is hence different and stronger.

While the present paper studies the EPR paradox associated with a
macroscopic superposition state constructed from coherent states,
similar EPR paradoxes have been formulated for other types of macroscopic
superposition states, e.g for NOON states \cite{teh-noon-epr-steer,herald-noon-1}
and Greenberger-Horne-Zeilinger (GHZ) states \foreignlanguage{australian}{\cite{macro-pointer-interpretation-jphysa,gHZ}}.
However, these paradoxes give inconsistencies for local causality,
or local realism. Less has been done on paradoxes that illustrate
the inconsistency between weak macroscopic realism and the incompleteness
of quantum mechanics, although related examples were given for number-state
superpositions in \cite{macro-coherence-paradox}. We expect such
paradoxes may also be possible for NOON and GHZ states, and for the
higher dimensional GHZ extensions with multiple particles at each
site \foreignlanguage{australian}{\cite{ghz-multi-mdr-wjmunro,son-lee-kim-ghz-multi}}.

The method of ``irrealism'' gives a powerful way to detect the incompleteness
of quantum mechanics, along the lines proposed by EPR \cite{irrealism},
which could be applied to the examples considered here. In fact, recent
work uses irrealism to analyse the incompleteness of the state for
the double-slit experiment \cite{irrealism-fringes}.

\section{Conclusion}

In this paper, we have illustrated how one may perform delayed-choice
experiments using superpositions of two coherent states. We map the
original proposals involving spin qubits ($|\uparrow\rangle$ and
$|\downarrow\rangle$) onto macroscopic tests, where the qubits are
coherent states $|\alpha\rangle$ and $|-\alpha\rangle$ ($\alpha\rightarrow\infty$).
The choice of measurement settting corresponds to a choice of a particular
unitary interaction. This gives a mapping between the rotations for
the spin qubits and those for coherent-state qubits. In order to
counter interpretations of the gedanken experiments that would suggest
macroscopic retrocausality, we have demonstrated consistency of the
predictions with the concept of weak macroscopic realism (wMR).

In Section III, we have presented a version of the delayed-choice
quantum eraser experiment, using entangled cat states. The loss of
which-way information shows as interference fringes in distributions
for the quadrature phase amplitude $P$. We argued the signature is
at the microscopic level of $\hbar$ (since the fringes must be finely
resolved) and hence that there is no evidence of macroscopic retrocausality.

Motivated further, in Sections IV we examined a delayed-choice version
of a macroscopic Leggett-Garg test for the entangled cat states. Here,
the test explicitly demonstrates failure of macrorealism, thus suggesting
an apparent macroscopic retrocausality. The violations of the Leggett-Garg
inequalities were then explained by introducing the concept of deterministic
macroscopic realism (dMR), which for entangled systems may also be
defined as macroscopic local realism. The premise dMR is stricter
than that of wMR. We showed that the violations of Leggett-Garg inequalities
falsify dMR, but can be viewed consistently with wMR. We thus avoid
interpretations of macroscopic retrocausality, by noting the failure
of dMR where one has unitary dynamics (in the form of basis rotations
that determine the measurement settings), at both sites.

In Section V, the apparent macroscopic retrocausality of the Leggett-Garg
set-up is demonstrated in a rigorous way, by showing violation of
the dimension witness inequality as in the work of Chaves, Lemos and
Pienaar \cite{delayed-choice-causal-model-chaves}. This implies failure
of all two-dimensional non-retrocausal models. One may avoid the conclusion
of macroscopic retrocausality, however, because of the higher dimensions
evident in the phase-space solutions.

We further showed in Section VI that, although the macroscopic experiments
are consistent with weak macroscopic realism (wMR), EPR paradoxes
exist for measurements giving a microscopic resolution. The paradoxes
indicate incompatibility between local causality (and wMR) with the
completeness of quantum mechanics. The latter is a strong paradox,
because wMR has not yet been falsified.

It is interesting to consider the prospect of an experiment. The two-mode
entangled cat states have been generated \cite{cat-bell-wang-1,Milman-cat,Leghtas-cat}.
The significant challenge is to realise the unitary rotation, which
is given by the Hamiltonian $H_{NL}=\Omega\hat{n}^{4}$ with a quartic
dependence on the field boson number. The quantum eraser can be
carried out more straightforwardly, using the interaction $H_{NL}=\Omega\hat{n}^{2}$,
which has been experimentally achieved as a Kerr nonlinearity \cite{collapse-revival-bec-2,collapse-revival-super-circuit-1}.
Realisations may also be possible using mesoscopic NOON states and
the nonlinear $N$-scopic beam splitter interaction for $N$ bosons,
described in \cite{macro-bell-lg}.

\section*{Acknowledgements}

This research has been supported by the Australian Research Council
Discovery Project Grants schemes under Grant DP180102470.

\begin{widetext}

\section*{Appendix}

\subsection{Quantum eraser and EPR calculation}

Here, we give details for the superposition $U_{\pi/8}|\pm\alpha\rangle=e^{-i\pi/8}\{\cos\pi/8|\pm\alpha\rangle+i\sin\pi/8|\mp\alpha\rangle\}$
examined in Section III. The calculations for the superposition $U_{\pi/4}|\pm\alpha\rangle$
are similar.

It is straightforward to evaluate $P(P_{A})_{+}=\left|\langle P_{A}|U_{\pi/8}|\alpha\rangle\right|^{2}$
and $P(P_{A})_{-}=\left|\langle P_{A}|U_{\pi/8}|-\alpha\rangle\right|^{2}$
for the simple case. For the accurate calculation based on the actual
measurements that would be used, one considers $|\psi(t_{4})\rangle$
and evaluates $P(P_{A},X_{B})=\left|\langle X_{B}|\langle P_{A}|\psi(t_{4})\rangle\right|^{2}$\textcolor{red}{}
\begin{eqnarray}
P(P_{A},X_{B}) & = & 2\frac{\exp(-P_{A}^{2}-X_{B}^{2}-2|\beta|^{2})}{\pi{\color{blue}{\normalcolor \left(1-e^{-2|\alpha|^{2}-2|\beta|^{2}}\right)}}}\{\sin^{2}(\sqrt{2}P_{A}|\alpha|)+\sinh^{2}(\sqrt{2}X_{B}|\beta|)\nonumber \\
 &  & -\frac{\sqrt{2}}{4}\sin(2\sqrt{2}P_{A}|\alpha|)\sinh(2\sqrt{2}X_{B}|\beta|)\}\label{eq:ap8}
\end{eqnarray}
This gives the result (\ref{eq:full-fringe}) using $P(P_{A})_{\pm}=P(P_{A}|X_{B}\gtrless0)$\textcolor{blue}{{}
 }and 
\begin{eqnarray}
{\normalcolor {\color{blue}{\normalcolor P(X_{B})}}} & {\normalcolor {\color{blue}{\normalcolor =}}} & {\normalcolor {\color{blue}{\normalcolor \int P(P_{A},X_{B})dP_{A}}}={\normalcolor \frac{\exp(-X_{B}^{2}-2|\beta|^{2})}{\sqrt{\pi}\left(1-e^{-2|\alpha|^{2}-2|\beta|^{2}}\right)}}{\normalcolor \left(1-e^{-2|\alpha|^{2}}+2\sinh^{2}(\sqrt{2}X_{B}|\beta|)\right)}}\label{eq:ap7}
\end{eqnarray}

To evaluate the EPR correlations, we calculate the variance of $P(P_{A})_{\pm}$.
We find for the simple analysis\textcolor{red}{{} }
\begin{align}
\int P_{A}P(P_{A})_{\pm}dP_{A} & =\frac{1}{\pi^{1/2}}\{\int P_{A}e^{-P_{A}^{2}}dP_{A}{\color{blue}{\normalcolor \mp}}\frac{1}{\sqrt{2}}\int P_{A}e^{-P_{A}^{2}}\sin(2\sqrt{2}P_{A}|\alpha|)dP_{A}\}\nonumber \\
 & =\frac{1}{\pi^{1/2}}\{0{\normalcolor {\color{blue}{\normalcolor \mp}}\frac{1}{\sqrt{2}}}\sqrt{2}\sqrt{\pi}|\alpha|e^{-2|\alpha|^{2}}\}={\color{blue}{\normalcolor \mp}}|\alpha|e^{-2|\alpha|^{2}}\nonumber \\
\int P_{A}^{2}P(P_{A})_{\pm}dP_{A} & =\frac{1}{\pi^{1/2}}\{\int P_{A}^{2}e^{-P_{A}^{2}}dP_{A}{\color{blue}{\normalcolor \mp}}\frac{1}{\sqrt{2}}\int P_{A}^{2}e^{-P_{A}^{2}}\sin(2\sqrt{2}P_{A}|\alpha|)dP_{A}\}\nonumber \\
 & =\frac{1}{\pi^{1/2}}\{\frac{\sqrt{\pi}}{2}{\color{blue}{\normalcolor \mp}}0\}=\frac{1}{2}\label{eq:ap9}
\end{align}
which gives the result (\ref{eq:var-p-simple-epr}). For the complete
measurement, we use the full result (\ref{eq:full-fringe}) for $P(P_{A})_{\pm}$.
Integration gives
\begin{align}
\int P_{A}P(P_{A})_{\pm}dP_{A} & ={\color{blue}{\normalcolor \mp\frac{|\alpha|erf(\sqrt{2}|\beta|)}{\{e^{2|\alpha|^{2}}-e^{-2|\beta|^{2}}\}}}}\nonumber \\
\int P_{A}^{2}P(P_{A})_{\pm}dP_{A} & =\frac{1}{2}+\frac{2|\alpha|^{2}}{\left(e^{2|\alpha|^{2}+2|\beta|^{2}}-1\right)}\label{eq:ap10}
\end{align}
leading to (\ref{eq:var-epr-pfull}).

\subsection{Calculation of EPR correlations}

We first evaluate $\Delta_{inf}^{2}X_{A}$ for the state (\ref{eq:qe}).
The \textcolor{red}{}inferred variance is defined as
\begin{align}
\Delta_{inf}^{2}X_{A} & =P(X_{B}>0)\Delta_{+}^{2}X_{A}+P(X_{B}\leq0)\Delta_{-}^{2}X_{A}\label{eq:var-app}
\end{align}
where clearly $P(X_{B}>0)=1/2$ . The conditional distributions are
defined
\begin{eqnarray}
P_{+}(X_{A})=P(X_{A}|X_{B}>0) & =\frac{\int_{0}^{\infty}P(X_{A},X_{B})dX_{B}}{\int_{0}^{\infty}P(X_{B})dX_{B}}\label{eq:pcond-ap}
\end{eqnarray}
and similarly $P_{-}(X_{A})=P(X_{A}|X_{B}\leq0)$, which, after evaluation
of $P(X_{A},X_{B})$ for the entangled cat state, gives
\begin{align}
P(X_{A})_{\pm} & =\frac{2\mathcal{N}^{2}e^{-X_{A}^{2}-2\left|\alpha\right|^{2}}}{\sqrt{\pi}}\left\{ \cosh(2\sqrt{2}\left|\alpha\right|X_{A})\mp erf(\sqrt{2}\left|\beta\right|)\sinh(2\sqrt{2}\left|\alpha\right|X_{A})-e^{-2\left|\beta\right|^{2}}\Biggl\}\right.\label{eq:pfull-ap}
\end{align}
The variance of these distributions are $\Delta_{\pm}^{2}\hat{X}_{A}=\left\langle \hat{X}_{A}^{2}\right\rangle -\left\langle \hat{X}_{A}\right\rangle ^{2}$where
\begin{eqnarray}
\left\langle \hat{X}_{A}\right\rangle _{\pm} & =\int P_{\pm}(X_{A})X_{A}dX_{A} & =\frac{\mp\sqrt{2}\left|\alpha\right|erf(\sqrt{2}\left|\beta\right|)}{\left(1-e^{-2\left|\alpha\right|^{2}-2\left|\beta\right|^{2}}\right)}\nonumber \\
\left\langle \hat{X}_{A}^{2}\right\rangle _{\pm} & =\int P_{\pm}(X_{A})X_{A}^{2}dX_{A} & =\frac{1}{2}+\frac{2\left|\alpha\right|^{2}}{\left(1-e^{-2\left|\alpha\right|^{2}-2\left|\beta\right|^{2}}\right)}
\end{eqnarray}
This leads to the result
\begin{align}
\Delta_{inf}^{2}X_{A} & =\frac{1}{2}+4\mathcal{N}^{2}\left|\alpha\right|^{2}-8\mathcal{N}^{4}\left|\alpha\right|^{2}erf(\sqrt{2}\left|\beta\right|)^{2}\nonumber \\
 & =\frac{1}{2}+\frac{2\left|\alpha\right|^{2}}{\left(1-e^{-2\left|\alpha\right|^{2}-2\left|\beta\right|^{2}}\right)}-\frac{2\left|\alpha\right|^{2}erf(\sqrt{2}\left|\beta\right|)^{2}}{\left(1-e^{-2\left|\alpha\right|^{2}-2\left|\beta\right|^{2}}\right)^{2}}\label{eq:full-var-ap}
\end{align}
Similarly, we evaluate $\Delta_{inf}^{2}P_{A}$ for the state (\ref{eq:aftertransP}).
Here,
\begin{align*}
\Delta_{inf}^{2}P_{A} & =P(X_{B}>0)\Delta_{+}^{2}P_{A}+P(X_{B}\leq0)\Delta_{-}^{2}P_{A}
\end{align*}
We first evaluate evaluate the conditional distributions of
\begin{eqnarray}
P_{+}(P_{A})=P(P_{A}|X_{B}>0) & =\frac{\int_{0}^{\infty}P(P_{A},X_{B})dX_{B}}{\int_{0}^{\infty}P(X_{B})dX_{B}}\label{eq:p-full-ap}
\end{eqnarray}
and similarly, $P_{+}(P_{A})=P(P_{A}|X_{B}\leq0)=\frac{\int_{0}^{\infty}P(P_{A},X_{B})dX_{B}}{\int_{-\infty}^{0}P(X_{B})dX_{B}}$
using
\begin{eqnarray}
P(P_{A},X_{B}) & = & \left|\langle X_{B}|\langle P_{A}|\psi(t_{4})\rangle\right|^{2}\nonumber \\
 & = & \frac{e^{-P_{A}^{2}}}{\sqrt{\pi}\{1-e^{-2|\alpha|^{2}-2|\beta|^{2}}\}}\left\{ 1-e^{-2|\beta|^{2}}\cos(2\sqrt{2}P_{A}|\alpha|)-\frac{\sqrt{2}}{2}erf(\sqrt{2}|\beta|)\sin(2\sqrt{2}P_{A}|\alpha|)\right\} \label{eq:pjoint-ap}
\end{eqnarray}
This gives
\begin{align}
P_{\pm}(P_{A}) & =\frac{2\mathcal{N}^{2}e^{-P_{A}^{2}}}{\sqrt{\pi}}\left\{ 1-e^{-2|\beta|^{2}}\cos(2\sqrt{2}P_{A}|\alpha|)\mp\frac{\sqrt{2}}{2}erf(\sqrt{2}|\beta|)\sin(2\sqrt{2}P_{A}|\alpha|)\Biggl\}\right.\label{eq:pp-ap}
\end{align}
Hence
\begin{eqnarray}
\left\langle \hat{P}_{A}\right\rangle _{\pm} & =\int P_{\pm}(P_{A})P_{A}dP_{A} & =\mp\frac{|\alpha|e^{-2|\alpha|^{2}}erf(\sqrt{2}|\beta|)}{\{1-e^{-2|\alpha|^{2}-2|\beta|^{2}}\}}\nonumber \\
\left\langle \hat{P}_{A}^{2}\right\rangle _{\pm} & =\int P_{\pm}(P_{A})P_{A}^{2}dP_{A} & =\frac{1}{2}+\frac{2|\alpha|^{2}}{\{e^{2|\alpha|+2|\beta|^{2}}-1\}}
\end{eqnarray}
which leads to
\begin{align}
\Delta_{inf}^{2}P_{A} & =\frac{1}{2}+4\mathcal{N}^{2}|\alpha|^{2}e^{-2|\alpha|-2|\beta|^{2}}-4\mathcal{N}^{4}|\alpha|^{2}e^{-4|\alpha|^{2}}erf(\sqrt{2}|\beta|)^{2}\nonumber \\
 & =\frac{1}{2}+\frac{2|\alpha|^{2}e^{-2|\alpha|-2|\beta|^{2}}}{\{1-e^{-2|\alpha|^{2}-2|\beta|^{2}}\}}-\frac{|\alpha|^{2}e^{-4|\alpha|^{2}}erf(\sqrt{2}|\beta|)^{2}}{\{1-e^{-2|\alpha|^{2}-2|\beta|^{2}}\}^{2}}\label{eq:onf-p-ap}
\end{align}
\textcolor{red}{}

\subsection{Cat state dynamics for the Dimension Witness test}

In this section we consider the two state solution for our dynamically
evolved macroscopic cat states under a non-linear interaction. Considering
$\alpha$ to be real, for an initial coherent state $|\alpha\rangle$
undergoing an evolution with a non-linear interaction $H_{NL}$, the
state created after an interaction time $t_{\theta}$ can be written
as,

\begin{align}
\left|\alpha,t_{\theta}\right\rangle  & =\exp[-\frac{\left|\alpha\right|^{2}}{2}]\sum_{n=0}^{\infty}\alpha^{n}\frac{\exp(-i\varOmega t_{\theta}n^{k})}{\sqrt{n!}}\left|n\right\rangle 
\end{align}
\textcolor{brown}{}We restrict to $k=4$. Let us constrain to $t_{\theta}=m\pi/8$
where $m$ is an integer and choose the units of time such that $\Omega=1$.
To obtain the two-state solution in terms of $|\alpha\rangle$ and
$|-\alpha\rangle$, we require solutions of type

\begin{align}
\exp[-\frac{\left|\alpha\right|^{2}}{2}]\sum_{n}\alpha^{n}\frac{\exp(-im\frac{\pi}{8}n^{4})}{\sqrt{n!}}\left|n\right\rangle  & =\exp[-\frac{\left|\alpha\right|^{2}}{2}]\sum_{n}A\frac{\alpha^{n}}{\sqrt{n!}}\left|n\right\rangle +\exp[-\frac{\left|\alpha\right|^{2}}{2}]\sum_{n}B\frac{(-1)^{n}\alpha^{n}}{\sqrt{n!}}\left|n\right\rangle \label{eq:13}
\end{align}
where $A$ and $B$ are constants. Now since the summation indexes
are the same, this requires $\exp(-im\frac{\pi}{8}n^{4})=A+(-1)^{n}B$.
By assigning $n=0,1$ we find $A+B=1$ and $e^{-im\frac{\pi}{8}}=A-B$,
giving the solutions as
\begin{eqnarray}
A & = & e^{-im\frac{\pi}{16}}\cos(m\frac{\pi}{16})\nonumber \\
B & = & ie^{-im\frac{\pi}{16}}\sin(m\frac{\pi}{16})\label{eq:16}
\end{eqnarray}
Hence we propose that for all integers $n$ such that $n=0,1,2,..$

\begin{align}
\exp(-im\frac{\pi}{8}n^{4}) & =e^{-im\frac{\pi}{16}}\left(\cos(m\frac{\pi}{16})+(-1)^{n}i\sin(m\frac{\pi}{16})\right)\label{eq:eqn}
\end{align}
We now prove this to be true. For even $n$, we see that the right
side of equation (\ref{eq:eqn}) satisfies $RHS=1$. We can write
$n=2J$ where $J=1,2,..$ in which case $n^{4}=(2J{\color{black})^{4}{\normalcolor =16J^{4}}}$
. Then we see that the left side ($LHS$) of equation (\ref{eq:eqn})
satisfies $LHS=1$, since $m$ is an integer. Next we consider odd
$n$. We see that $RHS=e^{-im\frac{\pi}{8}}$. We can write $n=2J+1$,
where $J$ is an integer, $J\geq1$. We now show that $n^{4}=(2J+1)^{4}=16M+1$,
where $M$ is integer. This is proved by considering $(2J+1)^{4}=16J^{4}+32J^{3}+24J^{2}+8J+1$
from which we see that the condition holds if $J$ is even. Then
also, $(2J+1)^{4}-1=16\{J^{4}+2J^{3}+\frac{J}{2}(3J+1)\}$. This gives
the result, since $3J+1$ is even if $J$ is odd and the term $\{J^{4}+2J^{3}+\frac{J}{2}(3J+1)\}$
becomes an integer for all values of $J$ \textcolor{blue}{}. Hence,
$LHS=\exp(-im\frac{\pi}{8})$. Hence we can write a two state solution
for time multiples of $\pi/8$, as 

\begin{align}
\left|\alpha,t_{\theta}\right\rangle  & =e^{-it_{\theta}/2}\left(\cos(t_{\theta}/2)\left|\alpha\right\rangle +i\sin(t_{\theta}/2)\left|-\alpha\right\rangle \right)\label{eq:state-result}
\end{align}
where $t_{\theta}=m\frac{\pi}{8}$.

\end{widetext}


\begin{thebibliography}{References}
\bibitem{wheeler-retro}J. A. Wheeler, \textquotedblleft The \textquoteleft past\textquoteright{}
and the \textquoteleft delayed-choice\textquoteright{} double-slit
experiment,\textquotedblright{} in Mathematical Foundations of Quantum
Theory, edited by A. R. Marlow (Academic Press, New York, 1978) pp.
9--48.

\bibitem{wheeler-retro-2}J. A. Wheeler in ``Quantum Theory and Measurement''
by J. A. Wheeler and W. H. Zurek (Princeton University Press pp. 192-213,
1984).\textcolor{red}{}

\bibitem{ma-zeilinger-rmp-delay-choice}\foreignlanguage{australian}{Xi.
Ma, J. Kofler, and A. Zeilinger, ``Delayed-choice gedanken experiments
and their realizations'', Rev. Mod. Phys. \textbf{88,} 015005 (2016)
and references therein.\textcolor{red}{}}

\bibitem{delayed-choice-qubit-scully-druhl}\foreignlanguage{australian}{M.
O. Scully and K. Drühl, Quantum eraser: A proposed photon correlation
experiment concerning observation and 'delayed choice' in quantum
mechanics, Phys. Rev. A \textbf{25,} 2208 (1982).}

\bibitem{scully-englert-walther}\foreignlanguage{australian}{M. O.
Scully, B.-G. Englert, and H. Walther, Quantum optical tests of complementarity,
Nature (London) \textbf{351}, 111 (1991).\textcolor{blue}{}}

\bibitem{herzog-zeilinger-quantum-eraser-complementarity-exp}\foreignlanguage{australian}{T.
J. Herzog, P. G. Kwiat, H. Weinfurter, and A. Zeilinger, Complementarity
and the quantum eraser, Phys. Rev. Lett. \textbf{75}, 3034 (1995).}

\bibitem{kim-scully-entangled-quantum-eraser-experiment}\foreignlanguage{australian}{Y.-H.
Kim, R. Yu, S. P. Kulik, Y. Shih, and M. O. Scully, Delayed choice
quantum eraser, Phys. Rev. Lett. \textbf{84}, 1 (2000).}

\bibitem{walborn-double-slit-quantum-eraser-experiment}\foreignlanguage{australian}{S.
P. Walborn, M. O. Terra Cunha, S. Pádua, and C. H. Monken, Double-Slit
Quantum Eraser, Phys. Rev. A \textbf{65}, 033818 (2002).\textcolor{blue}{}}

\bibitem{jacques-aspect-experiment-wheeler}\foreignlanguage{australian}{V.
Jacques, E. Wu, F. Grosshans, F. Treussart, P. Grangier, A. Aspect,
and J.-F. Roch, ``Experimental realization of wheelers delayed-choice
gedanken experiment''. Science \textbf{315}, 5814 (2007).\textcolor{blue}{}}

\bibitem{truscott-atoms-delayed-choice}\foreignlanguage{australian}{A.
G. Manning, R. I. Khakimov, R. G. Dall, and A. G. Truscott, ``Wheeler\textquoteright s
delayed-choice gedanken experiment with a single atom''. Nat. Phys.
\textbf{11,} 539--542\textcolor{red}{{} }(2015).\textcolor{blue}{}}

\bibitem{tang-exp-wheeler-delayed-choice}\foreignlanguage{australian}{J.-S.
Tang, Y.-L. Li, X.-Y. Xu, G.-Y. Xiang, C.-F. Li, and G.-C. Guo, Realization
of quantum Wheelers delayed-choice experiment, Nat. Photon. \textbf{6},
600 (2012).}

\bibitem{Ma-zeilinger-exp-causal-delayed-choice-interpretation}\foreignlanguage{australian}{Xiao-Song
Ma et al., Quantum erasure with causally disconnected choice, Proceedings
of the National Academy of Sciences \textbf{110} (4) (2013).}

\bibitem{englert-scully-walther-double-slit-am-journ-phy}\foreignlanguage{australian}{B.
G. Englert, M. O. Scully, and H. Walther, Quantum erasure in double-slit
interferometers with which-way detectors, Am. J. Phys. \textbf{67},
325 (1999).\textcolor{red}{}}

\bibitem{mohrhoff-delayed-choice-interpret-am-journ-phy}\foreignlanguage{australian}{U.
Mohrhoff, Objectivity, retrocausation, and the experiment of Englert,
Scully, and Walther, Am. J. Phys. \textbf{67}, 330 (1999).}

\bibitem{kastner-quantum-eraser-interpret-found-phy}\foreignlanguage{australian}{R.
E. Kastner, The \textquoteleft delayed choice quantum eraser\textquoteright{}
neither erases nor delays,\textquotedblright{} Found. Phys. \textbf{49},
717 (2019).}

\bibitem{ingraham-nonlocality-delay-choice}\foreignlanguage{australian}{R.
L. Ingraham, Quantum nonlocality in a delayed-choice experiment with
partial, controllable memory erasing, Phys. Rev. A \textbf{50}, 4502
(1994). R. L. Ingraham,\textquotedblleft Erratum: Phys. Rev. A \textbf{51},
4295 (1995).}\textcolor{red}{}

\bibitem{faetti-quantum-eraser-ent-interpretation-1}\foreignlanguage{australian}{S.
Faetti, ``An alternative analysis of the delayed-choice quantum eraser'',
arXiv:1912.04101 {[}quant-ph{]}.\textcolor{red}{{} }}

\bibitem{la-cour-yudichak-classical-quantum-eraser-dim-wit}\foreignlanguage{australian}{Brian
R. La Cour and Thomas W. Yudichak, Classical model of delayed-choice
quantum eraser, Phys. Rev. A \textbf{103}, 062213 (2021).}

\bibitem{Ionicioiu-terno}\foreignlanguage{australian}{R. Ionicioiu
and D. Terno, Proposal for a quantum delayed-choice experiment, Phys.
Rev. Lett. \textbf{107}, 230406 (2011).}

\bibitem{IT-ent1}\foreignlanguage{australian}{R. Ionicioiu, T. Jennewein,
R. B. Mann, and D. R. Terno, Is wave-particle objectivity compatible
with determinism and locality?, Nat. Commun. \textbf{5}, 4997 (2014).}

\bibitem{IT-ent-2}\foreignlanguage{australian}{R. Rossi, Restrictions
for the causal inferences in an interferometric system, Phys. Rev.
A \textbf{96}, 012106 (2017).}

\bibitem{ITent3}\foreignlanguage{australian}{A. S. Rab, E. Polino,
Z.-X. Man, N. Ba An, Y.-J. Xia, N. Spagnolo, R. Lo Franco, and F.
Sciarrino, Entanglement of photons in their dual wave-particle nature,
Nat. Commun. \textbf{8}, 915 (2017).}

\bibitem{quantum-bs-I-terno-experiment}\foreignlanguage{australian}{A.
Peruzzo, P. Shadbolt, N. Brunner, S. Popescu and J. L. O\textquoteright Brien.
A Quantum Delayed-Choice Experiment, Science \textbf{338}, 634 (2012).}

\bibitem{kaiser-coudreau-milman-experiment-ent-delayed-choice}\foreignlanguage{australian}{F.
Kaiser, T. Coudreau, P. Milman, D. B. Ostrowsky, and S. Tanzilli,
Entanglement-enabled delayed-choice experiment. Science \textbf{338},
637 (2012).}

\bibitem{zheng-quantum-delayed-choiceBS-exp}\foreignlanguage{australian}{S.
B. Zheng, Y. P. Zhong, K. Xu, Q. J. Wang, H. Wang, L. T. Shen, C.
P. Yang, J. M. Martinis, A. N. Cleland, and S. Y. Han, Quantum delayed-choice
experiment with a beam splitter in a quantum superposition, Phys.
Rev. Lett. \textbf{115}, 260403 (2015).}

\bibitem{delayed-choice-causal-model-chaves}R. Chaves, G. B. Lemos
and J. Pienaar, Causal Modeling the Delayed-Choice Experiment,\textit{\emph{
Phys. Rev. Lett. }}\textbf{120}, 190401 (2018).

\bibitem{delayed-choice-experiment-chaves}E. Polino, I. Agresti,
D. Poderini, G. Carvacho, G. Milani, G. B. Lemos, R. Chaves and F.
Sciarrino, Device-independent test of a delayed choice experiment,
Phys. Rev. A \textbf{100}, 022111 (2019).

\bibitem{huang-delayed-choice-causal-model-compatibility}H.-L. Huang,
Y.-H. Luo, B. Bai, Y.-H. Deng, H.Wang, Q. Zhao, H.-S. Zhong, Y.-Q.
Nie,W.-H. Jiang, X.-L.Wang et al., Compatibility of causal hidden-variable
theories with a delayed-choice experiment, Phys. Rev. A \textbf{100},
012114 (2019).

\bibitem{yurke-stoler-1}\foreignlanguage{australian}{B. Yurke and
D. Stoler, }Generating quantum mechanical superpositions of macroscopically
distinguishable states via amplitude dispersion\foreignlanguage{australian}{,
Phys. Rev. Lett. \textbf{57}, 13 (1986).}

\bibitem{manushan-cat-lg}\foreignlanguage{australian}{M. Thenabadu
and M. D. Reid, Leggett-Garg tests of macrorealism for dynamical cat
states evolving in a nonlinear medium, Phys. Rev. A \textbf{99}, 032125}\textcolor{red}{{}
}(2019)\textcolor{black}{.}\textcolor{red}{}\foreignlanguage{australian}{}

\bibitem{macro-bell-lg}M. Thenabadu, G-L. Cheng, T. L. H. Pham, L.
V. Drummond, L. Rosales-Zárate and M. D. Reid, Testing macroscopic
local realism using local nonlinear dynamics and time settings, Phys.
Rev. A \textbf{102}, 022202 (2020).

\bibitem{manushan-bell-cat-lg}M. Thenabadu and M. D. Reid, Bipartite
Leggett-Garg and macroscopic Bell inequality violations using cat
states: distinguishing weak and deterministic macroscopic realism
arXiv:2012.14997; MD Reid and M Thenabadu, Weak versus deterministic
macroscopic realism, arXiv:2101.09476\textcolor{red}{}

\bibitem{legggarg-1}A. Leggett and A. Garg, Quantum mechanics versus
macroscopic realism: is the flux there when nobody looks? Phys. Rev.
Lett. \textbf{54}, 857 (1985).

\bibitem{bell-3}J. S. Bell, On the Einstein-Podolsky-Rosen paradox,
Physics \textbf{1},195 (1964).

\bibitem{epr-1}A. Einstein, B. Podolsky, and N. Rosen, Can Quantum-Mechanical
Description of Physical Reality Be Considered Complete?, Phys. Rev.
\textbf{47}, 777 (1935).

\bibitem{macro-coherence-paradox}M. D. Reid, Criteria to detect macroscopic
quantum coherence, macroscopic quantum entanglement, and an Einstein-Podolsky-Rosen
paradox for macroscopic superposition states, Phys. Rev. A \textbf{100},
052118 (2019).\textcolor{red}{}

\bibitem{cat-bell-wang-1}C. Wang et al., A Schrödinger cat living
in two boxes, Science \textbf{352}, 1087 (2016).

\bibitem{collapse-revival-bec-2}M. Greiner, O. Mandel, T. Hånsch
and I. Bloch, Collapse and revival of the matter wave field of a Bose-Einstein
condensate, Nature \textbf{419}, 51 (2002).

\bibitem{collapse-revival-super-circuit-1}G. Kirchmair et al., Observation
of the quantum state collapse and revival due to a single-photon Kerr
effect, Nature \textbf{495}, 205 (2013).

\bibitem{weak-solid-qubits-williams-jordan}\foreignlanguage{australian}{\textcolor{black}{N.
S. Williams and A. N. Jordan,  Weak Values and the Leggett-Garg Inequality
in Solid-State Qubits, Phys. Rev. Lett. }\textbf{\textcolor{black}{100}}\textcolor{black}{,
026804 (2008).}}

\selectlanguage{australian}%
\bibitem{jordan_kickedqndlg2-2} A. N. Jordan, A. N. Korotkov, and
M. Buttiker, Leggett-Garg Inequality with a Kicked Quantum Pump, Phys.
Rev. Lett.\textbf{ 97}, 026805 (2006). 

\selectlanguage{english}%
\bibitem{nst}\foreignlanguage{australian}{L. Clemente and J. Kofler,
Necessary and sufficient conditions for macroscopic realism from quantum
mechanics, Phys. Rev. A \textbf{91}, 062103 (2015).}

\bibitem{halliwell-lg-multidimension}\foreignlanguage{australian}{J.
J. Halliwell and C. Mawby, Conditions for Macrorealism for Systems
Described by Many-Valued Variables, Phys. Rev. A \textbf{102}, 012209
(2020).}

\bibitem{noon-dowling}J. P. Dowling, Quantum optical metrology --
the lowdown on high-N00N states, Contemporary Physics \textbf{49},
125 (2008).

\bibitem{bognoon-1}B. Opanchuk, L. Rosales-Z\'arate, R. Y Teh, and
M. D. Reid, \foreignlanguage{australian}{Quantifying the mesoscopic
quantum coherence of approximate NOON states and spin-squeezed two-mode
Bose-Einstein condensates,} Phys. Rev. A \textbf{94,} 062125 (2016).

\bibitem{emary-review}\foreignlanguage{australian}{C. Emary, N. Lambert,
and F. Nori, Leggett-Garg inequalities, Rep. Prog. Phys \textbf{77},
016001 (2014).}

\selectlanguage{australian}%
\bibitem{NSTmunro-1-1}\textcolor{black}{G. C. Knee}, K. Kakuyanagi,
M.-C. Yeh, Y. Matsuzaki, H. Toida, H. Yamaguchi, S. Saito, A. J. Leggett
and W. J. Munro, A strict experimental test of macroscopic realism
in a superconducting flux qubit, Nat. Commun. \textbf{7}, 13253 (2016).

\selectlanguage{english}%
\bibitem{experiment-lg-2}\foreignlanguage{australian}{A. Palacios-Laloy,
F. Mallet, F. Nguyen, P. Bertet, Denis Vion, Daniel Esteve and Alexander
N. Korotkov, Experimental violation of a Bell's inequality in time
with weak measurement, Nature Phys. \textbf{6,} 442 (2010).\textcolor{black}{}}

\bibitem{dressel-bell-hybrid}\foreignlanguage{australian}{J. Dressel
and A. N. Korotkov, Avoiding loopholes with hybrid bell-leggett-garg
inequalities, Phys. Rev. A \textbf{89}, 012125 (2014).}

\selectlanguage{australian}%
\bibitem{lgexpphotonweak-1-2}J. Dressel, C. J. Broadbent, J. C. Howell
and A. N. Jordan, Experimental Violation of Two-Party Leggett-Garg
Inequalities with Semiweak Measurements, Phys. Rev. Lett. \textbf{106},
040402 (2011).

\selectlanguage{english}%
\bibitem{goggin-1}\foreignlanguage{australian}{M. E. Goggin, et al.,
Violation of the Leggett-Garg inequality with weak measurements of
photons, Proc. Natl. Acad. Sci. \textbf{\textcolor{black}{108}}, 1256
(2011).}

\selectlanguage{australian}%
\bibitem{massiveosci-1-1-1}A. Asadian, C. Brukner, and P. Rabl, Probing
Macroscopic Realism via Ramsey Correlation Measurements, Phys. Rev.
Lett. \textbf{112}, 190402 (2014).

\bibitem{Mitchell-1-1}C. Budroni, G. Vitagliano, G. Colangelo, R.
J. Sewell, O. Gühne, G. Tóth, and M. W. Mitchell, Quantum Nondemolition
Measurement Enables Macroscopic Leggett-Garg Tests, Phys. Rev. Lett.
\textbf{115}, 200403 (2015).\textcolor{red}{}

\selectlanguage{english}%
\bibitem{lauralg-1}L. Rosales-Z\'arate, B. Opanchuk, Q. Y. He, and
M. D. Reid, \foreignlanguage{australian}{Leggett-Garg tests of macrorealism
for bosonic systems including two-well Bose-Einstein condensates and
atom interferometers, Phys. Rev. A \textbf{97}, 042114 (2018).}

\bibitem{leggett-garg-recent-1}\foreignlanguage{australian}{\textcolor{red}{}}R.
Uola, G. Vitagliano and C. Budroni, Leggett-Garg macrorealism and
the quantum nondisturbance conditions, Phys. Rev. A \textbf{100},
042117 (2019).

\bibitem{halliwell-leggett-garg-double-slit}J. Halliwell, A. Bhatnagar,
E. Ireland, H. Nadeem and V. Wimalaweera, Leggett-Garg tests for macrorealism:
interference experiments and the simple harmonic oscillator, Phys.
Rev. A \textbf{103}, 032218 (2021).

\bibitem{pan-leggett-garg-weak-value-interference-exp}\foreignlanguage{australian}{A.
K. Pan, ``Interference experiment, anomalous weak value, and Leggett-Garg
test of macrorealism'', Phys. Rev. A \textbf{102}, 032206 (2020).}

\bibitem{Dimension-Witness-Brunner}N. Brunner, S. Pironio, A. Acin,
N. Gisin, A. A. M$\acute{e}$thot and V. Scarani, ``Testing the Dimension
of Hilbert Spaces'', \textit{\emph{Phys. Rev. Lett. }}\textbf{100},
210503 (2008).

\bibitem{Quantum-Dimension-Witness}R. Gallego, N. Brunner, C. Hadley
and A. Ac$\acute{\imath}$n, ``Device-Independent Tests of Classical
and Quantum Dimensions'', \textit{\emph{Phys. Rev. Lett. }}\textbf{105},
230501 (2010).

\bibitem{bowles-dimension-test-exp}J. Bowles, M. T. Quintino, and
N. Brunner, Certifying the Dimension of Classical and Quantum Systems
in a Prepare-and-Measure Scenario with Independent Devices, Phys.
Rev. Lett. \textbf{112}, 140407 (2014).

\bibitem{ahrens-exp-dimension-test-nat-phys}J. Ahrens, P. Badzi,
A. Cabello, and M. Bourennane, Experimental device-independent tests
of classical and quantum dimensions, Nat. Phys. \textbf{8}, 592 (2012).

\bibitem{Yu-causal-model-exp}S. Yu, Y.N. Sun, W. Liu, Z.D. Liu, Z.J.
Ke, Y.T. Wang, J.S. Tang, C.F. Li, and G.C. Guo, Realization of a
causal-modeled delayed-choice experiment using single photons, Phys.
Rev. A \textbf{100}, 012115 (2019).\foreignlanguage{australian}{}

\bibitem{laura-decoh-steer-josa}\textcolor{black}{L. Rosales-Zarate,
R. Y. Teh, S. Kiesewetter, A. Brolis, K. Ng, and M. D. Reid, Decoherence
of Einstein--Podolsky--Rosen steering, J. Opt. Soc. Am. B }\textbf{\textcolor{black}{32}}\textcolor{black}{{}
A82 (2015).}\textcolor{red}{}

\bibitem{macro-pointer-interpretation-jphysa}\foreignlanguage{australian}{M.
D. Reid, Interpreting the macroscopic pointer by analysing the elements
of reality of Schrodinger cat, J. Phys. A: Math. Theor. \textbf{50},
41LT01 (2017).}

\bibitem{cats-brune}M. Brune, E. Hagley, J. Dreyer, X. Maître, A.
Maali, C. Wunderlich, J. M. Raimond, and S. Haroche, Observing the
Progressive Decoherence of the \textquotedblleft Meter\textquotedblright{}
in a Quantum Measurement, Phys. Rev. Lett. \textbf{77}, 4887 (1996).

\bibitem{cats-monroe-1}C. Monroe, D. M. Meekhof, B. E. King, D. J.
Wineland, A ``Schrodinger cat'' superposition state of an atom,
Science \textbf{272}, 1131 (1996).

\bibitem{frowis-rmp}F. Fröwis, P. Sekatski, W. Dür, N. Gisin, and
N. Sangouard, Macroscopic quantum states: measures, fragility, and
implementations, Rev. Mod. Phys. \textbf{90}, 025004 (2018).

\bibitem{epr-reid-2}M. D. Reid, Demonstration of the Einstein-Podolsky-Rosen
Paradox using Nondegenerate Parametric Amplification, Phys. Rev. A
\textbf{40}, 913 (1989).

\bibitem{rmpepr-2}M. D. Reid, P. D. Drummond, W. P. Bowen, E. G.
Cavalcanti, P. K. Lam, H. A. Bachor, U. L. Andersen and G. Leuchs,
The Einstein-Podolsky-Rosen paradox: From concepts to applications,
Rev. Mod. Phys. \textbf{81}, 1727 (2009).

\bibitem{bohm-crit}E. G. Cavalcanti, P. D. Drummond, H. A. Bachor
and M. D. Reid, Spin entanglement, decoherence and Bohm\textquoteright s
EPR paradox, Optics Express \textbf{17} (21), 18693 (2009).\foreignlanguage{australian}{}

\bibitem{hw-1-2}H. M. Wiseman, S. J. Jones and A. C. Doherty, \textit{\emph{Steering,
Entanglement, Nonlocality and the Einstein-Podolsky-Rosen Paradox,
Phys. Rev. Lett. }}\textbf{98}, 140402 (2007).

\bibitem{sjonessteer-2-1}S. J. Jones\textcolor{black}{, H. M. Wiseman
and A. Doherty,}\textit{\emph{ Entanglement, Einstein-Podolsky-Rosen
correlations, Bell nonlocality, and steering, Phys. Rev. A}} \textbf{76},
052116 (2007).

\bibitem{eric-2}E. G. Cavalcanti, S. J. Jones, H. M. Wiseman and
M. D. Reid, Experimental criteria for steering and the Einstein-Podolsky-Rosen
paradox, Phys. Rev. A\textbf{ 80}, 032112 (2009).

\bibitem{uola-steer-review}R. Uola, A. C. S. Costa, H. C. Nguyen,
and O. Gühne, Quantum Steering, Rev. Mod. Phys. \textbf{92}, 015001
(2020).

\bibitem{bell-later-theorems}J. Clauser and A. Shimony, Bell's theorem:
experimental tests and implications, Rep. Prog. Phys. \textbf{41},
1881 (1978).

\bibitem{bell-rmp-review}Nicolas Brunner, Daniel Cavalcanti, Stefano
Pironio, Valerio Scarani and Stephanie Wehner, Bell nonlocality, Rev.
Mod. Phys. \textbf{86}, 419 (2014).

\bibitem{teh-noon-epr-steer}\foreignlanguage{australian}{R. Y. Teh,
L. Rosales-Zárate, B. Opanchuk, and M. D. Reid, Signifying the nonlocality
of NOON states using Einstein-Podolsky-Rosen steering inequalities,
Phys. Rev. A \textbf{94}, 042119 (2016).}

\bibitem{herald-noon-1}\foreignlanguage{australian}{S. Slussarenko,
Morgan M. Weston, Helen M. Chrzanowski, Lynden K. Shalm, Varun B.
Verma, Sae Woo Nam \& Geoff J. Pryde, Unconditional violation of the
shot-noise limit in photonic quantum metrology, Nature Photonics \textbf{11},
700 (2017).}

\bibitem{gHZ}D. M. Greenberger, M. A. Horne and A. Zeilinger, in
``Bell\textquoteright s Theorem, Quantum Theory, and Conceptions
of the Universe'' (Kluwer, Dordrecht, 1989), p. 69.

\bibitem{ghz-multi-mdr-wjmunro}M. D. Reid and W. J. Munro, Macroscopic
boson states exhibiting the Greenberger-Horne-Zeilinger contradiction
with local realism, Phys. Rev. Lett. \textbf{69}, 997 (1992).

\bibitem{son-lee-kim-ghz-multi}W. Son, Jinhyoung Lee, and M. S. Kim,
Generic Bell Inequalities for Multipartite Arbitrary Dimensional Systems,
Phys. Rev. Lett. \textbf{96}, 060406 (2006).

\bibitem{irrealism}A. L. O. Bilobran and R. M. Angelo, A measure
of physical reality, Europhys. Lett. \textbf{112}, 40005 (2015).\textcolor{blue}{}

\bibitem{irrealism-fringes}\foreignlanguage{australian}{F. R. Lustosa,
P. R. Dieguez, and I. G. da Paz, Irrealism from fringe visibility
in matter waves double-slit interference with initial contractive
states, Phys. Rev. A \textbf{102}, 052205} (2020).\textcolor{red}{}

\bibitem{Milman-cat}P. Milman, A. Auffeves, F. Yamaguchi, M. Brune,
J. M. Raimond, and S. Haroche, A proposal to test Bell\textquoteright s
inequalities with mesoscopic non-local states in cavity qed, Eur.
Phys. J. D \textbf{32}, 233 (2005).

\bibitem{Leghtas-cat}Z. Leghtas, G. Kirchmair, B. Vlastakis, M. H.
Devoret, R. J. Schoelkopf, and M. Mirrahimi, Deterministic protocol
for mapping a qubit to coherent state superpositions in a cavity,
Phys. Rev. A \textbf{87}, 042315 (2013).
\end{thebibliography}
\end{document}